\documentclass[a4paper,11pt]{article}
\usepackage{jheppubmodnew} % for details on the use of the package, please see the JINST-author-manual
\usepackage[T1]{fontenc} % if needed
\usepackage{physics}
\usepackage{mathrsfs}
\usepackage{float}
\usepackage{wrapfig}
\usepackage{appendix}
\usepackage{graphicx}
\usepackage{xcolor}
\usepackage{amsmath}
\usepackage{mathtools}
\usepackage{simpler-wick}
\usepackage{dsfont}
\usepackage{centernot}
\usepackage{stackengine}
\usepackage{caption}
\usepackage{tcolorbox}
\usepackage{cite}
\stackMath

\newcommand{\be}{\begin{equation}}
\newcommand{\ee}{\end{equation}}

\newcommand{\lm}{\lambda}
\newcommand{\del}{\partial}
\newcommand{\gra}{\vec{\nabla}}

\newcommand{\w}{\omega}
\newcommand{\g}{\delta}
\newcommand{\e}{\epsilon}
\newcommand{\da}{\dagger}
\newcommand{\bs}{\begin{split}}
\newcommand{\es}{\end{split}}

\newcommand{\la}{\langle}
\newcommand{\ra}{\rangle}

\newcommand{\vcp}{\vec{p}}
\newcommand{\vx}{\vec{x}}
\newcommand{\vy}{\vec{y}}

\newcommand{\vk}{\vec{k}}
\newcommand{\vv}{\vec{v}}
\newcommand{\ns}{N_*}

\newcommand{\vp}{\vec{p}}

\newcommand{\mc}{\mathcal{C}}

\newcommand{\mk}{\mathfrak{C}}

\title{\boldmath Mapping Tachyon effective field theory to a subsector of Klein-Gordon theory}

\author[a]{P. V. Athira}
\author[b]{, Ashik H}
 \author[c]{and Priyadarshi Paul}
 \affiliation[a]{Centre for High Energy Physics, Indian Institute of Science,\\
C. V. Raman Avenue, Bangalore 560012, India}
\affiliation[b]{International Center for Theoretical Sciences- Tata Institute of Fundamental Research \\
Bengaluru, India}
\affiliation[c]{
School of Mathematical Sciences, Queen Mary University of London,\\
Mile End Road, London E1 4NS, United Kingdom
}
\emailAdd{athirapv@iisc.ac.in, ashik.h@icts.res.in, priyadarshi.paul@qmul.ac.uk }

\abstract{ On an unstable D-brane, the rolling of the tachyon away from the maximum of its potential is described by time-dependent solutions in string theory. Subsequent analysis leads to an understanding of physics around the tachyon vacuum in terms of an effective field theory. The classical solutions of this effective field theory in the late time limit are in one-to-one correspondence with configurations of non-rotating, non-interacting dust particles. In this work, we map this effective theory near the minimum of the potential to a consistent quantum description using collective field theory methods. The Hilbert space description we obtain in this way consists of a coherent state of particles at rest and excitations on top, i.e., a subsector of Klein-Gordon theory. This is in accordance with known results where one considers a decaying D-brane as providing a time-dependent source for closed strings, and the final closed string state produced is a coherent state. It also suggests, at the quantum level, an equivalence between the open string description and the closed string description regarding the decay of an unstable D-brane. 
}

\begin{document}
\maketitle
\flushbottom

\newpage
\section{Introduction}
\begin{figure}[H]
	\centering
	\includegraphics[scale=0.23]{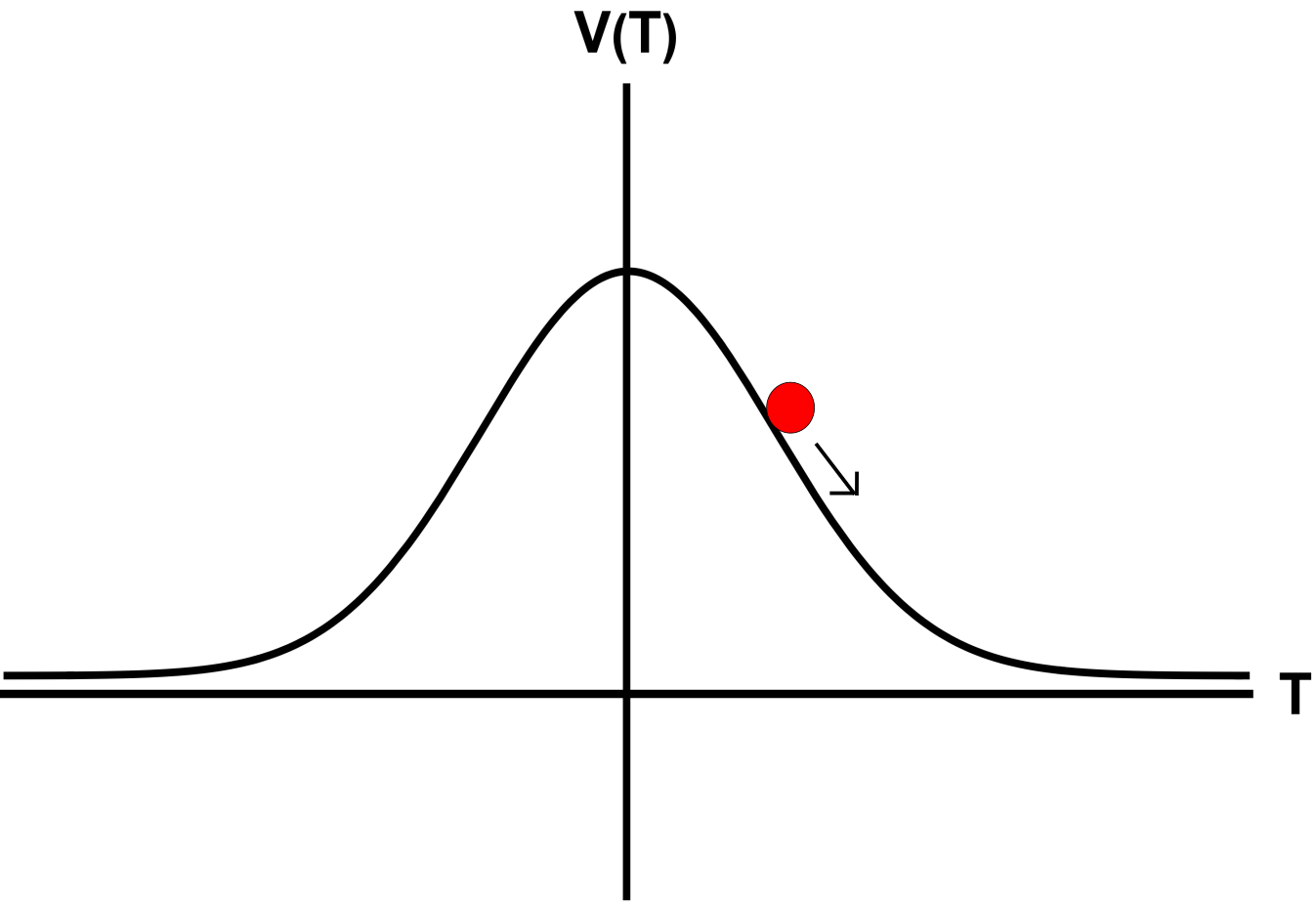}
	\caption{Qualitative form of the tachyon potential $V(T)$: A tachyon displaced away from the maximum will roll towards the minimum. However, the minimum is at $T\rightarrow \infty$, and the tachyon will reach the minimum only asymptotically. }
	\label{fig:rolling_tachyon}
\end{figure}

The dynamics of tachyon on an unstable D-p-brane in string theory can be described by the following effective action \cite{Sen:1999md, Sen:Tachyonic_matter, Sen:review, Garousi:2000tr}.
\be
\label{eq:action_of_tachyon}
S=-\int d^px ~V(T)\sqrt{1+\eta^{\mu\nu}\del_{\mu}T\del_{\nu}T}
\ee
Here $T$ is the tachyon field and $V(T)$ is the potential governing its dynamics. $V(T)$ has a maximum at $T=0$ and approaches zero in the limit $T\rightarrow \infty$ ( Figure \ref{fig:rolling_tachyon}). In this limit, the classical equations of motion obtained from \eqref{eq:action_of_tachyon} reduce to Euler and continuity equations for a relativistic non-interacting dust. Given such a simple classical description, it is natural to ask whether this theory can be quantised.  After all, at the classical level, the system behaves like a collection of particles. One might therefore attempt the usual perturbative quantisation methods used in scalar field theories. However, this approach fails because there are no plane wave solutions around the minimum of the potential. This suggests that a different approach to quantisation is needed. In this paper, we suggest a quantisation procedure based on collective field theory methods. 
Collective field theory arises as a natural candidate because classically the system looks like non-interacting dust, and in collective field theory, one starts with a system of particles and trades the particle degrees of freedom with the collective field. Since we know the quantum description of a collection of free particles, we could understand the quantum mechanics of the collective theory obtained from it. 

The collective theory analysis gives the simple result that the quantum theory of the tachyon effective action near the minimum of the potential describes a subsector of Klein-Gordon theory. More precisely, the Hilbert space consists of a coherent state $|\mc\ra$ of particles at rest and excitations on top of it, $a^{\da}_{\vk_1}a^{\da}_{\vk_2}....a^{\da}_{\vk_n}|\mc\ra$. This is schematically shown in figure \ref{fig:kg_vs_tachyon}.

One would have naturally expected some relation with the Klein-Gordon theory because its Fock space contains states having a large number of relativistic particles. However, a more precise understanding of the result involves concepts from string theory. There are two equivalent descriptions at play here. One is the open-string theoretic description \cite{Sen:rolling} and another is a closed string description \cite{Lambert:2003zr, Chen_2002, Rey:2003xs}. One gets the tachyon effective theory in the open-string side, whereas in the closed-string side, we get a radiation of massive closed strings, which can be described using Klein-Gordon theory. Earlier works have established classical agreement between the physics of these two descriptions \cite{Sen:2003matrix, Sen:open_closed_tree}. Our results advance this map and point in the direction of a quantum equivalence between open and closed descriptions.
\begin{figure}[H]
	\centering
	\includegraphics[scale=0.35]{images/kg_vs_tt_spectrum2.png}
	\caption{The spectrum in Klein-Gordon theory has a vacuum state $|0\ra$ and excitations on top. On the other hand, the Hilbert space of tachyon theory has a coherent state $|\mathcal{C}\ra$ and excitations on top.}
	\label{fig:kg_vs_tachyon}
\end{figure}

 To be more specific, note that an unstable D-p brane in $25+1$ dimensional bosonic string theory or $9+1$ dimensional superstring theory has a tachyonic mode in the open-string spectrum \cite{Sen:review}. This is due to the tachyon field ($T$) being at the maximum of the potential $V(T)$. One can displace the tachyon from the maximum of the potential and let it \emph{roll towards the minimum} of the potential \footnote{In case of bosonic string theory, $V(T)$ is not symmetric about $T=0$. Rolling it in the direction without the minimum will result in a singularity. Our analysis is only applicable to the case when tachyon rolls towards the minimum}. In a tree-level analysis, such a scenario is captured by time-dependent solutions in open-string theory \cite{Sen:rolling}. Here, one observes that the energy density of the system - which includes the D-brane tension as well as the kinetic and potential energies of the tachyon - stays constant during time evolution. However, the pressure goes to zero asymptotically ($x^0\rightarrow \infty$). Thus, in the late time limit, the system behaves like pressureless dust. 
 
 It is important to note that the tachyon approaches the minimum of the potential only asymptotically. The vacuum of the theory corresponds to a zero energy, zero pressure \emph{time independent} configuration where the tachyon is at rest at the minimum of the potential. This corresponds to a system without any D-branes, and hence physical open-string degrees of freedom are absent \cite{Sen:rolling}. This, in the effective field theory, translates to the statement that there are no plane wave solutions around the minimum of the potential \cite{Sen:Tachyonic_matter}. The action given in \eqref{eq:action_of_tachyon} is not obtained from a direct string theory calculation; it is an effective description that captures the physics of a rolling tachyon \cite{Sen:1999md, Sen:Tachyonic_matter, Garousi:2000tr, Lambert:2001fa, Lambert:2002hk, Kluson:2000iy, Kutasov:2003er}.

In the closed string side, one can turn on $g_s$, and study the closed strings resulting from the decay of the unstable D-brane  \cite{Lambert:2003zr, Klebanov:2003km, McGreevy:2003kb, Gaiotto:2003rm, Rey:2003xs}. It turns out that the D brane completely decays away, leaving its energy into closed strings, which are mostly localised along the plane of the original D brane. Also, most of the energy of the D brane gets transferred to closed string states of mass $\sim g_s^{-1}$. One can see that the decaying D-brane acts like a time-dependent source for closed strings, and the final state obtained in this process is a coherent state \cite{Klebanov:2003km}. We note that these closed string states follow the relativistic dispersion relation $E=\sqrt{p^2+m^2}$. So an effective field-theoretic description could be given by Klein-Gordon theories having different masses. 

Classically, one could see similarities between the two descriptions. We get pressureless dust in the late time limit from open-string theory and a closed string radiation localised along the plane of the original D-brane from the closed string analysis. We are advancing this further in this work. The final closed string state obtained from decaying a D-brane is a coherent state of the closed string mode \cite{Lambert:2003zr}. What we obtain from quantising the tachyon effective theory using collective theory methods is also a coherent state $|\mc\ra$ (and excitations on top of it). \emph{This suggests a duality between the open and closed string theory descriptions} \cite{Sen:open_closed_tree, Sen:2003matrix, Yee:2004ec}.

The paper is organised as follows,
\begin{itemize}
	\item In section \ref{sec:tachyon_effective_theory}, we introduce the Tachyon effective action and describe its classical dynamics near the minimum of the potential. We see that the classical dynamics of a tachyon are like non-interacting, non-rotating dust. Analysing the classical equations of motion, we further establish the absence of plane wave solutions around the minimum.
	\item In section \ref{sec:collective_theory}, we review the collective field theory methods for a collection of free particles. We first discuss it in the simple context of non-relativistic particles in \ref{sec:non_rel_collective_theory} and then generalise the results to the relativistic case in \ref{sec:rel_collective_theory}. 
	\item In section \ref{sec:collective_to_tachyon} we use collective theory methods to quantise the tachyon effective field theory. The Hilbert space description of the Tachyon theory is obtained and is compared with states in a Klein-Gordon theory.
	\item We conclude with discussions and future directions in \ref{sec:discussion}.
\end{itemize}

\section{Tachyon effective theory}
\label{sec:tachyon_effective_theory}
In this section, we briefly review the tachyon effective action proposed in \cite{Sen:Time_Tachyon}, which describes the classical dynamics of the tachyon field. We will also explain the absence of plane wave solutions around the vacuum.
\subsection{Classical dynamics}
\label{sec:tachyon_dynamics}
Following \cite{Sen:Time_Tachyon}, the tachyon field is described by the effective action,
\be
\label{eq:T_action}
S=-\int~d^{p+1}x~V(T)\sqrt{1+\eta^{\mu\nu}\del_{\mu}T\del_{\nu}T}
\ee
The potential $V(T)$ has a maximum at $T=0$ and decays exponentially for large $T$. Near the minimum of the potential at $T\to\infty$, it behaves as $V(T)=e^{-{\alpha T\over 2}}$ where $\alpha=1$ for bosonic string theory and $\alpha=\sqrt{2}$ for superstring theory. The exponential form of the potential is chosen so that the late time dynamics of the tachyon field reproduce the expected behaviour of a rolling tachyon \cite{Sen:Tachyonic_matter, Sen:rolling, Minahan_2002, Sugimoto_2002}. In particular, as the tachyon rolls towards the minimum of the potential, the pressure approaches zero while the energy density remains finite. Due to this constant energy density, one can see that this asymptotic limit is different from the minimum energy configuration where the tachyon is at rest at the minimum of the potential with zero energy and pressure.
Therefore, in all the analysis that follows, even though we work in the large $T$ limit, we never truly reach the vacuum solution.

To study the classical dynamics in the $T\rightarrow \infty$ limit, it is convenient to use the Hamiltonian formalism. Since the Lagrangian is proportional to the potential $V(T)$, which vanishes in this limit, the Lagrangian description becomes less useful. The Hamiltonian formulation, however, remains well defined and allows us to analyze the dynamics more easily. 

The momentum conjugate to $T$ is given by
\begin{equation}
\Pi(x)=\frac{\delta S}{\delta \partial_0 T(x)}
      =\frac{V(T)\partial_0 T}{\sqrt{1+(\partial T)^2}},
\end{equation}
which leads to the Hamiltonian
\begin{equation}
H=\int d^p x\,\sqrt{\Pi^2+V(T)^2}\sqrt{1+(\nabla T)^2}.
\end{equation}
In the $T\rightarrow\infty$ limit, near the minimum of the potential,  the $V(T)^2$ term can be neglected, and the Hamiltonian reduces to
\begin{equation}
\label{eq:H_tachyon}
H=\int d^p x\, |\Pi(x)|\sqrt{1+(\nabla T)^2}.
\end{equation}
In what follows we drop the absolute value sign and assume $\Pi>0$. 
The Hamilton's equations give
\be
\begin{split}
 	&\del_0T=\sqrt{1+(\gra T)^2},
 	\\& \del_0\Pi(x)=\gra\cdot \left({\Pi(x)\gra T\over \sqrt{1+(\gra T)^2}}\right).
\end{split} 
\ee

Now, let us define $u_{\mu}=-\partial_{\mu}T$ and $\epsilon(x)=\frac{\Pi(x)}{\sqrt{1+(\nabla T)^2}}$. 
In terms of these variables, the equations of motion take the form
\be
u^2=-1, \qquad \partial_\mu \left(\epsilon(x) u^\mu \right)=0 .
\ee
These are precisely the equations governing a relativistic, non-rotating, non-interacting dust with four-velocity $u^\mu$ and energy density $\epsilon(x)$. This implies that the classical dynamics of the tachyon field near the minimum of the potential is that of pressureless dust, where $\partial_\mu T$ plays the role of the velocity field, and $\Pi(x)$ corresponds to the local energy density\footnote{This physical interpretation of the field $\Pi$ will become important later when we try to quantise the system using collective theory methods }. Although the classical dynamics becomes simple in this limit, the analysis of quantum fluctuations around this background is more subtle. In particular, the usual perturbative quantisation breaks down, as we now discuss.

\subsection{Failure of perturbative quantization}
\label{sec:failure_of_pert}
Given the tachyon effective action and its simple classical dynamics, one might expect that the usual perturbative quantization can be carried out. However, as was shown in \cite{Sen:Tachyonic_matter}, the absence of plane wave solutions around the minima of the potential implies that such a perturbative quantisation is not possible. We summarize this property below. 
Expanding the square-root in \eqref{eq:T_action}, we get 
\be
S = -\int d^{p+1}x~e^{-{\alpha T\over 2}}\left(1+{\eta^{\mu\nu}\over 2}\del_{\mu}T\del_{\nu}T+....\right)
\ee
Now, defining a new field $\phi=e^{-{\alpha T\over 4}}$, the above expression reduces to 
\be
S = \int d^{p+1}x~{16\over \alpha^2} \left( -{\eta^{\mu\nu}\over 2}\del_{\mu}\phi\del_{\nu}\phi-{\alpha^2\over 16}\phi^2+.... \right)
\ee
The minimum of the potential is at $\phi = 0$.
By ignoring the higher derivative terms, we see that $\phi=a~e^{ik\cdot x}$ is a solution of the equation of motion, with $k^2=-{\alpha^2\over 8}$. However, this ceases to be a solution when one takes into account all the higher derivative corrections. To see this, we rewrite the action in terms of $\phi$ without expanding the square root,
\be
\begin{split}
	S=-\int d^{p+1}x \phi^2 \sqrt{1+{16\over \alpha^2}\phi^{-2}\eta^{\mu\nu}\del_{\mu}\phi\del_{\nu}\phi}
\end{split}
\ee
The corresponding equation of motion is
\be
\del_{\mu}\left({\del^{\mu}\phi\over \sqrt{1+{16\over \alpha^2}\phi^{-2} \eta^{\mu\nu}\del_{\mu}\phi\del_{\nu}\phi }}\right)={\alpha^2\over 8}\left({\phi+{8\over \alpha^2 }\phi^{-1} {\eta^{\mu\nu}\del_{\mu}\phi\del_{\nu}\phi }\over \sqrt{1+{16\over \alpha^2}\phi^{-2} \eta^{\mu\nu}\del_{\mu}\phi\del_{\nu}\phi }}\right)
\ee 
Substituting $\phi=a e^{ik\cdot x}$ gives
\be
{a\over \sqrt{1-{16\over \alpha^2} k^2 }}=0
\ee
This is not satisfied by any choice of $k^{\mu}$  and therefore we must have $a=0$. Thus, there are no plane-wave solutions around the minimum of the potential. This result agrees with the expected properties of the tachyon vacuum. The vacuum corresponds to the configuration where the tachyon is at rest at the minima of the potential, and the system has zero energy and zero pressure. This means the absence of D-branes and 
physical open string degrees of freedom. 

Since perturbative quantization relies on expanding the field in terms of plane-wave modes, the usual perturbative approach cannot be applied. This motivates the search for an alternative quantization method. In the following, we will use insights from collective field theory to quantize the tachyon effective theory.

\section{Collective field theory}
\label{sec:collective_theory}
The basic idea of collective field theory is to trade discrete quantum mechanical degrees of freedom with a continuous field degree of freedom. Works by B.Sakita, A. Jevicki, and collaborators \cite{Gervais:1975yg, Jevicki:1979mb, Jevicki:1980zg} are good resources for collective field theory concepts relevant to the context we are working on. We will see that collectivising a system of free particles gives an alternative quantum description in terms of field variables, which in the $\hbar\rightarrow0$ limit will also give the classical equations of motion governing pressureless dust.  We will briefly review the collective field description for a collection of $N_*$ free non-relativistic particles, as this will make some of the analysis coneptually easier to grasp. We will then generalise it to relativistic particles. 

\subsection{Non-relativistic free particles}
\label{sec:non_rel_collective_theory}
Consider a system of $N_*$ free non-interacting non-relativistic particles in $p$ spatial dimensions, with the  Hamiltonian,
\be
\label{eq:H_QM_non_rel}
H=\sum_{i=1}^{N_*} {\vcp_i^{~2}\over 2m}
\ee
In the usual particle description, the dynamics of this system is written in terms of the phase space coordinates $\{\vx_i,\vp_i\}$, where $\{\vx_i;i=1,2,..,N_*\}$ denotes the positions of the particles and $\vcp_i=-i\hbar \gra_{\vx_i}$ is the conjugate momenta in the position space.  In the collective description, instead of working with the particle coordinates  $\{\vx_i,\vcp_i\}$ we describe the system using a field $\phi(\vec{y})$ and its conjugate momentum $\pi(\vec{y})$, where $\pi(\vec{y})=-i\hbar\,\delta/\delta\phi(\vec{y})$. The quantum theory is then formulated in terms of these fields. In particular, the Hamiltonian and the Hilbert space will be expressed in terms of these variables. Following \cite{Gervais:1975yg, Jevicki:1979mb, Jevicki:1980zg}, the field is related to the particle positions through
\be
\label{eq:phi_def}
\phi(\vy)=m\sum_{i=1}^{N_*}\g(\vy-\vx_i)
\ee
\emph{However, this relation should not be viewed as a strict definition of $\phi(\vy)$.} In the collective description, $\phi(\vy)$ is treated as an independent field variable. The relation above is instead implemented through a non-trivial Jacobian that appears in the inner product of the field theory. We will see how this arises below.

The Jacobian is defined such that the quantum mechanical inner product between two QM states $\psi_1$ and $\psi_2$ is equal to the field theory inner product between corresponding wavefunctionals $\Phi_1[\phi]$ and $\Phi_2[\phi]$.
\be
\label{eq:QM_non_rel_innerpdt}
\begin{split}
	\left(\psi_1,\psi_2\right)&=\int \left(\prod_{i=1}^Nd^p\vx_i\right) \psi_1^*\left(\vx_1,..,\vx_{N_*}\right)\psi_2\left(\vx_1,..,\vx_{N_*}\right)
	\\&=\int \left(\prod_{i=1}^Nd^p\vx_i\right) \left(\int D\phi~ \g\left[\phi(\vy)-m\sum_i\g(\vy-\vx_i)\right]\right) \Phi_1^*[\phi]\Phi_2[\phi]
	\\&=\int D\phi~J_{\ns}[\phi] \Phi_1^*[\phi]\Phi_2[\phi]
\end{split}
\ee 
The steps performed above require some explanation. The first equality is the usual QM inner-product for an $\ns$ particle system. In the second line, we inserted an identity 
\begin{equation}
\label{eq:identity_Jacobian}
1=\int D\phi~ \g\left[\phi(\vy)-m\sum_i\g(\vy-\vx_i)\right].
\end{equation}
The wavefunctionals $\Phi_1$ and $\Phi_2$ are defined by demanding that enforcing \eqref{eq:phi_def} on $\Phi_1$ and $\Phi_2$ should give $\psi_1$ and $\psi_2$.
\begin{equation}
	\label{eq:def_Phi_1_Phi_2}
	\begin{split}
		&\Phi_1[\phi]\bigg|_{\phi(\vy)=m\sum_{i=1}^{\ns}\g(\vy-\vx_i)}=\psi_1(\vx_1,\vx_2,...,\vx_{\ns})
		\\&\Phi_2[\phi]\bigg|_{\phi(\vy)=m\sum_{i=1}^{\ns}\g(\vy-\vx_i)}=\psi_2(\vx_1,\vx_2,...,\vx_{\ns})
	\end{split}
\end{equation}
Note that there might be different wavefunctionals $(\Phi)$ which will reduce to the same QM wavefunction $(\psi)$ under \eqref{eq:phi_def}. So, one could say that the map from the collective theory state to the quantum mechanics state is many-to-one. We will comment more on this later. 

The final equality in \eqref{eq:QM_non_rel_innerpdt} defines the Jacobian as,
\be
J_{\ns}[\phi]=\int \left(\prod_{i=1}^Nd^p x_i\right) \g\left[\phi[\vy]-m\sum_{i=1}^{\ns}(\vy-\vx_i)\right].
\ee
For computations, it is convenient to write this Jacobian as,
\be
\label{eq:non_rel_J}
J_{\ns}[\phi]=\int D\lm ~e^{i\int d^py~\lm(\vy)\phi(\vy)} \left(\int d^px~ e^{-im\lm(\vx)}\right)^{\ns}
\ee
by introducing an auxiliary field $\lambda(y)$, which allows the functional delta function to be written in Fourier form.

Thus, in the collective description, the quantum states are represented by wavefunctionals $\Phi[\phi]$, and the inner product between these states is defined using the Jacobian that encodes the relation between $\phi$ and the particle coordinates. With this setup, we now analyze the Hamiltonian in terms of the collective fields and study the resulting eigenstates.  We will also compute the inner product between these eigenstates and check that it agrees with the result obtained in the original quantum mechanical description. 
\subsubsection{Hamiltonian and eigenstates}
In this section, we rewrite the Hamiltonian \eqref{eq:H_QM_non_rel}, in terms of the collective fields $\phi$ and $\pi$, and study the corresponding eigenstates and the spectrum of the theory. 

It can be shown, using \eqref{eq:phi_def} and application of the chain rule of differentiation, that operators invariant under exchanging any two particles could be written in terms of the collective field and its conjugate. So, the Hamiltonian could be written in terms of $\phi$ and $\pi$. We will first show how to write $\sum_{i=1}^{\ns}\vp_i$ in terms of $\phi$ and $\pi$. Generalising this method to $H=\sum_{i=1}^{\ns}{\vp_i^2\over 2m}$ is straight forward and we derive this in the Appendix \ref{app:non_rel_collective}.
 \be
 \label{eq:sum_p_i_der}
\begin{split}
	\sum_i \vcp_i&=-i\hbar \sum_i \gra_{\vx_i}
	\\&=-i\hbar \sum_i \int d^py \left(\gra_{\vx_i} \phi(\vy)\right) {\g \over \g \phi(\vy)} 
	\\&=-i\hbar\sum_i \int d^py ~\gra_{\vx_i} m\g(\vy-\vx_i) {\g\over \g \phi(\vy)}
	\\&=-i\hbar\int d^py \left(\sum_i m\g(\vy-\vx_i)\right) \gra_{\vy}{\g\over \g\phi(\vy) }
	\\&=\int d^py~\phi(\vy)  \gra_{\vy} \pi(\vy)
\end{split}
\ee
In going to the second line, we applied the chain rule to rewrite derivatives with respect to $x_i$ in terms of the collective variable $\phi(\vy)$. The third line follows from the definition of $\phi(\vy)$ given in  \eqref{eq:phi_def}, which will be imposed by the Jacobian while computing any matrix element. In the last step, we again used the defining expressions for $\phi(\vy)$ and $\pi(\vy)$. 

One should be cautious about the derivation given in \eqref{eq:sum_p_i_der}. Here we used the chain rule to go from $\{\vx_i;~i=1,2,..,\ns\}$, having finite degrees of freedom, to $\phi(\vy)$, which has infinite degrees of freedom. Strictly speaking, such a change of variables is ill-defined. A more precise derivation is given in Appendix \ref{app:field_theory_ops}\textcolor{blue}{.0}, where we start from the quantum mechanical expression for $\left(\psi_1,\left(\sum_{i=1}^{\ns} \vp_i\right)~\psi_2\right)$ and then use \eqref{eq:identity_Jacobian} to get to a field theoretic expression. The final answer will be the same as what we got in \eqref{eq:sum_p_i_der} using the naive chain rule for differentiation.

Similar manipulations can be performed to get the expression for $\sum_i p_i^2$, in terms of $\phi$ and $\pi$. So, the Hamiltonian is obtained as,
 \be
 \label{eq:H_non_rel}
 H=\sum_{i=1}^{N_*}{p_i^2\over 2m}={1\over 2} \int d^{p}y~\phi(\vy) \left(\gra \pi(\vy)\right)^2-{i\hbar \over 2m} \int d^{p}y~\phi(\vy) \gra^2 \pi(\vy)
 \ee
Note that the second term is linear in $\pi$, and therefore carries an explicit factor of $\hbar$. It may thus be interpreted as a quantum correction. In particular, if this term is neglected, the resulting equations of motion reduce to those of a fluid. We will show this explicitly in section \ref{sec:classical_limit_nonrel}.

Given the Hamiltonian in the collective description, we now determine its energy eigenfunctionals and the corresponding spectrum. Using the relation \eqref{eq:def_Phi_1_Phi_2}, which relates the quantum mechanical wavefunctions $\psi$ to the collective theory wavefunctionals $\Phi$, we construct the corresponding states in the collective description. The simplest case to consider is $N_*=1$, i.e., the one-particle quantum mechanical system. (From now on, we set $\hbar=1$, restoring it whenever necessary.)

For single-particle QM, the energy eigenstate is just a plane wave,
\be
\label{eq:QM_1_particle_eigenstate}
\begin{split}
&\psi_{\vk}(\vx)=e^{i\vk\cdot\vx}
\\&H\psi_{\vk}(\vx)={\vk^2\over 2m} ~\psi_{\vk}(\vx)
\end{split}
\ee
Following the reasoning explained earlier, the corresponding wavefunctional is obtained by requiring that substituting the definition of the collective field \eqref{eq:phi_def} reproduces the original quantum-mechanical wavefunction. This leads to
\be
\label{eq:CT_1_particle_eigenstate}
\Phi_{\vk}[\phi]={1\over m}\int d^py~e^{i\vk\cdot\vy}\phi(\vy)
\ee
One can see that substituting $\phi(\vy)=m\g(\vy-\vx)$ will give the QM energy eigenstate \eqref{eq:QM_1_particle_eigenstate} as required.  One can also check that \eqref{eq:CT_1_particle_eigenstate} is an eigenstate of the Hamiltonian.
\be
\begin{split}
H~\Phi_{\vk}[\phi]&=-{1\over 2m }\int d^py~' \phi(\vy~') \gra_{\vy~'}^2 {\g\over \g \phi(\vy~')}~{1\over m}\int d^py~e^{i\vk\cdot\vy}\phi(\vy)
\\&=-{1\over 2m^2}\int d^py'd^py e^{i\vk\cdot\vy} \phi(\vy~') \gra_{\vy~'}^2 \g(\vy~'-\vy)
\\&={\vk^2\over 2m}~\Phi_{\vk}[\phi]
\end{split}
\ee
Here, we did an integration by parts to get the last step. This simple example makes the key idea involved in finding the eigenstates clear. We will later introduce a systematic way to write down energy eigenstates by means of a creation operator. For the purposes of this section, we will work with a small value for $\ns$, and this is enough to understand the main concepts involved.
We now turn to the case $\ns=2$. The two-particle eigenfunction is
\be
\begin{split}
	\psi_{(\vk_1,\vk_2)}(\vx_1,\vx_2)=e^{i\vk_1\cdot\vx_1}e^{i\vk_2\cdot\vx_2}+\vx_1\leftrightarrow \vx_2	
\end{split}
\ee
with, 
\be
\begin{split}
	 H \psi_{(\vk_1,\vk_2)}={\vk_1^2+\vk_2^2\over 2m} \psi_{(\vk_1,\vk_2)}
\end{split}
\ee
Using the same prescription, the corresponding collective wavefunctional becomes
\be
\label{eq:CT_2_particle_state}
\begin{split}
	 \Phi_{(\vk_1,\vk_2)}[\phi]={1\over m^2}\int d^py_1d^py_2~e^{i\vk_1\cdot\vy_1}e^{i\vk_2\cdot\vy_2} \phi(\vy_1)\phi(\vy_2)-\frac{1}{m}\int d^py ~e^{i(\vk_1+\vk_2)\cdot \vy} \phi(\vy)
\end{split}
\ee
and one can verify that
\be
\begin{split}
	 H \Phi_{(\vk_1,\vk_2)}[\phi]={\vk_1^2+\vk_2^2\over 2m} \Phi_{(\vk_1,\vk_2)}[\phi]
\end{split}
\ee
For simplicity, we have omitted the conventional normalization factor ${1\over \sqrt{2}}$. One can verify the last equation given above by the usual techniques of functional differentiation. Have a look at Appendix \ref{app:non_rel_collective} for detailed calculations. 

Alternatively, one could directly solve the time-independent Schrödinger equation for the collective theory Hamiltonian. This is analogous to the Schrödinger picture treatment of the Klein-Gordon theory, and it leads to the same energy eigenstates. The details are given in Appendix \ref{app:direct_solve_hamiltonian}.

\subsubsection{Computing Inner product}
It remains to check that the inner product in the collective theory reproduces the usual quantum-mechanical result. One can verify that this is indeed the case. We will see that \eqref{eq:QM_non_rel_innerpdt} holds in general as expected. We will start with the simple case of $\ns=1$, where the QM inner product gives. 
\be
\left( \psi_{\vk_1},\psi_{\vk_2} \right)=(2\pi)^p\g(\vk_1-\vk_2)
\ee
One can use the explicit form of the Jacobian given in  \eqref{eq:non_rel_J} to compute the field theory inner product as follows, 
\be
\label{eq:inner_pdt_one_particle}
\begin{split}
\left(\Phi_{\vk_1},\Phi_{\vk_2}\right)_1&=\int D\phi~J_1[\phi] \Phi_{\vk_1}^*[\phi]\Phi_{\vk_2}[\phi]
\\&={1\over m^2}\int d^py_1d^py_2e^{-i\vk_1\cdot\vy_1}e^{i\vk_2\cdot\vy_2}~\left(\int D\phi~J_{1}[\phi]\phi(\vy_1)\phi(\vy_2)\right)
\end{split}
\ee
The subscript $"1"$ in the LHS is to denote that we are in $\ns=1$. The term in the bracket can be computed easily. We will do it for general $\ns$ as it will be useful later. 
\be
\label{eq:2_pt_fn}
\begin{split}
	\int D\phi ~J_{\ns}[\phi]\phi(y_1)\phi(y_2)&=\int D\phi D\lm e^{i\int \lm \phi} \left(\int d^px e^{-im\lm(\vx)}\right)^{\ns} \phi(\vy_1)\phi(\vy_2)
	\\&=\int D\lm D\phi \left(-{\g^2\over \g \lm(\vy_1)\g\lm(\vy_2)} e^{i\int d^py ~\lm(\vy)\phi(\vy)}\right)
	\left(\int d^px e^{-im\lm(\vx)}\right)^{\ns}
	\\&=-\int D\lm {\g^2\g \left[ \lm \right]\over \g \lm(\vy_1)\g\lm(\vy_2)} 	\left(\int d^px e^{-im\lm(\vx)}\right)^{\ns}
	\\&=m^2\ns(\ns-1)V^{\ns-2}+m^2\ns V^{\ns-1} \g^p(\vy_1-\vy_2)
\end{split}
\ee
In the last step, we did an integration by parts, followed by performing the $\lm$ integral using the $\g$ functional. $V$ is the spatial volume of the $D-p$ brane ($V=\int d^p y $).  Substituting \eqref{eq:2_pt_fn} for $\ns=1$ in \eqref{eq:inner_pdt_one_particle} gives,
\be
\left(\Phi_{\vk_1},\Phi_{\vk_2}\right)_1=(2\pi)^p\g(\vk_1-\vk_2)
\ee 
We see that this is exactly the QM inner product. This is expected from the equality of inner products in collective theory and quantum mechanics \eqref{eq:QM_non_rel_innerpdt} which arises from the definition of the Jacobian. 

One can also perform similar computations for $\ns=2$ states given in \eqref{eq:CT_2_particle_state}  and arrive at the same conclusion (See \ref{app:inner_pdt_computation} for calculations). 
\be
\begin{split}
	& \left(\psi_{(\vk'_1,\vk'_2)},\psi_{(\vk_1,\vk_2)}\right)=2\{2\pi \g(\vk_1-\vk'_1) 2\pi \g(\vk_2-\vk'_2)+\vk_1\leftrightarrow \vk_2 \}
	\\&\left( \Phi_{(\vk'_1,\vk'_2)}, \Phi_{(\vk_1,\vk_2)} \right)_2=2\{2\pi \g(\vk_1-\vk'_1) 2\pi \g(\vk_2-\vk'_2)+\vk_1\leftrightarrow \vk_2 \}
\end{split}
\ee
The results we have discussed so far suggest that collective field theory is effectively capturing $\ns$-particle quantum mechanics. However, trading finite degrees of freedom in QM to infinite degrees of freedom of a field in collective theory means we expect some sort of many-to-one mapping. This is indeed the case, but the Jacobian constraints the field by imposing \eqref{eq:phi_def} such that the Hilbert space structure is the same as that of $\ns$-particle quantum mechanics. 

\subsubsection{Hilbert space structure}
Having discussed the Hamiltonian, the eigenstates, and the inner product, we now describe the Hilbert space structure of the collective theory.

Consider collective field theory with $\ns=1$. We saw in the previous section that the one-particle energy eigenstates in quantum mechanics can be mapped to the eigenfunctionals $\Phi_{\vk}$ defined in \eqref{eq:CT_1_particle_eigenstate}. However, the collective theory also admits states such as $\Phi_{\vk_1,\vk_2}$. Since the Hamiltonian does not depend on $\ns$, even in the $\ns=1$ sector the state $\Phi_{\vk_1,\vk_2}$ is an eigenstate of the Hamiltonian with eigenvalue $(\vk_1^2+\vk_2^2)/2m$. Thus, the interpretation of such states within the $\ns=1$ theory requires some clarification.
These states are in fact null states of the $\ns=1$ theory. To see this, note that the Jacobian $J_1$ imposes $\phi(\vy)=m\g(\vy-\vx)$, and substituting this in the expression for $\Phi_{\vk_1,\vk_2}$ evaluates to zero.
\be
\begin{split}
	\Phi_{\vk_1,\vk_2}&={1\over m^2}\int d^py_1d^py_2~e^{i\vk_1\cdot\vy_1+i\vk_2\cdot\vy_2}~\phi(\vy_1)\phi(\vy_2)-{1\over m}\int d^py~e^{i(\vk_1+\vk_2)\cdot \vy}\phi(\vy)
	\\&={1\over m^2}\int d^py_1d^py_2 e^{i\vk_1\cdot\vy_1+i\vk_2\cdot\vy_2} \phi(\vy_1)\bigg( \phi(\vy_2)-m\g(\vy_2-\vy_1)\bigg)
	\\&\stackrel{\ns=1}{=}{1\over m^2}\int d^py_1d^py_2 e^{i\vk_1\cdot\vy_1+i\vk_2\cdot\vy_2}~m \g(\vy_1-\vx)\bigg( m\g(\vy_2-\vx)-m\g(\vy_2-\vy_1)\bigg)
	\\&=0
\end{split}
\ee
Here the superscript $\ns=1$ above the 3rd equality means that we are imposing $\phi(\vy)=m\g(\vy-\vx)$. One can also show this using the Jacobian explicitly, and this is given in the Appendix \ref{app:higher_states_null}. The statement on the field theory side is that the inner product of $\Phi_{\vk_1,\vk_2}$ with any other state is zero, and hence the state is null.
\be
\left(\Phi,\Phi_{\vk_1,\vk_2}\right)_1=0~\forall~\Phi\Rightarrow \Phi_{\vk_1,\vk_2}=0
\ee
Similarly, one can show that the 3-insertion energy eigenstate $\Phi_{\vk_1,\vk_2,\vk_3}$ is also a null state. The result can be generalized.
Thus, all higher-insertion states are null in the $\ns=1$ theory. The only remaining possibility is the zero-insertion state. Such a state will just be a constant, independent of $\phi$. We choose this constant to be unity, as it is just a matter of normalization. \emph{This state will have important applications later, so we denote it by $\mc$, ie  $\mc[\phi]=1$.} 

It turns out that the state $\mc$ in the $\ns=1$ collective theory could be interpreted as a 1-particle state in QM with zero momenta. This also follows from the fact that the Jacobian enforces $\phi(\vy)=m\g(\vy-x)$. One can arrive at this result from a field theory perspective by noting that the Jacobian makes sure that the operator $\int d^py ~\phi(\vy)$ is equivalent to $\ns m$ (See \ref{app:proof_int_phi_nm}). 
\be
\Phi_{\vk=0}={1\over m}\int d^py~\phi(\vy)
\stackrel{\ns=1}{=}1
\ee
This tells you that the state $\Phi_{\vk=0}$ is the same as $\mc$! This is an instance of the many-to-one mapping we mentioned in earlier sections. 

These results point in the direction that the Hilbert space of $\ns=1$ collective field theory is the same as the $1$-particle QM Hilbert space denoted by $\mathcal{H}_1$. The energy eigenstates in QM are in one-to-one correspondence with the states $\Phi_{\vk}$. Naively, one might think that the collective theory has more (and different) states than QM, but the Jacobian constrains the field $\phi$ such that we get a one-particle Hilbert space effectively.  

Similar results go through for $\ns=2$ collective theory also. The states $\Phi_{\vk_1,\vk_2,..,\vk_n}$ for $n>2$ are all null states, whereas the state $\Phi_{\vk}$ should be identifield with $\Phi_{\vk,0}$ and the state $\mc$ should be identified with $\Phi_{\vk_1=0,\vk_2=0}$. These results can be generalised to any value of $\ns$.

We conclude this section by summarizing the main results:
\begin{itemize}
	\item Quantum Mechanics of $\ns$ number of particles can be reformulated in terms of the collective field $\phi$ and its conjugate momenta $\pi$.
	\item  Although the field $\phi$ appears to have infinitely many degrees of freedom, the nontrivial Jacobian in the inner product effectively constrains it so that the theory remains equivalent to ordinary quantum mechanics. 
    \item Hilbert space of collective theory is the same as that of $\ns$-particle QM, ie, $\mathcal{H}_{\ns}$.  However, the map from collective theory to QM is many-to-one. The Hilbert space is spanned by states $\Phi_{\vk_1,\vk_2,..,\vk_{\ns}}$ which are $\ns$-particle energy eigenstates with momenta $\vk_1,\vk_2,...,\vk_{\ns}$. Energy eigenstates having $\ns-s$ insertions ($\Phi_{\vk_1,\vk_2,...,\vk_{\ns-s}}$) are equivalent to $\ns$-insertion state ($\Phi_{\vk_1,\vk_2,..,\vk_{\ns}}$ ) with $s$ of the momenta put to zero whereas states having more than $\ns$ insertions ($\Phi_{\vk_1,\vk_2,..,\vk_{n}};~n>\ns$ ) are null.
\end{itemize}
Hence, the collective description provides an alternative quantum formulation of the system in terms of field variables. We will now show that, in the $\hbar \to 0$ limit, this description also gives the classical equations of motion of non-relativistic pressureless dust.
\subsubsection{Classical limit}
\label{sec:classical_limit_nonrel}
We now briefly examine the classical limit of the non-relativistic collective Hamiltonian given in \eqref{eq:H_non_rel}. In the limit $\hbar\rightarrow0$, the Hamiltonian reduces to,
\be
H={1\over 2}\int d^py~\phi(\vy) \left(\gra \pi(\vy)\right)^2
\ee
The corresponding Hamilton's equations are,
\be
\begin{split}
	&{\g H\over \g \phi(\vy)}=-\del_t \pi~\Rightarrow~{\del_t\pi}+{\left(\gra \pi\right)^2\over 2}=0~\stackrel{\gra}{\Rightarrow} \del_t(\gra\pi)+\gra \pi \cdot \gra (\gra \pi)=0
	\\&{\g H\over \g \pi(\vy)}=\del_t\phi\Rightarrow \del_t\phi+\gra\cdot\left(\phi\gra\pi\right)=0
\end{split}
\ee
Defining the velocity field $\vv=\gra \pi$, we get,
\be
\del_t\vv+\vv\cdot \gra \vv=0,~~~\del_t\phi+\gra \cdot \left(\phi \vv\right)=0
\ee
These are precisely the  Euler and the continuity equation for a pressureless (dust-like) non-relativistic fluid. Thus, in the classical limit, the collective field theory reproduces the dynamics of pressureless matter. This result is important because, in the large $T$ limit, the tachyon effective action also describes pressureless dust.
Thus, we expect that the relativistic collective theory will reproduce the correct classical tachyon dynamics and, moreover, provide a natural way to include quantum effects through systematic $\hbar$-corrections.  

\subsection{Relativistic free particles}
\label{sec:rel_collective_theory}
In this section, we analyze the relativistic free $\ns$-particle system in the collective description and study the corresponding Hilbert space structure. We will explicitly show how the classical part of the collective Hamiltonian maps to the tachyon effective Hamiltonian. This will also provide insights into how the tachyon effective action can be quantized by including the order $\hbar$ terms that arise from the collective theory.

For the system of relativistic free particles, collective field theory description can be obtained in an identical way to the non-relativistic case we discussed in previous sections. The steps involved are the same, and the conclusions we could draw from them are also analogous. The only difference is that the relativistic Hamiltonian is different from the non-relativistic Hamiltonian. This will reflect in the observation that the form of the Hamiltonian in collective theory as well as energy eigenvalues will be different. 
\subsubsection{Collective Hamiltonian}
We start with a system of $N_*$ relativistic free particles of mass $m$, given by the Hamiltonian,
\be
\label{eq:free_H}
H=\sum_{i=1}^{N_*} \sqrt{\vcp_i^2+m^2}
\ee
Let us introduce the collective field $\phi(x)$ as the same as in the non-relativistic case (see \eqref{eq:phi_def}).
\begin{equation*}
\phi(\vy)=m\sum_{i=1}^{N_*}\g(\vy-\vx_i)
\end{equation*}
As discussed in the previous section, one can use the chain rule of differentiation to write terms of the form $\sum_i \vp_i^{~n}$ in terms of the collective field $\phi$ and its conjugate momentum $\pi$. By expanding the  square root in \eqref{eq:free_H} and replacing each term in the expansion with the collective variables, we obtain the first few terms as,  (See  Appendix \ref{app:rel_collective} for details),
 \be
 \label{eq:H_first3_terms}
 \begin{split}
 	&N_*m=\int ~d^py ~\phi(\vy)
 	\\& \sum_{i=1}^{N_*}{\vec{p}^{~2}_i\over 2m}={1\over 2} \int d^{p}y~\phi(y) \left(\gra \pi(y)\right)^2-{i\hbar \over 2m} \int d^{p}y~\phi(\vy) \gra^2 \pi(\vy)
 	\\& -\sum_{i=1}^{N_*}{\vp_i^{~4}\over 8m^3}=-{1\over 8}\int d^py~\phi(\vy) \left(\gra \pi(\vy)\right)^4+{i\hbar\over 4m} \int d^py~\phi(\vy)\gra\pi(\vy)\cdot\gra\left(\gra\pi(\vy)\right)^2
 	\\&\hspace{2.4cm}+{i\hbar\over 4m}\int d^py~\phi(\vy)\gra^2\pi\left(\gra\pi(\vy)\right)^2+{\hbar^2\over 8m^2} \int d^py~\phi(\vy)~\gra\cdot\gra\left(\gra\pi(\vy)\right)^2
 	\\&\hspace{2.4cm}+{\hbar^2\over 8m^2}\int d^py~\phi(\vy)\gra^2\pi(\vy)\gra^2\pi(\vy)+{\hbar^2\over 4m^2}\int d^py~\phi(\vy)\gra\pi(\vy)\cdot\gra \left(\gra^2\pi(\vy)\right)\
 	\\&\hspace{2.4cm}-{i\hbar^3\over 8m^3}\int d^py~\phi(\vy) \gra^4\pi(\vy)
 \end{split}
 \ee
 One can see a general pattern here. $\sum_i \vp_i^n$ will have terms up to $O(\hbar^{n-1})$.
 % We know from the non-relativistic case that the $\hbar\rightarrow 0$ limit will give the classical dynamics of a fluid.
 Examining the $O(\hbar^0)$ terms, we see that they can be resummed to get the classical part of the Hamiltonian.
 \be
 \label{eq:rel_collective_hbar_0_limit}
 \begin{split}
 H&=\int d^py ~\phi(\vy)+{1\over 2} \int d^{p}y~\phi(\vy) \left(\gra \pi(\vy)\right)^2-{1\over 8} \int d^py \phi(\vy) \left(\gra \pi(\vy)\right)^4+....+O(\hbar)
 \\&=\int d^py~\phi(\vy)\left(1+{1\over 2} \left(\gra \pi(\vy)\right)^2-{1\over 8}\left(\gra \pi(\vy)\right)^4 +....\right)+O(\hbar)
 \\&=\int d^py~\phi(\vy)\sqrt{1+\left(\gra \pi(\vy)\right)^2} +O(\hbar)
 \end{split}
 \ee
 If we identify $\phi=\Pi,~ \pi=- T$, then the $O(\hbar^0)$ term (classical part) matches presicely the $T \to \infty$ of the Tachyon effective hamiltonian  in \eqref{eq:H_tachyon}. Therefore, the equations of motion obtained from the leading term in \eqref{eq:rel_collective_hbar_0_limit} will describe the dynamics of a non-interacting, non-rotating fluid. And we can say that the classical part of collective theory and tachyon effective theory are related by a canonical transformation.
 The higher-order terms in $(\hbar)$ can be considered as quantum corrections to the tachyon effective theory Hamiltonian.

\subsubsection{Hilbert space structure}
Having obtained the Hamiltonian, one can now write down the energy eigenfunctionals and study the Hilbert space structure. It turns out that the eigenfunctionals have the same form as in non-relativistic theory, only that their energy eigenvalues change owing to the relativistic dispersion relation. The Hilbert space will be, $\mathcal{H}_{\ns}$, relativistic $\ns$-particle Hilbert space.

To see that the form of the wavefunctionals doesnt change, note that in quantum mechanics, plane waves are eigenstates of the relativistic Hamiltonian, having $\w_k=\sqrt{\vk^2+m^2}$ as the eigenvalue.
\be
 \begin{split}
 	\sqrt{\vcp^2+m^2}~e^{i\vk \cdot \vx}&=\sqrt{-\gra_{\vx}^2+m^2}~e^{i\vk\cdot\vx}
 	\\&=\sqrt{\vk^2+m^2} ~e^{i\vk\cdot\vx}
    \\&=\w_{\vk} ~e^{i\vk\cdot\vx}
 \end{split}
 \ee
So a general $\ns$-particle energy eigenstate is given as,
\be
\psi\left(\vx_1,\vx_2,...,\vx_{\ns}\right)=e^{i\vk_1\cdot\vx_1}e^{i\vk_2\cdot\vx_2}....e^{i\vk_{\ns}\cdot\vx_{\ns}}
+\text{permutations}
\ee
 Just like we did in the non-relativistic case, one can now write down the corresponding energy eigenfunctional $\Phi_{\vk_1,\vk_2,...,\vk_{\ns}}$ of the collective theory Hamiltonian. The guiding idea is that, when the relation \eqref{eq:phi_def} is imposed, the collective theory wavefunctionals should reproduce the corresponding quantum mechanical wavefunctions. This implies that the form of the eigenfunctionals remains the same as in the non-relativistic theory.  Thus, the structure of Hilbert space remains unchanged going from the non-relativistic to the relativistic case. The Hamiltonian and its eigenvalue, however, now follow the relativistic dispersion relation.
 
 For example, the $\ns=1$ case will have eigenstates,
  \be
 \begin{split}
 	&\psi_{\vk}(\vx)=e^{i\vk\cdot\vx}
 	\\& H\psi_{\vk}(\vx)=\w_{\vk} \psi_{\vk}(\vx)
 	\\& \Rightarrow \Phi_{\vk}={1\over m}\int d^py~e^{i\vk\cdot\vy}\phi(\vy)
 	\\& H\Phi_{\vk}=\left(m+{\vk^2\over 2m}-{\vk^4\over 8m^3}+...\right)\Phi_{\vk}=\w_{\vk}\Phi_{\vk}
 \end{split}
 \ee
 
 Similarly, the $\ns=2$ case will have eigenstates,
 \be
 \begin{split}
 	&\psi_{(\vk_1,\vk_2)}(\vx_1,\vx_2)=e^{i\vk_1\cdot\vx_1}e^{i\vk_2\cdot\vx_2}+\vx_1\leftrightarrow \vx_2
 	\\& H \psi_{(\vk_1,\vk_2)}=\left(\w_{\vk_1}+\w_{\vk_2}\right) \psi_{(\vk_1,\vk_2)}
 	\\& \Rightarrow \Phi_{(\vk_1,\vk_2)}[\phi]={1\over m^2}\int d^py_1d^py_2~e^{i\vk_1\cdot\vy_1}e^{i\vk_2\cdot\vy_2} \phi(\vy_1)\phi(\vy_2)-m\int d^py ~e^{i(\vk_1+\vk_2)\cdot \vy} \phi(\vy)
 	\\& H \Phi_{(\vk_1,\vk_2)}[\phi]=\left(2m+{\vk_1^2+\vk_2^2\over 2m}-{\vk_1^4+\vk_2^4\over 8m^3}+....\right) \Phi_{(\vk_1,\vk_2)}[\phi]
 	=\left(\w_{\vk_1}+\w_{\vk_2}\right) \Phi_{(\vk_1,\vk_2)}[\phi]
 \end{split}
 \ee
Note that the collective theory Hamiltonian was obtained by expanding the square root in the quantum mechanical Hamiltonian and then writing each term in terms of collective variables. Therefore, one must check term by term that the states $\Phi_{\vk},~\Phi_{\vk_1,\vk_2}$ defined above are indeed eigenstates (see appendix~\ref{app:rel_collective}). There is subtlety here regarding the first term in the Hamiltonian \eqref{eq:H_first3_terms}. Its action on a general state cannot be further simplified a priori.
\begin{equation*}
\int d^py \phi(\vy)~\Phi_{\vk_1,\vk_2,...,\vk_n}
\end{equation*}
This is in contrast to the other terms in the Hamiltonian. For example, acting with the field theory expression for $\sum_{i=1}^{N_*}{\vec{p}^{~2}_i\over 2m}$, we directly get the correct eigenvalue.
\begin{equation*}
    \left({1\over 2} \int d^{p}y~\phi(y) \left(\gra \pi(y)\right)^2-{i\hbar \over 2m} \int d^{p}y~\phi(\vy) \gra^2 \pi(\vy)\right)~\Phi_{\vk_1,\vk_2,...,\vk_n}=\left(\sum_{i=1}^n{\vk_i^2\over 2m}\right)~\Phi_{\vk_1,\vk_2,...,\vk_n}
\end{equation*}
So, we would expect the action of $\int d^py \phi(\vy)$ to give an eigenvalue equation. This is ensured by the Jacobian. One can show that working with collective field theory at fixed $\ns$, the operator $\int d^py~\phi(\vy)$ is equivalent to $\ns m$. (We discuss the proof in \ref{app:proof_int_phi_nm}).
\be
\begin{split}
&\left(\Phi_1,\int d^py~\phi(\vy)~\Phi_2\right)=\ns m~~\forall \Phi_1,\Phi_2
\\&\Rightarrow \int d^py~\phi(\vy)~\Phi_{\vk_1,\vk_2,...,\vk_n}=\ns m~\Phi_{\vk_1,\vk_2,...,\vk_n}
\end{split}
\ee
So, the Jacobian makes sure of the fact that the operator $\int d^py~\phi(\vy)$ is equivalent to $\ns m$, ie, $\int d^py~\phi(\vy)=\ns m$.  This result will have implications for understanding the Hilbert space of collective theory.

Once we have the states, we can now compute the inner product between the states to understand the Hilbert-space structure. Note that the form of the Jacobian remains unchanged when going from the non-relativistic case to the relativistic case. Since the Jacobian determines which states are null and which states are equivalent to each other, the conclusions we obtained in the non-relativistic case continue to hold here:

For \emph{a fixed $\ns$ relativistic collective theory}, 
	\begin{enumerate}
		\item The Hilbert space $\mathcal{H}_{\ns}$ is spanned by $\Phi_{\vk_1,\vk_2,..,\vk_{\ns}}$ which are $\ns$ particle energy eigenstates with momenta ${\vk_1,\vk_2,..,\vk_{\ns}}$.
		\item The state $\mc$ is to be understood as a $\ns$ particle state ($\Phi_{\vk_1,\vk_2,..,\vk_{\ns}}$) with all particles at rest and the state $\Phi_{\vk}$ is a $\ns$ particle state with one particle having momentum $\vk$ and the remaining $\ns-1$ particles at rest. Generalising this, one can say $\Phi_{\vk_1,\vk_2,..,\vk_n};~n<\ns$ corresponds to a $\ns$-particle state having $n$-of the particles with momenta $\vk_1,\vk_2,...,\vk_n$ and the remaining $\ns-n$ particles at rest.
		\item The states $\Phi_{\vk_1,\vk_2,..,\vk_n}$ for $n>\ns$ are all null states. 
	\end{enumerate}

We conclude the section by writing down different energy eigenstates. They are all $\ns$-particle states with different numbers of particles at rest. 
 \be
 \label{eq:list_ns_states}
 \begin{split}
 	&\mc=1
 	\\&\Phi_{\vk}={1\over m}\int d^py e^{i\vk\cdot\vy}{\phi(\vy)}
 	\\&\Phi_{\vk_1,\vk_2}={1\over m^2} \int d^py_1d^py_2~e^{i\vk_1\cdot \vy_1}e^{i\vk_2\cdot\vy_2} \phi(\vy_1)\phi(\vy_2)-{1\over m}\int d^py ~e^{i(\vk_1+\vk_2)\cdot\vy}\phi(\vy)
 	\\&.............
 	\\& \Phi_{\vk_1,\vk_2,...,\vk_n}={1\over m^n} \int \prod_{i=1}^n\left(d^py_i~e^{i\vk_i\cdot\vy_i}\right)\phi(\vy_1)\prod_{j=2}^n 
 	\left(\phi(\vy_j)-m\sum_{i=1}^{j-1}\g(\vy_j-\vy_i)\right)
 \end{split}
 \ee
 
 \section{From Collective theory to Tachyon effective theory}
 \label{sec:collective_to_tachyon}
 In the previous sections, we showed that the collective Hamiltonian for relativistic particles agrees with the tachyon effective Hamiltonian at the classical level. This suggests that the collective description may provide a useful way to study the quantum structure of the tachyon theory. In particular, the matching of the classical Hamiltonians indicates that the collective variables $(\phi,\pi)$ and the tachyon variables $(\Pi,T)$ are related by a canonical transformation,
\be
\label{eq:CT_TT_canonical}
\phi=\Pi, \qquad \pi=-T .
\ee
In this section, we study this map in more detail. As we will see, the situation is more subtle in tachyon theory. Imposing physical requirements, we see that: the Jacobian appearing in the inner product is modified, the naive collective field theory eigenstates are not eigenstates of the tachyon Hamiltonian, and the interpretation of the states is different. To understand these differences more clearly, we begin by examining an important distinction between the collective description and the tachyon effective theory.

\subsection{Difference between tachyon theory and collective theory}
There is an important difference between the $\ns$-particle collective theory and the tachyon effective theory. In the collective description, the field $\phi$ satisfies the constraint
\[
\int d^p y\, \phi(\vy) = \ns m .
\]
In contrast, the corresponding quantity
\[
\int d^py\,\Pi(\vy)
\]
in the tachyon theory represents the total rest mass of the system and is not fixed; it can take different values depending on the energy of the original D$p$-brane configuration. Therefore, in order to pass from the collective theory to the tachyon theory, we must relax the constraint $\int d^py\,\phi(\vy)=\ns m$. Relaxing this condition effectively allows $N_*$ to vary. This has the following consequences,
\begin{itemize}
	\item The condition $\int d^py ~\phi(\vy)=\ns m$ is enforced by the Jacobian $J_{\ns}$. If this condition is relaxed, the Jacobian must also be modified.
	\item In collective theory, the states listed in  \eqref{eq:list_ns_states}, are $\ns$-particle energy eigenstates. However, these states no longer remain eigenstates of the tachyon Hamiltonian. 
\end{itemize}
This also changes the way we interpret these states: they can now be viewed as states with different particle numbers rather than having a fixed $N_*$-number of particles. In ordinary quantum field theory, different particle-number sectors are connected by creation and annihilation operators. These operators change the eigenvalue of the number operator and allow us to move from one particle-number sector to another. As we will see, an analogous set of well-defined operators also emerges in tachyon theory once this condition is relaxed.

To understand how such a structure may arise in the present case, let us first recall how this works in the Klein–Gordon theory.
In the  Klein–Gordon theory, the number operator is given by
\[
N=\int {d^pk\over (2\pi)^p}\, a^{\dagger}_{\vk}a_{\vk} \, .
\]
It satisfies the commutation relations
\be
 \label{eq:N_a_ada_commutation}
[N,a_{\vk}]=-a_{\vk}~, \qquad [N,a^{\dagger}_{\vk}]=a^{\dagger}_{\vk} \, .
\ee
These relations show that $a_{\vk}$ and $a^{\dagger}_{\vk}$ change the eigenvalue of $N$: the operator $a_{\vk}$ lowers the particle number, while $a^{\dagger}_{\vk}$ raises it. In this way, they connect different particle-number sectors.

In terms of these operators, the Hamiltonian takes the simple form
\be
\label{eq:H_Klein_Gordon}
H=\int {d^pk\over (2\pi)^p} \, \omega_{\vk} a^{\dagger}_{\vk}a_{\vk} \, .
\ee

We now ask whether an analogue of $N$, $a_{\vk}$ and $a^{\dagger}_{\vk}$ exists in the tachyon theory. The answer is yes. Moreover, it will allow us to write the tachyon Hamiltonian in a compact form similar to \eqref{eq:H_Klein_Gordon}. We now construct these operators explicitly.
 
Motivated by the relation $\int d^py\,\phi(\vy)=\ns m$ in collective theory, we define the number operator in tachyon theory as
 \be
 N={1\over m} \int d^py~ \Pi(\vy)
 \ee
We now look for operators $a_{\vk}$ and $a^{\dagger}_{\vk}$ satisfying
\eqref{eq:N_a_ada_commutation}.
With some effort, one can verify that the following operators satisfy these relations:
 \be
 \label{eq:a_and_a_da}
 \begin{split}
 	& a_{\vk}=\int d^py~e^{-i\vk\cdot\vy} e^{m{\g\over \g \Pi(\vy)}}
 	\\& a^{\da}_{\vk}={1\over m}\int d^py~e^{i\vk\cdot\vy} ~\Pi(\vy) e^{-m{\g\over \g \Pi(\vy)}}
 \end{split}
 \ee
 \footnote{Restoring $\hbar$, we get, $a_{\vk}={1\over \hbar^p}\int d^py~e^{-{i\vk\cdot\vy}\over \hbar} e^{m{\g\over \g \Pi(\vy)}}$ and $a^{\da}_{\vk}={1\over m}\int d^py~e^{{i\vk\cdot\vy}\over \hbar} ~\Pi(\vy) e^{-m{\g\over \g \Pi(\vy)}}$. The expression for $H$, and $N$ remains unchanged in terms of $a_{\vk}$ and $a^{\da}_{\vk}$.  }
 For example, one finds that, 
 \be
 \begin{split}
 	[N,a_{\vk}]&=\left[  {1\over m}\int d^py' \Pi(\vec{y'}),~  \int d^py~e^{-i\vk\cdot\vy} e^{m{\g\over \g \Pi(\vy)}}\right]
 	\\&= {1\over m}\int d^py' d^py e^{-i\vk\cdot\vy} \left[\Pi(\vec{y'}), e^{m{\g\over \g \Pi(\vy)}} \right]
 	\\&= -\int d^py' d^py e^{-i\vk\cdot \vy} \g(\vy-\vy') e^{m{\g\over\g \Pi(\vy)}}
 	\\&=-a_{\vk}
 \end{split}
 \ee
 In the 3rd equality, we used the result,
 \be
 \left[ \Pi(\vy '), e^{m {\g \over \g \Pi(\vy~')}}\right]=m\g(\vy-\vy~') e^{m {\g \over \g \Pi(\vy)}},
 \ee
 which can be easily derived from the BCH formula,
 \be
 e^{m{\g\over \g \Pi(\vy)}} \Pi(\vec{y'}) e^{-m{\g\over \g \Pi(\vy)}}=\Pi(\vec{y'})+m\g^p(\vy-\vec{y'}).
 \ee
 Similarly, one can check that $a^{\da}_{\vk}$ satisfies the required commutation relation. 
 Moreover, these operators obey
 \be
 \label{eq:commutation_a_and_ada}
 \left[a_{\vk_1},a^{\da}_{\vk_2}\right]=(2\pi)^p\g^p\left(\vk_1-\vk_2\right).
 \ee
 We have therefore relaxed the constraint $\int d^py ~\phi(\vy)=\ns m$ when going from collective theory to tachyon theory. In the tachyon theory, the corresponding quantity  $ \int d^py~ \Pi(y)$ is proportional to the number operator and is unrestricted. This allows us to introduce creation and annihilation operators. As we mentioned earlier, this also means that the Jacobian in tachyon theory must differ from the one in collective theory. 
 It turns out that demanding consistency of the structure we introduced so far uniquely determines the Jacobian. 
 We will see this explicitly in the next subsection.
\subsection{Jacobian of the tachyon theory}
\label{sec:J_of_tachyon}
 Recall that in collective theory, the inner product, for a fixed $N_*$, is defined as
\be
(\Phi_1,\Phi_2)=\int D\phi~J_{N_*}[\phi]\Phi_1^*\Phi_2 .
\ee
In tachyon theory, however, we are not working within a fixed $N_*$ sector. As we have seen, the operators $a_{\vk}$ and $a_{\vk}^{\dagger}$ do not commute with the number operator $\int d^p y\, \Pi(y)$, and therefore they allow us to move between different particle sectors. For this reason, we cannot use the Jacobian $J_{N_*}$ that appears in the collective theory inner product.
Instead, we define the tachyon theory inner product with an arbitrary Jacobian and then determine it uniquely by requiring that the creation and annihilation operators are Hermitian conjugates of each other. We explain this construction in this subsection.
 
 Consider a  general inner product of the form,
 \be
 \left( \Phi_1[\Pi],\Phi_2[\Pi] \right) = \int D\Pi~J[\Pi]\Phi_1[\Pi]^*\Phi_2[\Pi].
 \ee
 Note that under this inner product, the hermitian conjugate of ${\g \over \g \Pi(\vy)}$ is given by,
 \be
 \left({\g\over \g \Pi(\vy)}\right)^{\da}=-{\g\over \g \Pi(\vy)}-{1\over J[\Pi]}{\g J[\Pi]\over \g \Pi(\vy)}.
 \ee
 So, the condition that $a^{\da}_{\vk}$ and $a_{\vk}$ are hermitian conjugates of each other can be written as,
 \be
 \label{eq:hermiticity_step}
 \begin{split}
 	&\left(a^{\da}_{\vk}\right)^{\da}=a_{\vk}
 	\\&\Rightarrow {1\over m} \int d^p \vy e^{-i\vk\cdot \vy} e^{m\left({\g\over \g \Pi(\vy)}+{1\over J}{\g J[\Pi]\over \g \Pi(\vy)}\right)} \Pi(\vy)=\int d^py e^{-i\vk\cdot\vy} e^{m{\g\over \g \Pi(\vy)}}
 	\\&\Rightarrow {1\over m}e^{-m{\g\over \g \Pi(\vy)}} e^{m\left({\g\over \g \Pi(\vy)}+{\g \ln J[\Pi]\over \g \Pi(\vy)} \right)} \Pi(\vy)=1
 \end{split}
 \ee
One can simplify this expression further. First, note that the argument of the second exponential can be written as,
\be
m{\g \over \g \Pi(\vy)}+m{\g \ln J[\Pi]\over \g \Pi(\vy)}=e^{-\ln J[\Pi]} ~m {\g \over \g \Pi(\vy)}~e^{\ln J[\Pi]}
\ee
This is of the form $BAB^{-1}$, and it can be exponentiated using $e^{BAB^{-1}}=Be^AB^{-1}$. This leads to,
\be
\label{eq:exponentiation_of_BABinv}
e^{m\left({\g\over \g \Pi(\vy)}+{\g \ln J[\Pi]\over \g \Pi(\vy)} \right)}=e^{-\ln J[\Pi]} e^{m {\g \over \g \Pi(\vy)}} e^{\ln J[\Pi]}
\ee
Substituting this in \eqref{eq:hermiticity_step} will simplify it further if one uses the fact that the adjoint action of $e^{-m{\g \over \g \Pi(\vy)}}$ on a functional $F[\Pi]$ will shift the argument of $F$ by a delta function peaked at $\vy$ having strength $m$.
\be
\label{eq:conjugation_by_pi}
e^{-m{\g \over \g \Pi(\vy)}} F[\Pi(\vy~')] e^{m{\g \over \g \Pi(\vy)}}=F[\Pi(\vy~')-m\g(\vy~'-\vy)]\equiv F[\Pi-m\g_y]
\ee
We introduced a compact notation in the last step. One can apply \eqref{eq:exponentiation_of_BABinv} and \eqref{eq:conjugation_by_pi} in \eqref{eq:hermiticity_step} to obtain,

\be
 \label{eq:hermiticity_condn}
\begin{split}
	&{1\over m}e^{-m{\g\over \g \Pi(\vy)}} e^{-\ln J[\Pi]} e^{m {\g \over \g \Pi(\vy)}} e^{\ln J[\Pi]}  \Pi(\vy)=1
	\\&\Rightarrow {1\over m} e^{-\ln J[\Pi-m\g_y]}  e^{\ln J[\Pi]}  \Pi(\vy)=1
	\\&\Rightarrow{1\over m} e^{\ln J[\Pi]-\ln J[\Pi-m\g_y]}\Pi(\vy)=1
	\\& \Rightarrow J[\Pi-m\g_y]={\Pi(\vy)\over m} J[\Pi]
\end{split}
\ee
 To solve for the final expression, go to functional momentum space and write,
 \be
 J[\Pi]=\int D\lm ~e^{i\int d^py~\lm(\vy)\Pi(\vy)} J[\lm]
 \ee
 In momentum space, the condition \eqref{eq:hermiticity_condn} reduces to,
 \be
 {i\over m}{\g J[\lambda]\over \g \lm (\vy)}=e^{-im\lm(\vy)} J[\lm]
 \ee
 This can be solved to get,
 \be
 J[\lm]=\exp{\int d^py~ e^{-im\lm(\vy)}}
 \ee
 Writing down the series expansion of the exponential puts the Jacobian in a form familiar from the collective field theory discussions. 
 \be
 \label{eq:soln_J_tachyon}
 \begin{split}
 J[\Pi]&=\int D\lm ~e^{i\int d^py~\lm(\vy) \Pi(\vy)} \exp{\int d^px~ e^{-im\lm(\vx)}}
 \\&=\int D\lm ~e^{i\int d^py~\lm(\vy) \Pi(\vy)} \sum_{n=0}^{\infty} {1\over n!}\left(\int d^px~ e^{-im\lm(\vx)}\right)^n
 \\&=\sum_{n=0}^{\infty} {J_n[\Pi]\over n!}
 \end{split}
 \ee
 where $J_n[\Pi]$ corresponds precisely to the Jacobian associated with the 
$\ns=n$ sector of the collective theory.
Thus, the Jacobian in tachyon theory is naturally a sum over all $\ns$-sector Jacobians. 
This is consistent with the operator structure: the operators 
$a_{\vk}$ and $a^{\dagger}_{\vk}$ connect different $\ns$ sectors.

The Jacobian will play an important role in the Hilbert space structure of the tachyon theory. On a technical note, one can immediately see that using the Jacobian \eqref{eq:soln_J_tachyon}, the states listed in \eqref{eq:list_ns_states} (under the exchange of $\phi$ with $\Pi$) will have an inner product different from collective theory.  Here, one has to compute the collective field theory inner product using $J_n$ and then perform the sum over $n$ weighted by ${1\over n!}$. We will address the implications of these changes in section \ref{sec:hilbert_space_of_tachyon}.

\subsection{Tachyon theory Hamiltonian}
\label{sec:tachyon_H_and_states}
We now discuss the Hamiltonian in tachyon theory. We know from the classical limit of collective field theory that the tachyon theory variables and collective theory variables are related by a canonical transformation \eqref{eq:CT_TT_canonical}. We further suggest that the Hamiltonian obtained from the collective theory Hamiltonian under this transformation is the tachyon theory Hamiltonian. 
 \be
 \label{eq:tachyon_H_series}
 H_{tachyon}=\int d^py~\Pi(\vy)+{1\over 2}\int d^py~\Pi(\vy)\left(\gra T(y)\right)^2+{i\hbar\over 2m}\int d^py~\Pi(\vy)\gra^2T(y)+...
 \ee
The first term describes the correct classical dynamics of rolling tachyon near the minimum of the potential, while the $O(\hbar)$ and higher order terms can be considered as quantum corrections. In other words, we are adding $O(\hbar)$ and higher-order terms to the Hamiltonian given in Section \ref{sec:tachyon_effective_theory},  to incorporate quantum corrections. To show the effectiveness of these $\hbar$-corrections, we first show, using the creation and annihilation operators introduced in the last section \eqref{eq:a_and_a_da}, that the Hamiltonian \eqref{eq:tachyon_H_series} has a compact form analogous to the Klein-Gordon Hamiltonian given in \eqref{eq:H_Klein_Gordon} \footnote{One has to restore $\hbar$ factors in $a_{\vk}$ and $a^{\da}_{\vk}$ to get the precise matching with \eqref{eq:tachyon_H_series}, however, we will let $\hbar=1$ in the derivations for simplicity. }.
 \begin{equation*}
 \begin{split}
 	H_{tachyon}&=\int {d^pk\over (2\pi)^p}~\w_{\vk}a^{\da}_{\vk}a_{\vk}
 	\\&=\int {d^pk\over (2\pi)^p}~\left\{m+{\vk^2\over 2m}-{\vk^4\over 8m^3}+..\right\}
 	\left({1\over m}\int d^py~e^{i\vk\cdot\vy} ~\Pi(\vy) e^{-m{\g\over \g \Pi(\vy)}}\right)\times
    \\& \hspace{7cm}\left(\int d^py'~e^{-i\vk\cdot\vy~'} e^{m{\g\over \g \Pi(\vy~')}}\right)
 \end{split}
 \end{equation*}
 Consider the first term in the curly brackets. It is easy to show that this matches precisely with the first term in \eqref{eq:tachyon_H_series}.
 \be
 \begin{split}
 	\int {d^pk\over (2\pi)^p} m&
 	\left({1\over m}\int d^py~e^{i\vk\cdot\vy} ~\Pi(\vy) e^{-m{\g\over \g \Pi(\vy)}}\right)\left(\int d^py'~e^{-i\vk\cdot\vy~'} e^{m{\g\over \g \Pi(\vy~')}}\right)
 	\\&= \int d^pyd^py' (2\pi)^p\g(\vy-\vy~')\left({1\over m} ~\Pi(\vy) e^{-m{\g\over \g \Pi(\vy)}}\right)\left( e^{m{\g\over \g \Pi(\vy~')}}\right)
 		\\&=\int d^py~\Pi(\vy)
 \end{split}
 \ee
 By similar manipulations, one can show that the other terms will also match (see Appendix \ref{app:rel_collective} for details). It is then trivial to check that the Hamiltonian will have the required commutation relations.
 \be
[H_{tachyon},a_{\vk}]=-\w_ka_{\vk}~,~~~[H_{tachyon},a^{\da}_{\vk}]=\w_ka^{\da}_{\vk}
\ee
Thus, we find a close similarity between the tachyon theory and the Klein–Gordon theory. In particular, one can introduce creation and annihilation operators in terms of which both the number operator and the Hamiltonian can be written. These operators satisfy the usual commutation relations, just as in the Klein–Gordon theory. 

\subsection{Theory A and Theory B}
A lot has happened in the last few sections. It might be a good point to pause and reflect on the developments. 
\begin{enumerate}
\item In the $\hbar\rightarrow 0$ limit, the collective theory Hamiltonian \eqref{eq:rel_collective_hbar_0_limit} agrees with the tachyon effective Hamiltonian \eqref{eq:H_tachyon} (with the identification: $\phi=\Pi,~\pi=-T$ ). \emph{One can say that this is the main motivation to study collective field theory, that the classical dynamics looks the same.}
\item When one considers all the $\hbar$-terms in the collective theory Hamiltonian, and takes into account the non-trivial Jacobian $J_{\ns}$, the theory becomes a consistent quantum theory describing $\ns$-relativistic free particles. The Hilbert space of the theory is $\mathcal{H}_{\ns}$.
\item One would've then suggested to add identical $\hbar$-terms to the tachyon effective theory Hamiltonian to get $H_{tachyon}$ \eqref{eq:tachyon_H_series}, define Jacobian of the theory to be $J_{\ns}$ and claim that the tachyon effective theory is also describing quantum mechanics of $\ns$-relativistic particles. \emph{But this is not what we are doing.} 
\item What we are doing instead is to take $H_{tachyon}$ as the Hamiltonian, but change the Jacobian to $J[\phi]=\sum_{n=0}^{\infty}{J_n[\phi]\over n!}$. The change in Jacobian results from imposing the physical condition that $\int d^py~\Pi(\vy)$ should be unconstrained.
\end{enumerate}

So we have two consistent quantum theories, Theory A and Theory B.
\begin{table}[h]
	\centering
	\begin{tabular}{|lccc|}
		\hline
		Name & Canonical variables & Hamiltonian & Jacobian \\
		\hline
		Theory A & $\Pi, T$ & $H_{\text{tachyon}}$ & $J_{\ns}[\phi]$ \\
		Theory B & $\Pi, T$ & $H_{\text{tachyon}}$ & $J[\phi]=\sum_{n=0}^{\infty} \frac{J_n[\phi]}{n!}$ \\
		\hline
	\end{tabular}
\end{table}

Theory A describes the quantum mechanics of $\ns$-relativistic free particles. In this theory, due to the Jacobian $J_{\ns}$, the value of $\int d^py ~\Pi(\vy)$ is fixed to be $\ns m$.  

Theory B on the otherhand has $\int d^py ~\Pi(\vy)=Nm$ where $N=\int{d^pk\over (2\pi)^p}~a^{\da}_{\vk}a_{\vk}$ is a number operator, having an unconstrained spectrum. The creation and annihilation operators take you from one eigensector of the number operator to another. We will study the Hilbert space of Theory B in Section \ref{sec:hilbert_space_of_tachyon}.

At the level of classical equations of motion, obtained in the $\hbar\rightarrow 0$ limit, Theory A and Theory B are the same. However, quantum mechanically, they are different. Physical properties of Theory B are more similar to what we expect in a quantum version of tachyon effective theory. \emph{Thus, we suggest that Theory B should be equated with tachyon effective theory. This is what we mean by quantising tachyon effective theory.} In other words, by adding the $\hbar$-terms to the Hamiltonian and introducing the Jacobian, we map the tachyon effective theory to Theory B, which is a consistent quantum theory as it stands.  

Since we spent a considerable amount of time studying operators, states, inner products, etc., in Collective theory (Theory A), it will be important to point out which of the results derived there will be valid/useful for Theory B (Tachyon theory).
\begin{itemize}
    \item Collective theory was used to get to Theory A. After that, we make a departure from collective theory by imposing physical requirements on the theory. This introduces more structure (like creation/annihilation operators), which in some sense "deforms" the theory to get to Theory B. This theory has the desired properties that one would expect to have in a quantum tachyon theory.
	\item The form of the operators (Hamiltonian, Momentum, etc) in Theory A and B will be the same
	\item As the Hamiltonian is the same for both Theory A and B, one would expect the list of states given in \eqref{eq:list_ns_states} to be the eigenstates (under, $\phi=\Pi,~\pi=-T$ ) of both Theory A and B. This is almost true, except for the first term in $H_{tachyon}$.  Given an energy eigenstate $\Phi_{\vk_1,\vk_2,...,\vk_n}$ from \eqref{eq:list_ns_states},
	\begin{equation*}
		\begin{split}
			&\text{Theory A:}~H_{tachyon}~\Phi_{\vk_1,\vk_2,...,\vk_n}=\left(\ns m+\sum_{i=1}^{n}{\vk_i^2\over 2m}+....\right) ~\Phi_{\vk_1,\vk_2,...,\vk_n}
			\\& \text{Theory B:}~H_{tachyon}~\Phi_{\vk_1,\vk_2,...,\vk_n}=\left(\int d^py~\Pi(\vy)+\sum_{i=1}^{n}{\vk_i^2\over 2m}+....\right) ~\Phi_{\vk_1,\vk_2,...,\vk_n}
		\end{split}
	\end{equation*}
	We used the fact that $\int d^py~\Pi(\vy)=\ns m$ in Theory A, but no such restriction is there in B. This difference is crucial, and it means that the states $\Phi_{\vk_1,\vk_2,...,\vk_n}$ will be $\ns$-particle energy eigenstates in Theory A, but in Theory B, they are no longer energy eigenstates and their physical interpretation will be different.
	\item The inner products will be different in Theory A and B. However, since the Jacobian in Theory B is $J[\phi]=\sum_{n=0}^{\infty}{J_n[\phi]\over n!}$, we can use methods developed in collective theory to compute the inner product between any two states using $J_{n}$, and then sum over $n$ with the ${1\over n!}$ to get the inner product in Theory B (See Appendix \ref{app:tachyon_computations}). This, along with the last point, implies that the Hilbert space structure of Theory A is different from that of Theory B.
\end{itemize}

In the next section, we will study the Hilbert space structure of Theory B (Or tachyon theory), and compare how it differs from Theory A (Or collective field theory).  We will also enquire about the similarities between Theory B and Klein-Gordon theory, because Theory B has creation and annihilation operators in terms of which the Hamiltonian looks exactly like Klein-Gordon Hamiltonian. 

\subsection{Hilbert space of tachyon theory}
\label{sec:hilbert_space_of_tachyon}

The tachyon theory has similarities with both collective theory and Klein-Gordon theory. Our starting point was the assertion that the tachyon theory Hamiltonian is the same as the collective theory Hamiltonian up to a canonical transformation. We then further imposed the physical requirement of the existence of different particle number sectors on the tachyon theory.  This led to the introduction of creation and annihilation operators, which expressed the number operator and Hamiltonian in a form analogous to Klein-Gordon theory. However, in this section, we will see that the Hilbert-space structure of tachyon theory is different from both collective theory and Klein-Gordon theory. 

In Klein-Gordon theory, the Hilbert space is spanned by states of the form,
\be
\label{eq:eigenstates_KG}
|\vk_1,\vk_2,...,\vk_{n}\ra=\left(\prod_{i=1}^{n}a^{\da}_{\vk_i}~\right)~|0\ra
\ee
Here $|0\ra$ denotes the vacuum state. These states form the usual Fock space and are eigenstates of the Hamiltonian \eqref{eq:H_Klein_Gordon}.
Since the tachyon Hamiltonian also has the form \eqref{eq:H_Klein_Gordon}, one might expect the eigenstates of tachyon theory to be of the same form \eqref{eq:eigenstates_KG}. However, this is not the case. In particular, the vacuum state $|0\ra$  is not accessible in the tachyon theory description.

Recall that the definition of the vacuum state is that it is annihilated by the operator $a_{\vk}$. Suppose $\Psi[\Pi]$ is the vacuum state in tachyon theory. Then,
 \be
 \begin{split}
 	&a_{\vk}\Psi[\Pi]=0
 	\\& \int d^py~e^{-i\vk\cdot\vy} e^{m{\g\over \g \Pi(\vy)}} \Psi[\Pi]=0
 	\\& \Rightarrow\int d^py ~e^{-i\vk\cdot \vy} \Psi[\Pi(\vec{y'})+m\g(\vy-\vec{y'})]=0
 	\\&\Rightarrow \Psi[\Pi(\vec{y'})+m\g(\vy-\vec{y'})]=0
 	\\&\Rightarrow \Psi[\Pi]=0
 \end{split}
 \ee
 Thus, the only solution is the trivial one. This means that one cannot express the vacuum state in the tachyon theory variables. This can be explained using the fact that the rolling tachyon solution in string theory is distinct from the vacuum configuration where the tachyon is placed at the minimum of the potential, having zero energy. This reflects in the tachyon theory side as the fact that the vacuum is inaccessible to the theory. We will comment more on this in \ref{sec:D_brane_energy}. 

The discussion above shows that even though the operators and their commutation relations are similar, the Hilbert space of tachyon theory is different from that of Klein–Gordon theory. In particular, the vacuum state is not a part of the tachyon theory Hilbert space. 

This leaves us with the set of states given in \eqref{eq:list_ns_states}, which are the energy eigenstates in collective theory. Since the Hamiltonian of collective theory and tachyon theory have the same form, one would expect that the eigenstates of collective theory (under the substitution $\phi=\Pi$ and $\pi=-T$) will be eigenstates of the tachyon theory. However, this is not the case as we will show below. \footnote{The collective theory states in \eqref{eq:list_ns_states} are given in the $\phi$ basis. When we talk about the same states in tachyon theory, it is implicit that the $\phi$ is replaced with $\Pi$.  }

First, note that the repeated action of creation operators on $\mc$ generates precisely the set of states given in \eqref{eq:list_ns_states}. This can be verified as follows,
 \be
 \begin{split}
 	a^{\da}_{\vk}~\mc&= {1\over m}\int d^py~e^{i\vk\cdot\vy} ~\Pi(\vy) e^{-m{\g\over \g \Pi(\vy)}} \cdot 1
 	\\&={1\over m}\int d^py~e^{i\vk\cdot\vy} ~\Pi(\vy)
 	\\&=\Phi_{\vk}
 \end{split}
 \ee
And, in general, an $n$-particle energy eigenstate of collective theory is compactly written as,
\be
\Phi_{\vk_1,\vk_2,...,\vk_n}=\prod_{i=1}^{n}a^{\da}_{\vk_i}~\mc
\ee
It is good to re-emphasise here that in collective theory, the state given above is an $\ns$-particle energy eigenstate having $n$ non-zero momenta, labelled by $\{\vk_1,\vk_2,\cdots,\vk_n\}$. However, in tachyon theory, they cannot be interpreted as such because the form of the Jacobian is different. 
\footnote{For example, see \eqref{eq:app_H_on_Phi_n}. Here, it is important that the Jacobian $J_{\ns}$ imposes $\int d^py~\phi(\vy)=\ns m$. This is not the case in tachyon theory as the Jacobian is $J[\phi]=\sum_{n=0}^{\infty} {J_n\over n!}$, leaving $\int d^py~\Pi(\vy)$ unconstrained. }

From the form of these states, it is clear that, if the states $\prod_{i=1}^{n}a^{\da}_{\vk_i}~\mc$ are to be eigenstates of $H$, then the state $\mc$ itself must be an eigenstate of $H$(or equivalently of $N$). But this is not the case. %To understand the nature of the state $\mc$%
This can be easily shown using the action of $a_{\vk}$ on $\mc$.
\be
\label{eq:a_on_C}
\begin{split}
	a_{\vk}~\mc&= \int d^py~e^{-i\vk\cdot\vy} e^{m{\g\over \g \Pi(\vy)}} \cdot 1
	\\&=\int d^py~e^{-i\vk\cdot\vy}
	\\&=(2\pi)^p\g^p(\vk)\cdot \mc.
\end{split}
\ee
This shows that $\mc$ is an eigenstate of the annihilation operator. Hence, it will not be an eigenstate of $N$ or $H$\footnote{As $\mc$ is an eigenstate of $a_{\vk}$, it should also be an eigenstate of $a^{\da}_{\vk}$ in order to be an eigenstate of $N$. But this is not possible as $a^{\da}_{\vk}$ and $a_{\vk}$ do not commute with each other and hence cannot have simultaneous eigenstates.}. This implies that the states $\prod_{i=1}^{n}a^{\da}_{\vk_i}~\mc$  are also not eigenstates of the tachyon Hamiltonian.
On the other hand, \eqref{eq:a_on_C} tells you that $\mc$ is a coherent state.  So the states $\prod_{i=1}^{n}a^{\da}_{\vk_i}~\mc$ should be interpreted as excitations built on top of this coherent state.

Formally, coherent states can be written as exponentials of creation operators acting on the vacuum. In the present case, the above relations imply that $\mc$ may be written as
 \be
 |\mc\ra=e^{a^{\da}_{\vec{0}}}~|0\ra
 \ee
 where the creation operator carries zero momentum. The state $|0\rangle$ here denotes the tachyon vacuum. Although we have argued earlier that the tachyon vacuum cannot be accessed within the effective theory, it is still useful to introduce it formally in order to express the coherent state in this way. Thus, the lowest energy state accessible to tachyon theory is $\mc$, a coherent state created by the zero-momentum creation operator acting on the tachyon vacuum, ie, a coherent state of particles at rest.

Let us now summarise the main points of this section. We started with the Hamiltonian of the tachyon theory \eqref{eq:tachyon_H_series}. This Hamiltonian is obtained from the collective theory Hamiltonian \eqref{eq:H_first3_terms} through the canonical transformations \eqref{eq:CT_TT_canonical}. Moreover, it has a form similar to the Klein–Gordon Hamiltonian \eqref{eq:H_Klein_Gordon}, upon using creation and annihilation operators \eqref{eq:a_and_a_da}. Despite these similarities, the Hilbert space of tachyon theory is different from both Klein–Gordon theory and collective theory. Comparing with Klein-Gordon, we see that the tachyon theory does not have an analogue of the vacuum state. So, one does not have the usual Fock space states of the form \eqref{eq:eigenstates_KG} in tachyon theory.  On the other hand, comparing with collective theory, we see that, owing to the change in the form of Jacobian \eqref{eq:soln_J_tachyon}, the lowest energy state $\mc$ becomes a coherent state of particles at rest and therefore not an eigenstate of the Hamiltonian. These results suggest that the Hilbert space of tachyon theory comprises of a coherent state $\mc$ of particles at rest, and then excitations on top of it.
\begin{equation}
	\label{eq:hilbert_space_span_tachyon}
	\mathcal{H}_{tachyon}=\text{span}\bigg\{ \Phi_{\vk_1,\vk_2,....,\vk_{n}}=a^{\da}_{\vk_1}a^{\da}_{\vk_2}....a^{\da}_{\vk_n}~\mc;~n\in\mathbb{Z}_{\geq 0}  \bigg\}
\end{equation}
The state $\mc$ is the lowest energy state as all the particles are at rest. One can then increase the energy and particle number by acting with creation operators on $\mc$. The inner product between any two states in \eqref{eq:hilbert_space_span_tachyon} can be obtained either by directly using the Jacobian \eqref{eq:soln_J_tachyon}, or by using commutation relations between creation and annihilation operators and the norm of $\mc$. Relevant, computations are compiled in \ref{app:tachyon_computations}.

This concludes our discussion of the tachyon theory Hilbert space. We have a coherent state $\mc$ and excitations on top of this. \emph{Thus, the tachyon effective theory captures a subsector of Klein-Gordon theory.} To see this more clearly, recall that in Klein–Gordon theory, one can also construct coherent states. In particular, a coherent state can be produced by acting with the creation operator with zero momentum. The excitations of the field can then be viewed as fluctuations on top of this coherent state. From this point of view, the tachyon theory can be understood as describing this sector of the Klein–Gordon theory: a coherent state corresponding to the zero-momentum mode together with the excitations built on top of it.

One can understand the physical meaning of this coherent state from the closed string analysis presented in \cite{Lambert:2003zr}. In string theory, the decaying D-brane acts as a time-dependent source for closed string radiation. In the case of homogeneous decay, where the tachyon field depends only on time, the final closed string state is a coherent state of closed string modes $e^{\alpha a^{\da}}|0\ra$. The coherent state $\mc$ that appears in the tachyon analysis can be understood similarly. Excitations on top of this coherent state take into account inhomogeneities in the initial D-brane configuration, corresponding to small spatial variations of the tachyon field on the brane. Given this interpretation, the energy of the coherent state can be related to the energy of the initial D-brane configuration. Moreover, one would expect the quantity $\int d^py ~\Pi(\vy)$ to have a continuous spectrum, whereas the number operator $N$ has a discrete spectrum.  So, one would expect the quantity $\la \int d^py ~\Pi(\vy)\ra$ to be related to the continuous parameter $\alpha$.  In the next section, we will see how to go about this.  

 \subsection{Coherent state and initial D-brane system} 
 \label{sec:D_brane_energy}
 In this section, we study the energy associated with the coherent state $\mc$. 
 We begin by writing down the expectation value of Hamiltonian $H$ in this state (\ref{app:tachyon_computations}),
 \be
 	{\left(\mc,H_{tachyon}~\mc\right)\over \left(\mc,\mc\right)}
 	=m V
 \ee
One would like to have the state $|\mc\ra$ to have energy parameterized by a real number. This is because the original D-brane system can have different values of energy, which is a sum of the D-brane tension and the energy (kinetic+potential) of rolling tachyon. This should be reflected in the energy of the coherent state in the tachyon theory. Also, the closed string analysis performed in \cite{Lambert:2003zr} suggests the coherent state of closed string mode has the form $e^{\alpha a^{\da}} |0\ra$ where $\alpha$ is the one-point disk amplitude of the closed string mode. So the coherent state in tachyon theory should also admit a similar parametrization. This can be achieved by redefining the creation and annihilation operators in tachyon theory \eqref{eq:a_and_a_da} by multiplying $a_{\vk}$ by a real number ${\mathfrak{C}}$ and $a^{\da}_{\vk}$ by ${1\over \mathfrak{C}}$ such that the commutation relations and expressions for $H, N$ are unchanged.  
 \be
 \label{eq:a_and_a_da_modified}
 \begin{split}
 	& a_{\vk}=\mathfrak{C}\int d^py~e^{-i\vk\cdot\vy} e^{m{\g\over \g \Pi(\vy)}}
 	\\& a^{\da}_{\vk}={1\over \mathfrak{C}}{1\over m}\int d^py~e^{i\vk\cdot\vy} ~\Pi(\vy) e^{-m{\g\over \g \Pi(\vy)}}
 \end{split}
 \ee  
 With this definition the action of the annihilation operator on the state $\mc$ becomes,
 \be
 a_{\vk}~\mc=2\pi \g(\vk)~\mathfrak{C}~\mc
 \ee
 In Klein–Gordon notation, the state can therefore be written as a coherent state of zero momentum mode.
 \be
|\mc\ra=e^{\mathfrak{C}a^{\dagger}_{0}}|0\ra
 \ee
 However, one should note that unlike in a Klein-Gordon theory, here $a_{\vk}$ and $a^{\da}_{\vk}$ depends on a real number $\mathfrak{C}$. What this essentially means is that the Jacobian will also have a $\mk$-dependence given by the following equation,
 \be
 J^{\mk}[\Pi]=\int D\lm ~e^{\int d^py\lm(\vy)\Pi(\vy)} \exp{\int d^px~\mk^2 e^{-im \lm(\vy)}}.
 \ee
This result can easily be seen from a slight modification of the calculations given in \ref{sec:J_of_tachyon}.  What we have done essentially is to change the Jacobian such that the wavefunctional $\mc[\Pi]$ now represents the state $e^{\mk a_0^{\da}}$ instead of $e^{a_0^{\da} }$. Alternatively, one could've stayed with the Jacobian $J[\Pi]$, in which case the state $\mc$ will represent $e^{ a^{\da}_{0}}|0\ra$, such that to get to another coherent state $e^{\mk a_0^{\da}}$ one has to act with the operator $e^{(\mk-1)a^{\da}_{0}}$ on $\mc$. The technical aspects of this fact is discussed in Appendix \ref{app:tachyon_basis}.
 
 The norm and expectation value of $H$ can now be computed to be, 
 \be
 \begin{split}
 & \left(\mc,\mc\right)=e^{\mathfrak{C}^2V}
 \\& {\left( \mc,~H_{tachyon}~\mc \right) \over  \left(\mc,\mc\right) }=\mathfrak{C}^2 m V
 \end{split}
 \ee
Since the coherent state $\mc$ has particles at rest, the expectation value of the number operator ($N$) times $m$ also coincides with the expectation value of energy given above.
 \be
 \la \int d^py ~\Pi(\vy)\ra\equiv {\left(\mc,~Nm~\mc\right)\over \left(\mc,\mc\right)}=\mathfrak{C}^2 m V
 \ee
 
So, the quantity $\la \int d^py ~\Pi(\vy)\ra$ is parametrised by a real number $\mathfrak{C}^2$ and physically it can be used to encode the energy of the original D-brane configuration so that different D-brane decay will result in different coherent states. Under such an interpretation, the quantity $\mathfrak{C}^2$ is proportional to the initial energy of the rolling tachyon configuration. Note that $\mathfrak{C}\rightarrow 0$ is a singular limit, the annihilation operator vanishes in this limit, and the creation operator blows up. One would've expected $\mathfrak{C}^2\rightarrow 0$ limit to describe the vacuum state where the tachyon is placed at the minimum of the potential. However, this is not the case, and this is in accordance with the fact that time-dependent rolling solutions are distinct from the vacuum solution \cite{Sen:rolling, Sen:tachyon_potential, Sen:1998min_of_V}. Even if the initial configuration has a slight positive energy $\e>0$, the tachyon will roll towards the minimum, and it will not reach the vacuum in finite time. This translates into the fact that the vacuum state is not accessible in the tachyon effective theory, as we see from our analysis.   

 \section{Discussion}
 \label{sec:discussion}
In this paper, we studied the quantization of the tachyon effective action proposed in \cite{Sen:Time_Tachyon} in the late time limit. At the classical level the tachyon system behaves like a non-interacting, non-rotating dust. This motivated us to ask whether a similar picture exists at the quantum level, and in particular, whether the spectrum resembles that of relativistic free particles. Because plane wave solutions are absent in this theory, the usual perturbative quantization cannot be applied. We therefore had to look for an alternative way to quantize the system. 
We thus turn to collective field theory, motivated by the fact that pressureless dust - described in terms of continuous variables such as energy density and velocity- microscopically corresponds to a collection of free relativistic particles. Collective field theory provides precisely this type of construction, where a system having discrete degrees of freedom is described in terms of continuous field variables.
Furthermore, we found that the classical part of relativistic collective field theory gives the dynamics of free relativistic dust, matching the tachyon effective theory in the large $T$ limit. This relation provides a natural way to incorporate systematic $\hbar$ corrections to the tachyon Hamiltonian so as to map it to a quantum theory of free particles. Using this framework, we could determine the spectrum of the tachyon theory. We found that the lowest energy state of the theory is a coherent state $\mc$, of particles at rest, and the
excited states are built on top of this coherent state. So, the Hilbert-space structure of tachyon theory consists of the coherent state $\mc$  and excitations on top of this, ie, a subsector of Klein-Gordon theory.

This structure admits a natural interpretation from the point of view of string theory. In string theory, tachyons are associated with unstable D-branes, and the rolling of tachyons on an unstable D-brane is described by classical time-dependent solutions in open string theory \cite{Sen:rolling} \footnote{Also see \cite{Cho:2023khj, Bernardes:2025zzu} for recent discussions on rolling tachyon solutions from the point of view of OSFT.}. Studying the energy-momentum tensor of this system, one observes that in the late time limit, it behaves like pressureless dust localised on the plane of the original D-brane. On the other hand, in a closed string description, the decaying D-brane acts as a time-dependent source producing closed string radiation. In the $g_s\rightarrow 0$ limit, one can see that the closed string radiation produced from the decay is localised along the plane of the initial D-brane \cite{Lambert:2003zr, Sen:review, Sen:2003bc, Mukhopadhyay:2002en, Gutperle:2004be, Strominger:2002pc, Ambjorn:2003ne}. So, in both open string analysis and closed string analysis, the classical properties of the final system resulting from decay of a D-brane are the same. In this article, we advance this equality to a quantum level. The analysis performed in \cite{Lambert:2003zr} suggests that the final closed string state produced from a homogeneous decay of D-brane is a coherent state of the closed string mode. The lowest energy state we obtain in the tachyon effective theory is also a coherent state. \emph{We thus see a duality between the open string description and closed string description.} In our description, apart from the coherent state, we also have excitations on top of the coherent state. This could be interpreted as the fact that there can be inhomogeneities due to open-string excitations on the initial D-brane. 

Another interesting result is that the vacuum state is not accessible in tachyon theory. This can also be understood from known results in string theory \cite{Sen:review, Sen:rolling, Gibbons:2000hf}. The vacuum configuration is a time-independent solution where the tachyon sits exactly at the minimum of its potential, and the system has zero energy and pressure \cite{Sen:1998min_of_V, Sen:tachyon_potential}. The rolling tachyon solution, on the other hand, is a time-dependent solution in open string theory where the tachyon is displaced from the maximum of the potential and rolls towards the minimum. In an ordinary scalar field theory, one would expect the field to oscillate around the minimum of the potential. However, this does not happen for the tachyon. Instead, the tachyon rolls down the potential while keeping its energy density constant, and it never reaches the minimum in finite time. Even if the tachyon has a small positive energy, it continues to roll towards the minimum \cite{Sen:rolling}. In our analysis, the energy of coherent state $|\mc\ra$ is proportional to $\abs{\mathfrak{C}}^2>0$. This means that our description can only describe time-dependent solutions of string theory where the initial system has at least a small positive energy to begin with. Hence, the vacuum state is not accessible in this description.
 
One immediate goal in the future would be to generalise the analysis of this paper to a collection of particles with different masses. This is because the closed string radiation contains strings of different masses and hence they are described by a collection of Klein-Gordon theories. Thus, we would expect the tachyon theory to accommodate particles having different masses. We expect such an analysis to be technically more involved than the simple case presented in this work. However, we don't expect any conceptual issues in performing this calculation, and we hope to pursue this in future work.
 
 Another interesting direction would be to understand the implications of this work in the context of Cosmology \cite{Sen:Time_Tachyon, Sen:tachyon_cosmo, Brown:1994py, PhysRevD.43.419, Gorini:2003wa, Felder:2002sv}. It was shown in \cite{Sen:Time_Tachyon} that the Tachyon field (T) could be used to provide an intrinsic notion of time in string theory. On coupling tachyon effective theory to gravity and writing down the Wheeler-DeWitt equation, one observes that the theory without the tachyon field is a subsector of the theory in which the tachyon is coupled to gravity. Such a "decoupling" does not happen for usual scalar field theory (Klein-Gordon, for example) coupled to gravity. In fact, what one usually observes is that coupling scalar field theory to gravity and imposing the Wheeler DeWitt equation further constraints the set of allowed states in the theory \cite{Suvrat_ads_wdw, Suvrat_ds1}. It would be interesting to see how the analysis of \cite{Sen:Time_Tachyon} changes if we include the $O(\hbar)$ corrections to the tachyon effective action suggested in this article.  
 
 In this work we have not made any comments on the 1+1-dimensional string theory and its dual formulation in terms of matrix model \cite{McGreevy:2003kb, Douglas:2003up, Takayanagi:2003sm, deBoer:2003hd, McGreevy:2003ep, Mandal:2003tj}. The decay of a D0-brane could be understood from matrix model calculations, and here also one obtains the final state to be a coherent state  \cite{Klebanov:2003km}. Furthermore, the matrix model could be understood in terms of a collective field theory of the eigenvalue density\cite{Das:1990kaa, Das:1995gd, Mandal:2024xqn}. It would be good to understand whether the work presented here has some interesting implications in this context.
\section*{Acknowledgements}
We are grateful to Ashoke Sen for guidance through out the project. We would also like to thank Mohd Ali, Jyotirmoy Barman, Nava Gaddam, Ritwick Kumar Ghosh, Anurag Kaushal, Alok Laddha, R Loganayagam, Raghu Mahajan, Kaustubh Singhi, Aninda Sinha, Farman Ullah, and Avi Wadhwa for helpful discussions. A.H and P.P are grateful to the organisers of the Asian Winter School - IISER Bhopal, where part of this work was completed. P.V.A acknowledges support from the IPDF-IISc fellowship program.  Research at ICTS-TIFR is supported by the Department of Atomic Energy, Government of India, under Project Identification No. RTI4019. P.P is supported by the UKRI Future Leaders Fellowship ``The materials approach to quantum spacetime'' under reference MR/X034453/1.
\vspace{5cm}
\raggedbottom
\appendix
\newpage
\section{Computations in collective theory}
In this appendix we compile calculations in collective field theory. Section \ref{app:field_theory_ops}\textcolor{blue}{.0} explains how to obtain a field theoretic expression for general Quantum mechanical operators, using $\sum_{i=1}^{\ns}$ as an example. The next two sections (\ref{app:non_rel_collective}\textcolor{blue}{.1}, \ref{app:direct_solve_hamiltonian}\textcolor{blue}{.2}) discuss collective theory Hamiltonian and their eigenstates for non-relativistic case, and in section \ref{app:rel_collective}\textcolor{blue}{.3}, we discuss that for the relativistic case. In \ref{app:proof_int_phi_nm}\textcolor{blue}{.4}, we prove $\int d^py~\phi(\vy)=\ns m$ from a field theory perspective. This is complimentary to discussions in the main text where we motivated this result by using $\phi(\vy)=m\sum_{i=1}^{\ns}\g(\vy-\vx_i)$. Also, we use this identity to show that energy eigenstates having $\ns-s$ insertions ($\Phi_{\vk_1,\vk_2,...,\vk_{\ns-s}}$) are equivalent to $\ns$-insertion state ($\Phi_{\vk_1,\vk_2,..,\vk_{\ns}}$ ) with $s$ of the momenta put to zero. By similar methods, in \ref{app:higher_states_null}\textcolor{blue}{.5}, we argue that the states $\Phi_{\vk_1,\vk_2,..,\vk_n}$ for $n>\ns$ are all null states in fixed $\ns$-collective theory. Inner product computation for the $\ns=2$ energy eigenstate is given in \ref{app:inner_pdt_computation}\textcolor{blue}{.6}.
\subsection*{0. Deriving field theoretic expression for QM operators}
\label{app:field_theory_ops}
We will derive the collective theory Hamiltonian for the non-relativistic system of free particles and discuss its eigenstates. However, we will start with finding the field theoretic equivalent of the operator $\sum_{i=1}^{\ns}\vp_i$. This is complimentary to the derivation given in \eqref{eq:sum_p_i_der} and also provides a justification for the same.

Consider the matrix element $\left(\psi_1,\left(\sum_{i=1}^{\ns} \vp_i\right)~\psi_2\right)$ in QM.
\be
\begin{split}
&\left(\psi_1,\left(\sum_{i=1}^{\ns} \vp_i\right)~\psi_2\right)=\left( \left(\sum_{i=1}^{\ns} \vp_i \right) \psi_1,\psi_2\right)
\\&=\int\left( \prod_{i=1}^{\ns} d^px_i \right) \left(i\hbar\sum_{i=1}^{\ns}{\gra_{x_i}}\psi^*_1(\vx_1,\vx_2,..,\vx_{\ns})\right) \psi_2(\vx_1,\vx_2,..,\vx_{\ns})
\\&=\int\left( \prod_{i=1}^{\ns} d^px_i \right)~\int D\phi~\g\bigg[\phi(\vy)-m\sum_{i=1}^{\ns}\g(\vy-\vx_i)\bigg]
 \left(i\hbar\sum_{i=1}^{\ns}{\gra_{x_i}}\psi^*_1(\vx_1,\vx_2,..,\vx_{\ns})\right) \Phi_2[\phi]
\\&=\int\left( \prod_{i=1}^{\ns} d^px_i \right)~\int D\phi~\left(-i\hbar\sum_{i=1}^{\ns}{\gra_{x_i}}\g\bigg[\phi(\vy)-m\sum_{i=1}^{\ns}\g(\vy-\vx_i)\bigg]\right)
 \psi^*_1(\vx_1,\vx_2,..,\vx_{\ns}) \Phi_2[\phi]
\end{split}
\ee
In the 3rd equality we inserted identity \eqref{eq:identity_Jacobian}, and replaced $\psi_2$ with $\Phi_2$. We cannot replace $\psi_1$ with $\Phi_1$ at this point because $\psi_1$ is being differentiated with $\vx_i$ \footnote{We tacitly chose the differential operator to act on $\psi_1$ as this will give the field theory operator that acts on $\Phi_2$.}. In the last equality we did an integration by parts in the $\vx_i$ variable, such that the derivative acts on the delta functional. This can be further simplified as follows,

\be
\label{eq:app_sum_p_i_der}
\begin{split}
	&\left(-i\hbar\sum_{i=1}^{\ns}{\gra_{x_i}}\g\bigg[\phi(\vy)-m\sum_{i=1}^{\ns}\g(\vy-\vx_i)\bigg]\right)=-i\hbar \sum_{i=1}^{\ns} {\gra_{x_i}} \int D\lm~e^{i\int d^py~\lm(\vy)\phi(\vy)}e^{-im\sum_{j=1}^{\ns}\lm(\vx_j)}
	\\&=\int D\lm~e^{i\int d^py~\lm(\vy)\phi(\vy)} e^{-im\sum_{j=1}^{\ns}\lm(\vx_j)}\left(-m\hbar \sum_{i=1}^{\ns}\gra_{\vx_i}\lm(\vx_i) \right)
	\\&=\int D\lm~e^{i\int d^py~\lm(\vy)\phi(\vy)} e^{-im\sum_{j=1}^{\ns}\lm(\vx_j)}\left(-\hbar\int d^py \gra_y\lm(\vy) \sum_{i=1}^{\ns}m\g(\vy-\vx_i) \right)
	\\&=i\hbar\int d^py \sum_{i=1}^{\ns}m\gra_{x_i}\g(\vy-\vx_i) {\g \over \g \phi(\vy)} \int D\lm~e^{i\int d^py~\lm(\vy)\phi(\vy)} e^{-im\sum_{j=1}^{\ns}\lm(\vx_j)}
	\\&=i\hbar\int d^py \sum_{i=1}^{\ns}m\gra_{x_i}\g(\vy-\vx_i) {\g \over \g \phi(\vy)} \g\bigg[\phi(\vy)-m\sum_{i=1}^{\ns}\g(\vy-\vx_i)\bigg]
\end{split}
\ee
Here, in the first equality we used the integral representation of the delta functional. To get the 4th equality, we did an integration by parts in the $\vy$ variable. One can substitute this expression into \eqref{eq:app_sum_p_i_der} to get,
\be
\begin{split}
	\left(\psi_1,\left(\sum_{i=1}^{\ns} \vp_i\right)~\psi_2\right)
	&=\int\left( \prod_{i=1}^{\ns} d^px_i \right)~\int D\phi~
	\psi^*_1(\vx_1,\vx_2,..,\vx_{\ns}) \Phi_2[\phi]
	\\&\left(i\hbar\int d^py \sum_{i=1}^{\ns}m\gra_{x_i}\g(\vy-\vx_i) {\g \over \g \phi(\vy)} \g\bigg[\phi(\vy)-m\sum_{i=1}^{\ns}\g(\vy-\vx_i)\bigg]\right)
	\\&=\int\left( \prod_{i=1}^{\ns} d^px_i \right)~\int D\phi~
	\psi^*_1(\vx_1,\vx_2,..,\vx_{\ns}) \g\bigg[\phi(\vy)-m\sum_{i=1}^{\ns}\g(\vy-\vx_i)\bigg]
	\\&\left(-i\hbar\int d^py \sum_{i=1}^{\ns}m\gra_{x_i}\g(\vy-\vx_i) {\g \over \g \phi(\vy)} \Phi_2[\phi]\right)
	\\&=\int D\phi~
	\Phi_1^*[\phi]\left(\int\left( \prod_{i=1}^{\ns} d^px_i \right)~\g\bigg[\phi(\vy)-m\sum_{i=1}^{\ns}\g(\vy-\vx_i)\bigg]\right)
	\\&\left(-i\hbar\int d^py \sum_{i=1}^{\ns}m\g(\vy-\vx_i) \gra_{y}{\g \over \g \phi(\vy)} \Phi_2[\phi]\right)
	\\&=\int D\phi~J[\phi]
	\Phi_1^*[\phi] \int d^py~\phi(\vy)\gra_{y}\pi(\vy)~\Phi_2[\phi]
\end{split}
\ee
In the second equality, we performed an integration by parts in the $\phi$ variable. The final equation is obtained using the defining equations of the conjugate momenta and the Jacobian. The final result implies that the operator $\left(\sum_{i=1}^{\ns} \vp_i\right)$ is equivalent to,
\be
\left(\sum_{i=1}^{\ns} \vp_i\right)\Rightarrow ~\int d^py~\phi(\vy)\gra_{y}\pi(\vy)
\ee
This is precisely what we got in \eqref{eq:sum_p_i_der} by naive application of the chain rule. However, a more precise derivation, like the one shown here, involves starting with the QM matrix element and inserting \eqref{eq:identity_Jacobian}. Since the final answer is the same, we will only use the chain rule method to get to field-theoretic expressions for relevant operators. 

\subsection*{1. Collective theory Hamiltonian and eigenstates; non-relativistic case}
\label{app:non_rel_collective}
We will derive the expression for non-relativistic Hamiltonian $H=\sum_{i=1}^{\ns}{\vp_i^{~2}\over 2m}$ in terms of collective variables $\phi$ and $\pi$.
\be
\label{eq:p2_term_full_derivation}
\begin{split}
	\sum_{i=1}^{\ns} {\vp_i^{~2}\over 2m}&=-{\hbar^2\over 2m} \sum_{i=1}^{\ns} \gra_{\vx_i}^2
	\\&=-{\hbar^2\over 2m} \sum_{i=1}^{\ns} \gra_{\vx_i}\cdot \int d^py~ \left(\gra_{\vx_i}\phi(\vy)\right){\g \over \g \phi(\vy)}
	\\&=-{\hbar^2\over 2m} \sum_{i=1}^{\ns} \left(\int d^py~\gra_{\vx_i}^2 \phi(\vy){\g\over \g \phi(\vy)}+\int d^py_1d^py_2~\gra_{\vx_i}\phi(\vy_1)\cdot\gra_{\vx_i}\phi(\vy_2){\g^2\over \g \phi(\vy_1)\g \phi(\vy_2)} \right)
	\\&=-{\hbar^2\over 2m} \sum_{i=1}^{\ns} \bigg(\int d^py~\gra_{\vx_i}^2 m\g(\vy-\vx_i)){\g\over \g \phi(\vy)}
	\\&\hspace{3cm}+\int d^py_1d^py_2~\gra_{\vx_i}m\g(\vy_1-\vx_i)\gra_{\vx_i}\cdot m\g(\vy_2-\vx_i){\g^2\over \g \phi(\vy_1)\g \phi(\vy_2)} \bigg)
	\\&=-{\hbar^2\over 2m} \sum_{i=1}^{\ns} \bigg(\int d^py~ m\g(\vy-\vx_i))\gra_{\vy}^2{\g\over \g \phi(\vy)}
	\\&\hspace{3cm}+\int d^py_1d^py_2~m\g(\vy_1-\vy_2) m\g(\vy_2-\vx_i) \gra_{\vy_1}{\g\over \g \phi(\vy_1)} \cdot \gra_{\vy_2}{\g\over \g \phi(\vy_2)}\bigg)
	\\&=-{\hbar^2\over 2m} \left(\int d^py~ \phi(\vy) \gra_{\vy}^2{\g\over \g \phi(\vy)} +m\int d^py~\phi(\vy) \left(\gra_{\vy}{\g\over \g \phi(\vy)}\right)^2 \right)
	\\&={1\over 2}\int d^py~\phi(\vy)\left(\gra\pi(\vy)\right)^2-{i\hbar\over 2m}\int d^py~\phi(\vy) \gra^2\pi(\vy)
\end{split}
\ee
Here we have used chain rule in the 2nd and 3rd equalities, ie, $\gra_{\vx_i}=\int dy \gra_{\vx_i}\phi(y){\g\over \g \phi(\vy)}$. This tacitly assumes that the Jacobian enforces $\phi(\vy)=m\sum_{i=1}^{\ns}\g(\vy-\vx_i)$. Thus, $\gra_{\vx_i}\phi(y)$ is a proxy for $\gra_{\vx_i}\sum_{j=1}^{\ns}m\g(\vy-\vx_j)$. This is made explicit in the 4th equality. Integration by parts is used in the 5th equality, such that derivatives are made to act on ${\g\over \g \phi}$. Then, we again used the definition of $\phi$ and $\pi$ to get to the final expression.  

Next we will show that the state $\Phi_{\vk_1,\vk_2}$ is an eigenstate of this Hamiltonian. We put $\hbar=1$ from now on. One can compute relevant terms one by one.  
\be
\begin{split}
	\gra_y^2 {\g\over \g \phi(\vy)} \Phi_{\vk_1,\vk_2}&={1\over m^2} \int d^py_1d^py_2~e^{i\vk_1\cdot\vx_1+i\vk_2\cdot\vx_2} \left(\phi(\vy_2)\gra_{y}^2\g(\vy-\vy_1)+ \phi(\vy_1)\gra_{y}^2\g(\vy-\vy_2)\right)
	\\& \hspace{4cm}-{1\over m}\int d^py' e^{i(\vk_1+\vk_2)\cdot {\vy~'}} \gra_{y}^2\g(\vy-\vy~')
	\\&={1\over m^2}\int d^py' ~\phi(\vy~') \left(-\vk_1^2 e^{i\vk_1\cdot\vy+i\vk_2\cdot \vy~'}+\vk_1\leftrightarrow \vk_2\right)+{\left(\vk_1+\vk_2\right)^2\over m} e^{i(\vk_1+\vk_2)\cdot \vy} 
\end{split}
\ee
Using this, we get the action of 2nd term in the Hamiltonian as,
\be
\label{eq:app_H_2nd_term}
\begin{split}
	-{1\over 2m}\int d^py ~\phi(\vy)\gra_y^2 {\g\Phi_{\vk_1,\vk_2}\over \g \phi(\vy)}&={\vk_1^2+\vk_2^2\over 2m}\int {d^py_1d^py_2\over m^2} \phi(\vy_1)\phi(\vy_2) e^{i\vk_1\cdot\vy_1+i\vk_2\cdot\vy_2} 
	\\&~~+{(\vk_1+\vk_2)^2\over 2m}\left({-1\over m}\right) \int d^py e^{i(\vk_1+\vk_2)\cdot \vy} \phi(\vy)
\end{split}
\ee
To compute the first term in the Hamiltonian we first have to compute $\gra_{y}{\g\over \g \phi(\vy)}\cdot \gra_{y}{\g\over \g \phi(\vy)} \Phi_{\vk_1,\vk_2} $.
\be
\begin{split}
	&\left(\gra_{y}{\g\over \g \phi(\vy)}\right)^2 \Phi_{\vk_1,\vk_2}
    \\&=	\gra_{y}{\g\over \g \phi(\vy)}\cdot
	{1\over m^2}\int d^py_1d^py_2 e^{i\vk_1\cdot \vy_1+i\vk_2\cdot\vy_2} \left( \phi(\vy_2)\gra_y\g(\vy-\vy_1)+\phi(\vy_1)\gra_y\g(\vy-\vy_2) \right)
	\\&\hspace{5cm}-\gra_{y}{\g\over \g \phi(\vy)}\cdot{1\over m}\int d^py' e^{i(\vk_1+\vk_2)\cdot \vy~'}\gra_y\g(\vy-\vy~')
	\\&={2\over m^2}\int d^py_1d^py_2 e^{i\vk_1\cdot\vy_1+i\vk_2\cdot \vy_2} \left( \gra_y \g(\vy_1-\vy)\cdot \gra_y\g(\vy_2-\vy)\right)
	\\&={1\over m^2}\left(-2\vk_1\cdot\vk_2 \right) e^{i(\vk_1+\vk_2)\cdot\vy}
\end{split}
\ee
So, the first term in Hamiltonian becomes,
\be
-{1\over 2}\int d^py \phi(\vy) \left(\gra_{y}{\g\over \g \phi(\vy)}\right)^2 \Phi_{\vk_1,\vk_2} =
{\vk_1\cdot\vk_2\over m^2}\int d^py~\phi(\vy) e^{i(\vk_1+\vk_2)\cdot \vy} \phi(\vy)
\ee
Adding this with \eqref{eq:app_H_2nd_term} we get the result that $\Phi_{\vk_1,\vk_2}$ is indeed an eigenstate of the Hamiltonian,
\be
\begin{split}
	H\Phi_{\vk_1,\vk_2}&={\vk_1^2+\vk_2^2\over 2m} \bigg( 
	{1\over m^2}\int d^py_1d^py_2e^{i\vk_1\cdot\vy_1+\vk_2\cdot\vy_2}\phi(\vy_1)\phi(\vy_2)
	-{1\over m}\int d^py e^{i(\vk_1+\vk_2)\cdot \vy} \phi(\vy)\bigg)
	\\&={\vk_1^2+\vk_2^2\over 2m}\Phi_{\vk_1,\vk_2}
\end{split}
\ee
\subsection*{2. Directly solving for eigenfunctionals of the Hamiltonian}
\label{app:direct_solve_hamiltonian}
In collective theory, we used the map \eqref{eq:phi_def} to obtain the energy eigenfunctionals in the collective theory, knowing their counterpart in quantum mechanics. Alternatively, one could directly try solving the time independent Schrodinger equation for the collective theory Hamiltonian. This is analogous to what we do in usual Klein-Gordon theory when working in the Schrodinger picture \cite{Hatfield:1992rz}. In this section we will explicitly do this for the relativistic Hamiltonian \eqref{eq:H_non_rel}. Generalising this to the relativistic Hamiltonian is straight forward in principle, but technically tedious.

The non-relativistic collective Hamiltonian in $\hbar=1$ units is,
\be
\label{eq:H_app_non_rel}
H=-{1\over 2}\int d^py ~\phi(\vy) \left(\gra {\g \over \g \phi(\vy)}\right)^2-{1\over 2m}\int d^py~\phi(\vy)\gra^2{\g \over \g \phi(\vy)}
\ee
If $\Phi_E$ is an energy eigenstate of energy $E$, the time-independent Schrodinger equation is given by,
\be
H\Phi[\phi]=E\Phi[\phi]
\ee 
Inspired by the Gaussian ground state wave functional in Klein-Gordon theory, we can also consider a Gaussian ansatz to begin with.
\be
\Phi[\phi]=\exp{-\int d^py_1d^py_2 ~G_2(\vy_1,\vy_2)\phi(\vy_1)\phi(\vy_2)}
\ee
The time independent Schrodinger equation will reduce to the following equation for $G_2$,
\be
\begin{split}
	&\int d^py_1d^py_2d^py_3g_3(\vy_1,\vy_2,\vy_3)\phi(\vy_1)\phi(\vy_2)\phi(\vy_3)
	+\int d^py_1d^py_2g_2(\vy_1,\vy_2)\phi(\vy_1)\phi(\vy_2)
	\\&\hspace{8cm}+\int d^py g_1(\vy)\phi(\vy)=E
	\\& \text{where}~g_1(\vy)=\gra_{\vy_1}\cdot \gra_{\vy_2}G_2(\vy_1,\vy_2)\bigg|_{\vy_1,\vy_2=\vy},~~g_2(\vy_1,\vy_2)=\gra_{\vy_1}^2\tilde{G}(\vy_1,\vy_2)
	\\& \text{and}~g_3(\vy_1,\vy_2,\vy_3)=\gra_{\vy_1}\tilde{G}_2(\vy_1,\vy_2)\cdot \gra_{\vy_1}\tilde{G}_2(\vy_1,\vy_3)
	~~\text{with}~\tilde{G}_2(\vy_1,\vy_2)=G_2(\vy_1,\vy_2)+G_2(\vy_2,\vy_1)
\end{split}
\ee 
One can compare the number of $\phi$s on both sides of the equation to conclude that only trivial solution exists, $G(\vy_1,\vy_2)=\mathfrak{s}\in \mathbb{R}$. So, the ground state having $E=0$ seems to be infinitely degenerate, parametrised by a real number $\mathfrak{s}$. However, all of these are the same state $\Phi=1$. Notice that the Jacobian $J_{\ns}$ enforces $\phi(\vy)=m\sum_{i=1}^{\ns}\g(\vy-\vx_i)$, which implies $\int d^py~\phi(y)=\ns m$ (See \ref{app:proof_int_phi_nm}\textcolor{blue}{.4}). Using this, we get,
\be
\Phi[\phi]=e^{\mathfrak{s}\int d^py_1d^py_2~\phi(\vy_1)\phi(\vy_2)}=e^{\mathfrak{s}\ns^2m^2}
\ee
As the $RHS$ is a number, we can identify this with the $\Phi=1$ state. This is also an instance where the many-to-one map from Collective theory to Quantum Mechanics is evident. 

Having obtained the ground state as $\Phi=1$, we can consider excitaations on top of this state. We propose the following anzatz for an energy eigenstate with non-zero energy.
\be
\Phi_f=\int d^py ~f(y)\phi(y)
\ee
The time-independent Schrodinger equation reduces to,
\be
\int d^py ~\phi(\vy) \left( -{\gra^2\over 2m}\right)f(y)=E\int d^py~\phi(\vy)~f(\vy)
\ee
This gives the usual plane wave solutions for $f$,
\be
\begin{split}
	& \left(\gra^2+2 m E\right) f(\vy)=0
	\\&\Rightarrow ~f(\vy)=e^{ i\vk \cdot \vy};~~E={\vk^2\over 2m}
\end{split}
\ee
These are exactly the states $\Phi_{\vk}$, obtained in the main text. Note that $H\Phi_{\vk}={\vk^2\over 2m}\Phi_{\vk}$, and these states are to be interpreted as $\ ns$-particle state with one of them moving with momenta $\vk$ and the remaining at rest.  However, if $\vk=0$, then owing to the Jacobian, $\Phi_{\vk=0}=\int d^py \phi(\vy)=\ns m$, such that it should be identified with the ground state.

Next, we will consider a general 2-insertion state.
\be
\Phi[\phi]=\int d^py_1d^py_2~f_2(\vy_1,\vy_2)\phi(\vy_1,\vy_2)+\int d^py f_1(\vy)\phi(\vy)
\ee
Acting with the Hamiltonian, we get,
\be
\begin{split}
&H\Phi[\phi]=\int d^py_1d^py_2~g_2(\vy_1,\vy_2)\phi(\vy_1)\phi(\vy_2)+\int d^py ~g_1(\vy)\phi(\vy)
\\& g_2(\vy_1,\vy_2)= -{1\over 2m}\left(\gra_{\vy_1}^2+\gra_{\vy_2}^2\right),~~g_1(\vy)=-{1\over 2m} \gra_{\vy}^2 f_1(\vy)-\gra_{\vy_1}\cdot \gra_{\vy_2}f_2(\vy_1,\vy_2)\bigg|_{\vy_1,\vy_2=\vy}
\end{split}
\ee
Using the Schrodinger equation $H\Phi=E\Phi$ we get the following conditions for $f_1$ and $f_2$,
\be
\begin{split}
	 &(i)~-{1\over 2m}\left(\gra_{\vy_1}^2+\gra_{\vy_2}^2\right)f_2(\vy_1,\vy_2)=E f_2(\vy_1,\vy_2)
	 \\& (ii)~~-{1\over 2m}\gra_{\vy}^2f_1(\vy)-\gra_{\vy_1}\cdot \gra_{\vy_2}f_2(\vy_1,\vy_2)\bigg|_{\vy_1,\vy_2=\vy}=Ef_1(\vy)
\end{split}
\ee
(i) has a unique solution in terms of plane waves.
\be
f_2(\vy_1,\vy_2)=e^{i\vk_1\cdot\vy_1}e^{i\vk_2\cdot\vy_2};~~E={\vk_1^+\vk_2^2\over 2m}
\ee
Substituting this in (ii) gives a differential equation for $f_1$ which can also be uniquely solved.
\be
\begin{split}
	&-{1\over 2m}{\gra_{\vy}^2f_1(\vy)}=\left({\vk_1^2+\vk_2^2\over 2m}\right) f_1(\vy)-\vk_1\cdot\vk_2 e^{i(\vk_1+\vk_2)\cdot \vy}
	\\&\Rightarrow f_1(\vy)=m e^{i(\vk_1+\vk_2)\cdot \vy}
\end{split}
\ee
So the state having energy eigenvalue ${\vk_1^2+\vk_2^2\over 2m}$ can be written as,
\be
\Phi=\int d^py_1d^py_2 ~e^{i\vk_1\cdot\vy_1} e^{i\vk_1\cdot\vy_1} \phi(\vy_1)\phi(\vy_2)-m\int d^py ~e^{i(\vk_1+\vk_2)\cdot \vy}\phi(\vy)
\ee
This state is the same as $\Phi_{\vk_1,\vk_2}$ in \eqref{eq:list_ns_states}, except for an overall ${1\over m^2}$ normalisation factor. 

One can generalise this procedure to obtain other energy eigenfunctionals. Start with the most general $N$ insertion state, 
\begin{equation}
\begin{split}
\Phi=\int \left(\prod_{i=1}^N d^py_i\phi(\vy_1)\right) f_N(\vy_1,..,\vy_N)+\int \left(\prod_{i=1}^{N-1} d^py_i\phi(\vy_1)\right)& f_{N-1}(\vy_1,..,\vy_{N-1})
\\&+...+\int dy~f_1(\vy)\phi(\vy)
\end{split}
\end{equation}
Writing down the time independent schrodinger equation, one can immediately solve for $f_N$ in terms of plane waves, just like the $2$-insertion case discussed above. Once $f_N$ is obtained, substituting it back in the equations will give a differential equation for $f_{N-1}$ that can be solved. Once $f_{N-1}$ is fixed, we can solve for $f_{N-2}$ and so on. One key point to remember is that the many-to-one mapping of collective theory to quantum mechanics also results in degeneracies of eigenstates. This we saw explicitly in the case of the zero-energy ground state. However, one take into account the Jacobian $J_{\ns}$ and the fact that it impliments $\phi(\vy)=m\sum_{i=1}^{\ns}\g(\vy-\vx_i)$ to make sense of the degeneracy.

\subsection*{3. Collective theory Hamiltonian and eigenstates; Relativistic case}
\label{app:rel_collective}
The Hamiltonian for relativistic free particles is given as,
\be
H=\sum_{i=1}^{\ns} \sqrt{\vp_i^2+m^2}=\ns m+\sum_{i=1}^{\ns} {\vp_i^{~2}\over 2m}-\sum_{i=1}^{\ns} {\vp_i^{~4}\over 8m^3}+....
\ee
Each term in this expansion can be expressed in terms of collective theory variables $\phi$ and $\pi$.  The first term is easy, as it follows from the definition of $\phi$ \eqref{eq:phi_def}.
\be
\int d^py ~\phi(\vy)=\ns m
\ee 
We already computed the second term in \eqref{eq:p2_term_full_derivation}. The $\sum_i\vp_i^4$ term can be calculated similarly.
A straightforward integration by parts gives,
\be
\begin{split}
\gra_{x_i}^4=&\int d^py~\gra_{x_i}^4\phi(\vy)~{\g \over \g \phi(\vy)}
+\int d^py_1d^py_2~\bigg(
2\gra_{x_i}\phi(\vy_1)\cdot \gra_{x_i}\left(\gra_{x_i}^2\phi(\vy_2)\right)
\\&+\gra_{x_i}^2\phi(\vy_1)\gra_{x_i}\phi(\vy_2)+\gra_{x_i}\cdot \gra_{x_i} \left(\gra_{x_i}\phi(\vy_1)\cdot\gra_{x_i}\phi(\vy_2)\right)
\bigg){\g^2\over \g\phi(\vy_1)\g\phi(\vy_2)}
\\&+\int d^py_1d^py_2d^py_3~\bigg(
2\gra_{x_i}\phi(\vy_1)\cdot \gra_{x_i} \phi(\vy_2)~\gra_{x_i}^2\phi(\vy_3)
\\&+2\gra_{x_i}\phi(\vy_1)\cdot \gra_{x_i}\left( \gra_{x_i}\phi(\vy_2)\cdot \gra_{x_i}\phi(\vy_3) \right)
\bigg){\g^2\over \g(\vy_1)\g(\vy_2)\g(\vy_3)}
\\&+\int d^py_1d^py_2d^py_3d^py_4 \gra_{x_i}\phi(\vy_1)\cdot \gra_{x_i}\phi(\vy_2)~\gra_{x_i}\phi(\vy_3)\cdot\gra_{x_i}\phi(\vy_4){\g^2\over \g(\vy_1)\g(\vy_2)\g(\vy_3)\g\phi(\vy_4)}
\end{split}
\ee
Recall that $\gra_{\vx_i}\phi(y)$ is a proxy for $\gra_{\vx_i}\sum_{j=1}^{\ns}m\g(\vy-\vx_j)$ just like in \eqref{eq:p2_term_full_derivation}. One can then use the delta function property ${\gra_{\vx_i}}\g(\vy-\vx_i){\g\over \g \phi(\vy)}=\g(\vy-\vx_i)\gra_{\vy}{\g\over \g \phi(\vy)}$ such that all the derivatives now act on ${\g \over \g \phi}$. The resulting term can be summed over from $i=1$ to $\ns$ to get the following answer.
\be
\label{eq:H_3rd_term_app}
\begin{split}
	-{\hbar^4\over 8m^3}\sum_{i=1}^{\ns}{\gra^4_{\vx_i}}=-{\hbar^4\over 8m^3}\int d^py \phi(\vy) \bigg\{&
	\gra_y^4{\g\over \g \phi(\vy)}+2m \gra_y{\g\over \g \phi(\vy)}\cdot \gra \left(\gra_y^2{\g\over \g \phi(\vy)}\right)
	\\&+m \gra_y^2{\g\over \g \phi(\vy)}\gra_y^2{\g\over \g \phi(\vy)}+m\gra_y\cdot \gra \left(\gra {\g\over \g \phi(\vy)}\right)^2
	\\&+2m^2 \left(\gra_y {\g\over \g \phi(\vy)}\right)^2\gra_y^2{\g\over \g \phi(\vy)}
	\\&+2m^2 \gra_y{\g\over \g \phi(\vy)}\cdot \gra_y \left(\gra_y{\g\over \g \phi(\vy)}\right)^2
	+m^3 \left(\gra_y{\g\over \g \phi(\vy)}\right)^4
	\bigg\} 
\end{split}
\ee 
Upon the substitution $\pi=-i\hbar {\g \over \g \phi(\vy)}$, one can verify that this is the same as what we wrote in \eqref{eq:H_first3_terms}. One can generalise this procedure to arbitrary higher orders in $\vp_i^2$. One would naturally ask the question, " is there a compact form for the relativistic Hamiltonian?". As we saw in \ref{sec:collective_to_tachyon}, the answer is yes, given that you introduce creation and annihilation operators.  But we know that for collective field theory we work with a fixed value of $\ns$, such that the creation and annihilation operators cannot be introduced with out changing the Jacobian. However, we can still use the compact form of the Hamiltonian in terms of $\phi$ and ${\g \over \g\phi}$, given below. 
\be
\begin{split}
	H&=\int {d^pk\over (2\pi)^p} \w_{\vk} a^{\da}_{\vk}a_{\vk}
	\\&=\int {d^pk\over (2\pi)^p} \sqrt{\vk^2+m^2} {1\over m}\int d^py_1~e^{i\vk\cdot\vy_1} \phi(\vy) e^{-m{\g\over \g \phi(\vy)}} \int d^py_2 e^{-i\vk\cdot \vy_2} e^{m{\g\over \g \phi(\vy)}}
	\\&=\int d^py_1d^py_2 {1\over m} \phi(\vy_1)e^{-m{\g\over \g\phi(\vy_1)}} e^{m{\g\over \g\phi(\vy_2)}} \int {d^pk\over (2\pi)^p} \sqrt{\vk^2+m^2} e^{i\vk\cdot(\vy_1-\vy_2)}
	\\&=\int d^py_1d^py_2 {1\over m} \phi(\vy_1)e^{-m{\g\over \g\phi(\vy_1)}} e^{m{\g\over \g\phi(\vy_2)}} \sqrt{-\gra_{y_2}^2+m^2}~ \g(\vy_1-\vy_2)
	\\&=\int d^py {1\over m}\phi(\vy)e^{-m{\g\over \g\phi(\vy)}} \sqrt{-\gra_y^2+m^2}~~ e^{m{\g\over \g\phi(\vy)}}
\end{split}
\ee

The term $\sqrt{-\gra_y^2+m^2}$ is to be understood as the series, 
\be
\sqrt{-\gra_y^2+m^2}=m-{\gra_y^2\over 2m}-{\left(\gra_y^2\right)^2\over 8m^3}+.....
\ee
We will compute the expression for the Hamiltonian term by term. Note that,
\be
\begin{split}
	&\gra_y^2 e^{f(\vy)}=e^{f(\vy)} \bigg\{ \left(\gra_yf(\vy)\right)^2+\gra_y^2f(\vy)  \bigg\}
	\\& \gra_y^4 e^{f(\vy)}=e^{f(\vy)} \bigg\{  
	\left(\gra_yf(\vy)\right)^4+2\gra_y^2f(\vy)\left(\gra_yf(\vy)\right)^2+2\gra_yf(\vy)\cdot \gra \left(\gra_yf(\vy)\right)^2
	\\&\hspace{1.8cm}+2\gra_yf(\vy)\cdot\gra\left(\gra_y^2f\right)+\gra_y^2f(\vy)\gra_y^2f(\vy)+\gra_y\cdot \left(\gra_yf(\vy)\right)^2+\gra_y^4f(\vy)
	 \bigg\}
\end{split}
\ee
We used compact notations such as, $(\gra_yf)^2=\gra_yf\cdot\gra_yf$. One can use $f(\vy)=m{\g\over \g \phi(\vy)}$ and evaluate the expression for $H$. The first three terms in the series expansion give,
\begin{equation*}
\begin{split}
	&\int d^py {1\over m}\phi(\vy)e^{-m{\g\over \g\phi(\vy)}} m e^{m{\g\over \g\phi(\vy)}}=\int d^py~\phi(\vy)
	\\& \int d^py {1\over m}\phi(\vy)e^{-m{\g\over \g\phi(\vy)}} \left(-{\gra_y^2\over 2m}\right) e^{m{\g\over \g\phi(\vy)}}=-\int d^py~\phi(\vy) \left( {1\over 2}\left(\gra_y {\g\over \g \phi(\vy)}\right)^2+{1\over 2m}\gra_y^2{\g\over \g\phi(\vy)} \right)
	\\&\int d^py {1\over m}\phi(\vy)e^{-m{\g\over \g\phi(\vy)}} \left(-{\gra_y^4\over 8m^3}\right) e^{m{\g\over \g\phi(\vy)}}
	=-\int d^py~\phi(\vy)
	\bigg\{
	{1\over 8}\left(\gra_y{\g\over\g\phi(\vy)}\right)^4
	\\&\hspace{4cm}+{1\over 4m}\gra_y^2{\g\over \g\phi(\vy)}\left(\gra_y{\g\over \g\phi(\vy)}\right)^2
	+{1\over 4m}\gra_{y}{\g\over \g\phi(\vy)}\cdot \gra \left({\g\over \g\phi(\vy)}\right)^2
	\\&\hspace{4cm}+{1\over 4m^2}\gra {\g\over \g\phi(\vy)} \cdot \gra \left(\gra^2{\g\over \g\phi(\vy)}\right)
	+{1\over 8m^2}\gra_y^2{\g\over \g\phi(\vy)}\gra_y^2{\g\over \g\phi(\vy)}
	\\&\hspace{4cm}+
	{1\over 8m^2}\gra\cdot\gra\left(\gra_y{\g\over \g\phi(\vy)}\right)^2
	+{1\over 8m^3}\gra_y^4{\g\over \g\phi(\vy)}
	\bigg\}
\end{split}
\end{equation*}
These precisely matches with the expressions \eqref{eq:p2_term_full_derivation}, \eqref{eq:H_3rd_term_app}. It is clear that all terms in the Hamiltonian will match with those obtained from collective field theory analysis. However, an all-order proof looks tedious, and we will not discuss it now.

Let us now show that the energy eigenstates given in \eqref{eq:list_ns_states} are indeed eigenstates of the relativistic Hamiltonian. We will use the compact form of the Hamiltonian,
\be
H=\int d^py {1\over m}\phi(\vy)e^{-m{\g\over \g\phi(\vy)}} \sqrt{-\gra_y^2+m^2}~~ e^{m{\g\over \g\phi(\vy)}}
\ee
Let us start with the simple state $\mc$.
\be
\begin{split}
H~\mc&=\int d^py {1\over m}\phi(\vy)e^{-m{\g\over \g\phi(\vy)}} \sqrt{-\gra_y^2+m^2}~~ e^{m{\g\over \g\phi(\vy)}}\cdot 1
\\&=\int d^py~\phi(\vy)
\\&=\ns m ~\mc
\end{split}
\ee
In the last step we input the fact that we are in a collective theory with fixed $\ns$ such that $\int d^py \phi(y)=\ns m$\footnote{Note that this is not true if you are in Tachyon theory. There it is just the number operator times the mass, $\int d^py~\Pi(\vy)=N m$}. 

Next, consider the 1-insertion states $\Phi_{\vk}$.
\be
\label{eq:app_H_Phi_k_proof}
\begin{split}
H~\Phi_{\vk}&=\int d^py {1\over m}\phi(\vy)e^{-m{\g\over \g\phi(\vy)}} \sqrt{-\gra_y^2+m^2}~~ e^{m{\g\over \g\phi(\vy)}} \left({1\over m}\int d^py' e^{i\vk\cdot \vy~'} \phi(\vy~')\right)
\\&=\int d^py {1\over m}\phi(\vy)e^{-m{\g\over \g\phi(\vy)}} \sqrt{-\gra_y^2+m^2}~~ \left({1\over m}\int d^py' e^{i\vk\cdot \vy~'} \left(\phi(\vy~')+m\g(\vy-\vy~')\right)\right)
\end{split}
\ee
Where the action of $e^{m{\g \over \g \phi(y)}}$ shifts the field $\phi(\vy~')$ by $m\g(\vy-\vy~')$. We will now simplify the terms in \eqref{eq:app_H_Phi_k_proof} one by one. The first term gives, 
\be
\begin{split}
	\int d^py {1\over m}\phi(\vy) &e^{-m{\g\over \g\phi(\vy)}} \sqrt{-\gra_y^2+m^2}~~ \left({1\over m}\int d^py' e^{i\vk\cdot \vy~'} \phi(\vy~') \right)
	\\&=\int d^py \phi(\vy) ~~ \left({1\over m}\int d^py' e^{i\vk\cdot \vy~'} \left(\phi(\vy~') -m\g(\vy-\vy~')\right)\right)
	\\&=\left(\int d^py \phi(\vy)-m\right)\Phi_{\vk}
\end{split}
\ee
Here also we used $e^{-m{\g \over \g \phi(y)}}$ to shift $\phi(\vy~')$. As for the second term in \eqref{eq:app_H_Phi_k_proof} we get,
\be
\begin{split}
	\int d^py {1\over m}\phi(\vy)&e^{-m{\g\over \g\phi(\vy)}} \sqrt{-\gra_y^2+m^2}~~ \left({1\over m}\int d^py' e^{i\vk\cdot \vy~'} m\g(\vy-\vy~')\right)
	\\&=\int d^py {1\over m}\phi(\vy)e^{-m{\g\over \g\phi(\vy)}} \sqrt{-\gra_y^2+m^2}~~ e^{i\vk\cdot \vy}
	\\&=\int d^py {1\over m}\phi(\vy) \sqrt{\vk^2+m^2}~~ e^{i\vk\cdot \vy}
	\\&=\w_{\vk}\Phi_{\vk}
\end{split}
\ee
Putting these together, we get,
\be
\begin{split}
H~\Phi_{\vk}&=\left(\int d^py \phi(\vy)+\w_{\vk}-m\right)\Phi_{\vk}
\\&=\left(\w_{\vk}+(\ns-1)m \right)\Phi_{\vk}
\end{split}
\ee
here we expanded the square root $\w_{\vk}=\sqrt{\vk^2+m^2}$ and also used $\int d^py~\phi(\vy)=\ns m$. This makes it clear that $\Phi_{\vk}$ is an eigenstate of the collective theory Hamiltonian, and it physically means an $\ns$-particle state with one particle moving with momentum $\vk$ and the remaining particles at rest.

We will show that the $2$-insertion state is also an eigenstate of the Hamiltonian. The calculation is straightforward and is very similar to the case we discussed above. 
\be
\label{eq:app_tilde_2_particle_def}
\begin{split}
\Phi_{\vk_1,\vk_2}&={1\over m^2}\int d^py_1 d^py_2~e^{i\vk_1\cdot\vy_1}e^{i\vk_2\cdot\vy_2} \phi(\vy_1)\phi(\vy_2)-{1\over m}\int d^py~e^{i(\vk_1+\vk_2)\cdot \vy}\phi(\vy)
\\&={1\over m^2}\int d^py_1d^py_2~e^{i\vk_1\cdot\vy_1+i\vk_2\cdot\vy_2} ~\phi(\vy_1)\bigg(\phi(\vy_2)-m\g(\vy_2-\vy_1)\bigg)
\end{split}
\ee
In the second line, we wrote it in a form that will be useful for generalisation. Acting with $H$ on this, we get,
\be
\begin{split}
H~\Phi_{\vk_1,\vk_2}&={1\over m}\int d^py~\phi(\vy)e^{-m{\g\over \g \phi(\vy)}} \sqrt{-\gra_{\vy}^2+m^2}~e^{m{\g \over \g \phi(\vy)}} 
\\&~~~~\times{1\over m^2}\int d^py_1d^py_2~e^{i\vk_1\cdot\vy_1+i\vk_2\cdot\vy_2}~\phi(\vy_1)
\bigg(\phi(\vy_2)-m\g(\vy_2-\vy_1)\bigg)
\\&={1\over m}\int d^py~\phi(\vy)e^{-m{\g\over \g \phi(\vy)}} \sqrt{-\gra_{\vy}^2+m^2}~{1\over m^2}\int d^py_1d^py_2~e^{i\vk_1\cdot\vy_1+i\vk_2\cdot\vy_2}
\\&\times~\bigg(\underline{\phi(\vy_1)}+m\g(\vy-\vy_1)\bigg)
\bigg(\underline{\phi(\vy_2)-m\g(\vy_2-\vy_1)}+m\g(\vy-\vy_2)\bigg)
\end{split}
\ee
The underlined terms are independent of $\vy$. We will denote the term having only the products of the underlined terms as $I_1$ and the rest of the terms as $I_2$. Let us first evaluate $I_1$.
\be
\begin{split}
	I_1& \equiv {1\over m}\int d^py~\phi(\vy)e^{-m{\g\over \g \phi(\vy)}} \sqrt{-\gra_{\vy}^2+m^2}~{1\over m^2}\int d^py_1d^py_2~e^{i\vk_1\cdot\vy_1+i\vk_2\cdot\vy_2} \phi(\vy_1)\times
    \\& \hspace{8cm}\bigg(\phi(\vy_2)-m\g(\vy_2-\vy_1)\bigg)
	\\&=\int d^p\vy ~\phi(\vy) ~\int {d^py_1d^py_2\over m^2}~e^{i\vk_1\cdot\vy_1+i\vk_2\cdot\vy_2} \bigg(\phi(\vy_1)-m\g(\vy_1-vy)\bigg) \times 
	\\&\hspace{7cm}\bigg(\phi(\vy_2)-m\g(\vy_2-\vy_1)-m\g(\vy_2-\vy)\bigg)
	\\&=\left(\int d^py ~\phi(\vy)-2m\right) {1\over m^2}\int d^py_1d^py_2~e^{i\vk_1\cdot\vy_1+i\vk_2\cdot\vy_2} ~\phi(\vy_1)\bigg(\phi(\vy_2)-m\g(\vy_2-\vy_1)\bigg)
	\\&=\left(\int d^py ~\phi(\vy)-2m\right) \Phi_{\vk_1,\vk_2}
\end{split}
\ee
Here we have made use of  $e^{\pm m{\g\over \g \phi(\vy)}} \phi(\vy~')=\phi(\vy')\pm m\g(\vy'-\vy)$.  Similarly, one can easily evaluate $I_2$. One just have to keep track of the fact that here there are terms that depend on $\vy$ on which the differential operator $\sqrt{-\gra_{\vy}^2+m^2}$ can act.
\be
\begin{split}
	&I_2=\int {d^py\over m}~\phi(\vy)e^{-m{\g\over \g \phi(\vy)}} \sqrt{-\gra_{\vy}^2+m^2}~{1\over m^2}\int d^py_1d^py_2~e^{i\vk_1\cdot\vy_1+i\vk_2\cdot\vy_2} \times
\\&\hspace{8cm}\bigg(\phi(\vy_1)m\g(\vy-\vy_2)+\vy_1\leftrightarrow \vy_2\bigg)
	\\&=\int {d^py}~\phi(\vy)e^{-m{\g\over \g \phi(\vy)}} ~\int {d^py_1d^py_2\over m^2}~e^{i\vk_1\cdot\vy_1+i\vk_2\cdot\vy_2}\bigg(\phi(\vy_1)\sqrt{-\gra_{\vy}^2+m^2}~\g(\vy-\vy_2)+\vy_1\leftrightarrow \vy_2\bigg)
	\\&=\int {d^py}~\phi(\vy)e^{-m{\g\over \g \phi(\vy)}} ~\int {d^py_1d^py_2\over m^2}~e^{i\vk_1\cdot\vy_1+i\vk_2\cdot\vy_2}\bigg(\w_{\vk_1}\phi(\vy_1)~\g(\vy-\vy_2)+
	\w_{\vk_2}\phi(\vy_2)~\g(\vy-\vy_1)\bigg)
	\\&=\int {d^py}~\phi(\vy) ~\int {d^py_1d^py_2\over m^2}~e^{i\vk_1\cdot\vy_1+i\vk_2\cdot\vy_2}\bigg(\w_{\vk_1}\left(\phi(\vy_1)-m\g(\vy-\vy_1)\right)~\g(\vy-\vy_2)+
	1\leftrightarrow2\bigg)
	\\&=\left(\w_{\vk_1}+\w_{\vk_2}\right) {1\over m^2}\int d^py_1d^py_2~e^{i\vk_1\cdot\vy_1+i\vk_2\cdot\vy_2} ~\phi(\vy_1)\bigg(\phi(\vy_2)-m\g(\vy_2-\vy_1)\bigg)
	\\&=\left(\w_{\vk_1}+\w_{\vk_2}\right) \Phi_{\vk_1,\vk_2}
\end{split}
\ee
Adding $I_1$ and $I_2$, we get,
\be
\begin{split}
	H~\Phi_{\vk_1,\vk_2}&=\left(\int d^py ~\phi(\vy)-2m+\w_{\vk_1}+\w_{\vk_2}\right)\Phi_{\vk_1,\vk_2}
	\\&=\left(\w_{\vk_1}+\w_{\vk_2}+\left(\ns-2\right)m\right)\Phi_{\vk_1,\vk_2}
\end{split}
\ee

This is also in accordance with our interpretation that $\Phi_{\vk_1,\vk_2}$ is an $\ns$ particle eigenstate of the collective Hamiltonian with two particles having momenta $\vk_1,\vk_2$ and the remaining at rest.
We close this section by generalising the results to $n$-insertion state $\Phi_{\vk_1,\vk_2,....,\vk_n}$, with $n\leq \ns$.
\be
\label{eq:app_H_on_Phi_n}
\begin{split}
	H~\Phi_{\vk_1,\vk_2,...,\vk_n}&= \left( \int d^py ~\phi(\vy)+\sum_{i=1}^n\w_{\vk_i}-nm\right)
	\\&=\left(\sum_{i=1}^{n}\w_{\vk_i}+(\ns-n)m\right)\Phi_{\vk_1,\vk_2,...,\vk_n}
\end{split}
\ee
This can be interpreted as an $\ns$-particle state with $n$ of the particles moving with  momenta $\{\vk_1,\vk_2,..,\vk_n\}$ and the remaining $\ns-n$ particles at rest. 

\subsection*{4. Proof of $\int d^py ~\phi(\vy)=\ns m$}
\label{app:proof_int_phi_nm}
In this section we will prove that $\int d^py ~\phi(\vy)=\ns m$. This is essentially a result of the definition of the Jacobian. 
\be
\label{eq:app_J_defn}
J_{\ns}[\phi]=\int \left(\prod_{i=1}^{\ns}d^px_i\right) \g\bigg[\phi(\vy)-m\sum_{i=1}^{\ns}\g(\vy-\vx_i)\bigg]
\ee
Consider two arbitrary states $\Phi_1$ and $\Phi_2$ in collective theory. The matrix element of $\int d^py ~\phi(\vy)$ between these two states can be shown to be equal to $\ns m$ as follows.
\be
\begin{split}
	\left(\Phi_1,\int d^py ~\phi(\vy)~\Phi_2\right)&=\int D\phi~J_{\ns}[\phi]\Phi_1^*[\phi] \left(\int d^py ~\phi(\vy)\right)~\Phi_2[\phi]
	\\&=\int D\phi~\int \left(\prod_{i=1}^{\ns}d^px_i\right) \g\bigg[\phi(\vy)-m\sum_{i=1}^{\ns}\g(\vy-\vx_i)\bigg]\times
    \\& \hspace{5cm}~\Phi_1^*[\phi] \left(\int d^py ~\phi(\vy)\right)~\Phi_2[\phi]
	\\&=\ns m~\int \left(\prod_{i=1}^{\ns}d^px_i\right) \psi_1^*(\vx_1,\vx_2,..,\vx_{\ns}) \psi_2(\vx_1,\vx_2,..,\vx_{\ns})
	\\&=\ns m \left(\Phi_1,\Phi_2\right)
\end{split}
\ee
We have used the definition of Jacobian in the 2nd equality, followed by defining the quantum mechanical wavefunctions $\psi_1$ and $\psi_2$ in the 3rd equality as
\begin{equation*}
	\begin{split}
		&\Phi_1[\phi]\bigg|_{\phi(y)=m\sum_{i=1}^{\ns}\g(\vy-\vx_i)}=\psi_1(\vx_1,\vx_2,..,\vx_{\ns}) \\&\Phi_2[\phi]\bigg|_{\phi(y)=m\sum_{i=1}^{\ns}\g(\vy-\vx_i)}=\psi_2(\vx_1,\vx_2,..,\vx_{\ns}).
	\end{split}
\end{equation*}
The final equality follows from the equivalence between quantum mechanics inner product and caollective theory inner product owing to the Jacobian \eqref{eq:app_J_defn}.

One can also prove these results without invoking any quantum mechanics variables and staying purely on the field theory side. This requires the use of the alternative form of Jacobian given in \eqref{eq:non_rel_J}. But the key idea is just the fact that the definition of $J_{\ns}$ implements $\phi(\vy)=m\sum_{i=1}^{\ns}\g(\vy-\vx_i)$.
\be
\begin{split}
	J_{\ns}[\phi] \int d^py~\phi(\vy)&=\int D\lm~ e^{i\int d^p y' ~\lm(\vy~')\phi(\vy~')} \left(\int dx ~e^{-im~\lm(\vx)}\right)^{\ns}\int d^py~\phi(\vy)
	\\&= \int d^py~\int D\lm~\left(\int dx ~e^{-im~\lm(\vx)}\right)^{\ns} \left(-i{\g  \over \g \lm(\vy)}\right) \left(e^{i\int d^p y' ~\lm(\vy~')\phi(\vy~')} \right)  
	\\&=\int d^py~\int D\lm~\left(e^{i\int d^p y' ~\lm(\vy~')\phi(\vy~')} \right)   \left(i{\g  \over \g \lm(\vy)}\right) \left(\int dx ~e^{-im~\lm(\vx)}\right)^{\ns}
	\\&=\int D\lm\left(e^{i\int d^p y' \lm(\vy~')\phi(\vy~')} \right)    \left(\int dx e^{-im\lm(\vx)}\right)^{\ns-1}\ns m \int d^py e^{-im\lm(\vy)}
	\\&=\int D\lm~\left(e^{i\int d^p y' ~\lm(\vy~')\phi(\vy~')} \right)    \left(\int dx ~e^{-im~\lm(\vx)}\right)^{\ns}\ns m 
	\\&=J_{\ns}[\phi]~\ns m
\end{split}
\ee
This makes it clear that working in collective field theory at fixed $\ns$ we get, $\int d^py ~\phi(\vy)=\ns m$.

This result can be put to use for proving the statement: Energy eigenstates having $\ns-s$ insertions ($\Phi_{\vk_1,\vk_2,...,\vk_{\ns-s}}$) are equivalent to $\ns$-insertion state ($\Phi_{\vk_1,\vk_2,..,\vk_{\ns}}$ ) with $s$ of the momenta put to zero. 

We will show this explicitly for $\ns=2$. We saw, in this case, that the state $\Phi_{\vk,0}$ should be identified with $\Phi_{\vk}$. This follows immediately from using $\int d^py ~\phi(\vy)=\ns m$. in the definition of $\Phi_{\vk,0}$.
\be
\begin{split}
   \Phi_{\vk,0}&={1\over m^2}\int d^py_1d^py_2 ~e^{i\vk\cdot\vy_2}~\phi(\vy_1)\phi(\vy_2)-{1\over m}\int d^py~e^{i\vk\cdot\vy}\phi(\vy)
   \\&={2\over m}\int d^py_2 ~e^{i\vk\cdot\vy_2}~\phi(\vy_2)-{1\over m}\int d^py~e^{i\vk\cdot\vy}\phi(\vy)
   \\&={1\over m}\int d^py~e^{i\vk\cdot\vy}\phi(\vy)
   \\&=\Phi_{\vk}
\end{split}
\ee
In the main text, we motivated this result by directly using $\phi(\vy)=m\sum_{i=1}^2\g(\vy-\vx_i)$. Here we see that one can independently derive this in a more field theoretic way, although the idea is the same, that the Jacobian enforces $\phi(\vy)=m\sum_{i=1}^2\g(\vy-\vx_i)$. The proof can be generalised for any value of $\ns$.

\subsection*{5. Higher insertion energy eigenstates are null }
\label{app:higher_states_null}
We will prove that when you work in collective theory at fixed $\ns$, an energy eigenstate $\Phi_{\vk_1,\vk_2,...,\vk_{n}}$ with $n>\ns$ is a null state in the theory. This is also a direct consequence of the definition of the Jacobian \eqref{eq:app_J_defn}. 

The two-insertion state is given by,
\be
\begin{split}
	\Phi_{\vk_1,\vk_2}&={1\over m^2}\int d^py_1d^py_2~e^{i\vk_1\cdot\vy_1+i\vk_2\cdot\vy_2}~\phi(\vy_1)\phi(\vy_2)-{1\over m}\int d^py~e^{i(\vk_1+\vk_2)\cdot \vy}\phi(\vy)
	\\&={1\over m^2}\int d^py_1d^py_2 e^{i\vk_1\cdot\vy_1+i\vk_2\cdot\vy_2} \phi(\vy_1)\bigg( \phi(\vy_2)-m\g(\vy_2-\vy_1)\bigg)
\end{split}
\ee
In the last line we wrote it in a form which will be useful for future generalisation.
Working with $\ns=1$, the Jacobian $J_{\ns=1}$ will enforce $\phi(\vy)=m\g(\vy-\vx)$. Substituting this in the above expression gives,
\be
\begin{split}
	\Phi_{\vk_1,\vk_2}\bigg|_{\phi(\vy)=m\g(\vy-\vx)}&={1\over m^2}\int d^py_1d^py_2 e^{i\vk_1\cdot\vy_1+i\vk_2\cdot\vy_2}~m \g(\vy_1-\vx)\bigg( m\g(\vy_2-\vx)-m\g(\vy_2-\vy_1)\bigg)
	\\&=0
\end{split}
\ee
This means that the state $\Phi_{\vk_1,\vk_2}$ is a null state for $\ns=1$. One can also arrive at this result from a field theory computation by using \eqref{eq:non_rel_J}. We will quickly sketch this below.
\be
\begin{split}
	J_1[\phi]\Phi_{\vk_1,\vk_2}&=\int D\lm e^{i\int d^py'~\lm(\vy~')\phi(\vy~')} \left(\int d^pxe^{-im\lm(\vx)}\right) 
	\\&\times{1\over m^2}\int d^py_1d^py_2 e^{i\vk_1\cdot\vy_1+i\vk_2\cdot\vy_2} \phi(\vy_1)\bigg( \phi(\vy_2)-m\g(\vy_2-\vy_1)\bigg)
	\\&=\int {d^py_1d^py_2\over m^2} e^{i\vk_1\cdot\vy_1+i\vk_2\cdot\vy_2}\bigg\{ \int D\lm \left(\int d^pxe^{-im\lm(\vx)}\right) 
	\\&\times \bigg(-{\g^2\over \g \lm(\vy_1)\g \lm(\vy_2)}+im\g(\vy_1-\vy_2){\g\over \g \lm(\vy_1)}\bigg)e^{i\int d^py'~\lm(\vy~')\phi(\vy~')}\bigg\}
\end{split}
\ee
Here in the 2nd equality we used $\phi(\vy)e^{i\int \lm \phi}=-i{\g\over \g \lm (\vy)} e^{i\int \lm \phi}$.
One can do integration by parts and show that the terms in the curly bracket evaluate to zero.
\be
\begin{split}
	\int &D\lm \left(\int d^pxe^{-im\lm(\vx)}\right) 
	\bigg(-{\g^2\over \g \lm(\vy_1)\g \lm(\vy_2)}+im\g(\vy_1-\vy_2){\g\over \g \lm(\vy_1)}\bigg)e^{i\int d^py'~\lm(\vy~')\phi(\vy~')}
	\\&= \int D\lm e^{i\int d^py'~\lm(\vy~')\phi(\vy~')}
	\bigg(-{\g^2\over \g \lm(\vy_1)\g \lm(\vy_2)}-im\g(\vy_1-\vy_2){\g\over \g \lm(\vy_1)}\bigg)\left(\int d^pxe^{-im\lm(\vx)}\right) 
	\\&= \int D\lm e^{i\int d^py'~\lm(\vy~')\phi(\vy~')}
	\ \bigg(m^2\g(\vy_1-\vy_2)e^{-im\lm(\vy_1)}-m^2\g(\vy_1-\vy_2)e^{-im\lm(\vy_1)}\bigg)
	\\&=0
\end{split}
\ee
This proves,
\be
J_1[\phi]\Phi_{\vk_1,\vk_2}=0,
\ee
which implies that for $\ns=1$, the state $\Phi_{\vk_1,\vk_2}$ is null.
\be
\left(\Phi,\Phi_{\vk_1,\vk_2}\right)\stackrel{\ns=1}{=}0~\forall ~\Phi
\ee
This argument can be generalised to any value of $\ns$. We will briefly show it for $\Phi_{\vk_1,\vk_2,\vk_3}$, and then generalise for arbitrary $\ns$. The 3-insertion energy eigenstates is given by,
\be
\begin{split}
	\Phi_{\vk_1,\vk_2,\vk_3}&={1\over m^3}\int d^py_1d^py_2d^py_3~e^{i\vk_1\cdot\vy_1+i\vk_2\cdot\vy_2+i\vk_3\cdot\vy_3} \phi(\vy_1)\bigg(\phi(\vy_2)-m\g(\vy_2-\vy_1)\bigg)
	\\&\hspace{5cm}\times~\bigg(\phi(\vy_3)-m\g(\vy_3-\vy_2)-m\g(\vy_3-\vy_1)\bigg)
\end{split}
\ee
If you put $\ns=3$, you will get the $3$-particle energy eigenstate in QM, labeled by momenta $\{\vk_1,\vk_2,\vk_3\}$. If you put $\ns=2,1$, you can easily see that the state reduces to zero as claimed. For example, putting $\ns=2$ in the above equation gives,
\be
\begin{split}
	\Phi_{\vk_1,\vk_2,\vk_3}&={1\over m^3}\int d^py_1d^py_2d^py_3~e^{i\vk_1\cdot\vy_1+i\vk_2\cdot\vy_2+i\vk_3\cdot\vy_3} 
	\bigg(m\g(\vy_1-\vx_1)+m\g(\vy_1-\vx_2) \bigg)
	\\&~~~~\times\bigg(m\g(\vy_1-\vx_1)+m\g(\vy_1-\vx_2)-m\g(\vy_2-\vy_1)\bigg)
	\\&~~~~~\times~
	\bigg(m\g(\vy_3-\vx_1)+m\g(\vy_3-\vx_2)-m\g(\vy_3-\vy_2)-m\g(\vy_3-\vy_1)\bigg)
	\\&=0
\end{split}
\ee
This result implies that in $\ns=2$ collective theory $\Phi_{\vk_1,\vk_2,\vk_3}$ is a null state.

It is clear from here that the general $\ns$ result will hold. We will briefly argue for this below. The key equation needed here is the expression for an $n$-insertion energy eigenstate. This is  given as,
\be
\begin{split}
	\Phi_{\vk_1,\vk_2,...,\vk_n}&={1\over m^n}\int \left(\prod_{i=1}^{n}d^p\vy_i\right)~e^{i\sum_{i=1}^n\vk_i\cdot\vy_i} ~\phi(\vy_1)
	\bigg(\phi(\vy_2)-m\g(\vy_2-\vy_1)\bigg)
	\\&\times \bigg(\phi(\vy_3)-m\g(\vy_3-\vy_2)-m\g(\vy_3-\vy_1)\bigg).......
	\\&\times \bigg(\phi(\vy_n)-m\g(\vy_n-\vy_{n-1})-m\g(\vy_n-\vy_{n-2})....-m\g(\vy_n-\vy_1)\bigg)
\end{split}
\ee
One can easily convince oneself that this expression for $\Phi_{\vk_1,\vk_2,...,\vk_n}; n>\ns$ reduces to zero if one substitutes $\phi(\vy)=\sum_{i=1}^{\ns}m\g(\vy-\vx_i)$.

\subsection*{6. Inner-product computation}
\label{app:inner_pdt_computation}
We want to compute the inner product
\begin{equation}
\left( \Phi_{(\vk'_1,\vk'_2)}, \Phi_{(\vk_1,\vk_2)} \right)
=
\int D\phi \,
\Phi^{*}_{(\vk'_1,\vk'_2)}[\phi]\,
\int D\lambda \,
e^{i\int \lambda \phi}
\left(
\int d^p x\, e^{- i m \lambda(x)}
\right)^2
\Phi_{(\vk_1,\vk_2)}[\phi] ,
\end{equation}
for the case $N_* = 2$.
The two-particle state is given by
\begin{equation}
\Phi_{(\vk_1,\vk_2)}[\phi]
=
\frac{1}{m^2}
\int d^p y_1 d^p y_2\,
e^{i \vk_1 \cdot \vy_1}
e^{i \vk_2 \cdot \vy_2}
\phi(\vy_1)\phi(\vy_2)
-
\frac{1}{m}
\int d^p y\,
e^{i (\vk_1+\vk_2)\cdot \vy}
\phi(\vy).
\end{equation}
Then we need to compute the following,
\begin{equation}
    \begin{split}
        &\left({1\over m^2}\int d^py'_1d^py'_2~e^{-i\vk'_1\cdot\vy'_1}e^{-i\vk'_2\cdot\vy'_2} \phi(\vy'_1)\phi(\vy'_2)-\frac{1}{m}\int d^py' ~e^{-i(\vk'_1+\vk'_2)\cdot \vy'} \phi(\vy')\right)
        \\& \int D\lm e^{i\int \lm \phi} \left(\int d^px e^{-im\lm(\vx)}\right)^{2} \times
        \\&\hspace{2cm}\left({1\over m^2}\int d^py_1d^py_2~e^{i\vk_1\cdot\vy_1}e^{i\vk_2\cdot\vy_2} \phi(\vy_1)\phi(\vy_2)
        -\frac{1}{m}\int d^py ~e^{i(\vk_1+\vk_2)\cdot \vy} \phi(\vy)\right)
    \end{split}
\end{equation}

Thus, the inner product contains three types of contributions:
\begin{itemize}
\item quartic in $\phi$,
\item cubic in $\phi$,
\item quadratic in $\phi$.
\end{itemize}
To compute the quartic term, we use the following trick,
\begin{equation}
    \begin{split}
        \int D \phi &\int D\lm \phi(\vy') \phi(\vy)e^{i\int \lm \phi} \left(\int d^px e^{-im\lm(\vx)}\right)^{2}  \\&=
        \int D\lm D\phi \left( \frac{1}{i^2} {\g^2\over\g \lm(\vy)\g\lm(\vy')} e^{i\int d^py ~\lm(\vy)\phi(\vy)}\right)
	\left(\int d^px e^{-im\lm(\vx)}\right)^2
	\\&=\frac{1}{i^2}\int D\lm {\g^2\g \left[ \lm \right]\over \g \lm(\vy)\g\lm(\vy')} 	\left(\int d^px e^{-im\lm(\vx)}\right)^{2}
	\\&= 2 m^2+2m^2 V \g^p(\vy-\vy')
    \end{split}
\end{equation}
In the last step, we performed an integration by parts and then carried out the  $\lm$ integral using the $\g$ function. Here, $V=\int d^p y$ denotes the spatial volume of the $D-p$ brane.
By using this trick, we can compute the following  general identities,
\begin{align*}
&\int D\lambda D\phi\,
e^{i\int \lambda \phi}
\left(\int dx\, e^{- i m \lambda(x)}\right)^\ns
= V^\ns, \\[6pt]
&\int D\lambda D\phi\,
\phi(y_1)\phi(y_2)
e^{i\int \lambda \phi}
\left(\int dx\, e^{- i m \lambda(x)}\right)^\ns
\\&\hspace{4cm}=
m^2\left[
\ns V^{\ns-1}\delta(y_1-y_2)
+
\ns(\ns-1)V^{\ns-2}
\right], \\[6pt]
&\int D\lambda D\phi\,
\phi(y_1)\phi(y_2)\phi(y_3)
e^{i\int \lambda \phi}
\left(\int dx\, e^{- i m \lambda(x)}\right)^\ns
\\ \nonumber
&\hspace{4cm}
=
m^3\Big[
\ns V^{\ns-1}\delta(y_1-y_2)\delta(y_1-y_3)
\\&\hspace{4cm}+
\ns(\ns-1)V^{\ns-2}
\sum_{\text{pairs}}
\delta(y_i-y_j)+\ns(\ns-1)(\ns-2)L^{\ns-3}
\Big], \\[6pt]
&\int D\lambda D\phi\,
\phi(y_1)\phi(y_2)\phi(y_3)\phi(y_4)
e^{i\int \lambda \phi}
\left(\int dx\, e^{- i m \lambda(x)}\right)^\ns
\\
&
=
m^4
\Big[\ns V^{\ns-1}\delta_{12}\delta_{34}\delta_{14}
\\&\hspace{2cm}+
\ns(\ns-1)V^{\ns-2}
\big(
\delta_{13}\delta_{24}
+
\delta_{14}\delta_{23}
+
\delta_{14}\delta_{13}+
\delta_{24}\delta_{23}+
\delta_{12}\delta_{14}+
\delta_{12}\delta_{13}+
\delta_{12}\delta_{34}
\big)
\\&\hspace{1cm}
+\ns(\ns-1)(\ns-2)(V)^{\ns-3}\sum_{\text{pairs}}
\delta(y_i-y_j)+ \ns(\ns-1)(\ns-2)(\ns-3)(V)^{\ns-4}
\Big].
\end{align*}
Here 
$\delta_{ij} \equiv \delta(y_i-y_j)$. For our problem we set $N_*=2$. Now we can compute each term in the inner product. 
\subsection*{ Quadratic Term}
The quadratic contribution gives
\begin{equation}
2\left[
(2\pi)^2
\delta(\vk_1+\vk_2)
\delta(\vk'_1+\vk'_2)
+
V (2\pi)
\delta(\vk_1+\vk_2-\vk'_1-\vk'_2)
\right],
\end{equation}
\subsection*{ Cubic Terms}
The cubic contributions combine to give
\begin{equation}
\begin{split}
&-4\Big[
(2\pi) V\delta(\vk_1+\vk_2-\vk'_1-\vk'_2)+(2\pi)^2
\delta(\vk_1+\vk_2)
\delta(\vk'_1+\vk'_2)
\Big]
\\& -2 (2 \pi)^2\Big[\delta(\vk_1)
\delta(\vk_2-\vk'_1-\vk'_2)+\delta(\vk_2)
\delta(\vk_1-\vk'_1-\vk'_2)
\\&\hspace{4cm}+\delta(\vk'_1)
\delta(\vk_1+\vk_2-\vk'_2)+\delta(\vk'_2)
\delta(\vk_1+\vk_2-\vk'_1)\Big].
\end{split}
\end{equation}
\subsection*{ Quartic Term}
The quartic contribution produces
\begin{equation}
    \begin{split}
        &2\Big[(2 \pi) V \delta(\vk_1+\vk_2-\vk_1'-\vk_2')+(2 \pi)^2(\delta(\vk_1-\vk'_1)\delta(\vk_2-\vk'_2)
        \\&\hspace{2cm}+\delta(\vk_1-\vk'_2)\delta(\vk_2-\vk'_1)+\delta(\vk_1+\vk_2)\delta(\vk'_1+\vk'_2)
      +\delta(\vk_1-\vk'_1-\vk'_2)\delta(\vk_2)\\&\hspace{2cm}+\delta(\vk_2-\vk'_1-\vk'_2)\delta(\vk_1)+\delta(\vk_1+\vk_2-\vk'_2)\delta(\vk'_1)+\delta(\vk_1+\vk_2-\vk'_1)\delta(\vk'_2))\Big]
    \end{split}
\end{equation}
\subsection*{Final Result}
After adding all quadratic, cubic, and quartic pieces, all volume-dependent terms cancel.

We are left with the expected symmetric result
\begin{equation}
\boxed{
\left( \Phi_{(\vk'_1,\vk'_2)}, \Phi_{(\vk_1,\vk_2)} \right)
=
2\Big[
(2\pi)\delta(\vk_1-\vk'_1)
(2\pi)\delta(\vk_2-\vk'_2)
+
(1 \leftrightarrow 2)
\Big].
}
\end{equation}
This is precisely the correctly normalized two-particle inner product.

\section{Tachyon theory computations}
\label{app:tachyon_computations}
In this section, we perform computations in tachyon theory, using the Jacobian derived in 
\ref{sec:J_of_tachyon}. The techniques are similar to those used in computations of collective theory inner product \ref{app:inner_pdt_computation}, with the only difference being that the tachyon theory Jacobian is a sum over collective theory Jacobians of different particle number. Also, unlike in collective theory, one can use the commutator algebra of creation and annihilation operators to further simplify the computations. 
\subsection*{Inner product between states}
We will start by computing the norm of the state $\mc$.
\be
\label{eq:norm_of_mc_app}
\begin{split}
	\left(\mc,\mc\right)&=\int D\Pi ~J[\Pi] 1\cdot1
	\\&=\int D\Pi \sum_{n=0}^{\infty} {J_n[\Pi]\over n!}
\end{split}
\ee
Note that the $\Pi$ integral can be performed easily,
\be
\begin{split}
	\int D\Pi ~J_n[\Pi]&=\int D\Pi ~\int D\lm e^{i\int d^py~\lm(\vy)\Pi(\vy)} \left(\int d^px~e^{-im\lm(\vx)}\right)^n
	\\&=\int D\lm~ \g\left[\lm\right] \left(\int d^px~e^{-im\lm(\vx)}\right)^n
	\\&=\left(\int d^px~\right)^n
	\\&=V^n
\end{split}
\ee
Here $V=\int d^px$ is the spatial volume of the D-brane.  Substituting this result in \eqref{eq:norm_of_mc_app}, we get,
\be
\left(\mc,\mc\right)=\sum_{n=0}^{\infty} {V^n\over n!}=e^V
\ee
Using this we can compute the inner product between any two states in the tachyon theory Hilbert space. For example, consider the excited states $\Phi_{\vk_1}$ and $\Phi_{\vk_2}$. Their inner product is computed as,
\be
\begin{split}
	\left(\Phi_{\vk_1}, \Phi_{\vk_2}\right)&=\left( a^{\da}_{\vk_1}\mc, a^{\da}_{\vk_2}\mc \right)
	\\&=\left(\mc, a_{\vk_1} a^{\da}_{\vk_2} \mc \right)
	\\&=\left( \mc, \left( (2\pi)^p\g^p(\vk_1-\vk_2)+ a^{\da}_{\vk_2} a_{\vk_1} \right) \mc \right)
	\\&=\left(\mc,\mc\right) \left( (2\pi)^p \g^p(\vk_1-\vk_2)+(2\pi)^p\g^p(\vk_1)(2\pi)^p \g^p(\vk_2) \right).
\end{split}
\ee
Here we used the commutation relation between the creation and annihilation operators, as well as the action of $a_{\vk}$ on $\mc$ given in \eqref{eq:a_on_C}.  Similarly, one can compute the inner product between any two states in the tachyon theory Hilbert space. 

As an aside, recall that in the main text, we obtained the result that $\mc$ is to be interpreted as a coherent state of particles at rest. 
\be
|\mc\ra=e^{a^{\da}_{\vec{0}}}~|0\ra
\ee
As a sanity check, we can compute the norm of this state in a Klein-Gordon like analysis, using properties of the vacuum state $|0\ra$.
\be
\begin{split}
	\la\mc |\mc \ra&= \la 0| e^{a_0} e^{a^{\da}_0}|0\ra
	\\&= \sum_{n,n'}{1\over n!n'!}\la 0|a_0^n (a^{\da}_0)^{n'}|0\ra
	\\&=e^V
\end{split}
\ee
Here in the intemediate step, we have used $\la0|a_0^n(a^{\da}_0)^{n'}|0\ra=\g_{n,n'}n!V^n$. 
This precicely matches the computation \eqref{eq:norm_of_mc_app} performed using the Jacobian.   

\subsection*{Energy of the coherent state}
In the last section, we discussed how to compute the inner product between different states in the tachyon theory. Using similar methods we can also compute matrix elements of operators. In this section we will briefly discuss the computation of the expectation value of the Hamiltonian in the coherent state. \footnote{We are working with the Jacobian $J[\Pi]$, so the state $|\mc\ra=e^{a^{\da}_{0}}|0\ra$, and $a_{\vk}~\mc=(2\pi)^p\g(\vk)~\mc$. One can easily generalise this computation to the case when the Jacobian is changed to $J^{\mk}[\Pi]$.}
\be
\begin{split}
	\left(\mc,H\mc\right)&=\left(\mc,~\int {d^pk\over (2\pi)^p} \w_{\vk} ~a^{\da}_{\vk}a_{\vk}~\mc\right)
	\\&=\int {d^pk\over (2\pi)^p} \w_{\vk} \left(a_{\vk}\mc,a_{\vk}\mc \right)
	\\&=\int {d^pk\over (2\pi)^p} \w_{\vk} (2\pi)^p\g(\vk) (2\pi)^p\g(\vk)~ \left(\mc,\mc\right)
	\\&=m V \left(\mc,\mc\right)
\end{split}
\ee
Here in the last step, we used the fact that $(2\pi)^p\g(0)=V$. Notice that here we used only the algebraic properties of the creation operator and its action on the coherent state. One can generalise this computation to compute the matrix element of Hamiltonian between any two states.

\subsection*{Change of basis between $J[\Pi]$ and $J^{\mk}[\Pi]$}
\label{app:tachyon_basis}
In this section we will explain the change in the wavefunctional representation of states when we go from the Jacobian $J[\Pi]$ to $J^{\mk}[\Pi]$. Since the case of $J^{\mk}[\Pi]$ is more general and putting $\mk=1$ will give you $J[\Pi]$, we will start with the coherent state $e^{\mk a^{\da}_{0}}|0\ra$ represented as $\mc[\Pi]=1$. The creation/annihilation operators and the Jacobian have the following explicit form. 
\be
\begin{split}
	& a_{\vk}=\mathfrak{C}\int d^py~e^{-i\vk\cdot\vy} e^{m{\g\over \g \Pi(\vy)}}
	\\& a^{\da}_{\vk}={1\over \mathfrak{C}}{1\over m}\int d^py~e^{i\vk\cdot\vy} ~\Pi(\vy) e^{-m{\g\over \g \Pi(\vy)}}
	\\& J^{\mk}[\Pi]=\int D\lm ~e^{\int d^py\lm(\vy)\Pi(\vy)} \exp{\int d^px~\mk^2 e^{-im \lm(\vy)}}.
\end{split}
\ee  
With this definition the action of the annihilation operator on the state $\mc$ becomes,
\be
\label{eq:app_action_a_on_C}
a_{\vk}~\mc=2\pi \g(\vk)~\mathfrak{C}~\mc
\ee
This tells you that the state $\mc$ is the coherent state of particles at rest, $|\mc\ra=e^{\mk a^{\da}_{0}}|0\ra$. One can immediately ask about the other coherent states in the theory. These can be obtained by the action of exponential of creation operators on $\mc$, or by solving a differential equation analogous to \eqref{eq:app_action_a_on_C}. We will follow the latter route, and one can show that it is equivalent to the former. 

A coherent state $\beta[\Pi]$ (of particles at rest) is defined as the eigenstate of the annihilation operator having eigenvalue $\beta$.
\be
a_{\vk}~\beta[\Pi]=\beta~(2\pi)^p\g(\vk)~\beta[\Pi]
\ee
Using the definition of annihilation operator in \eqref{eq:app_action_a_on_C}, this can be written as,
\be
\begin{split}
	&\mk \int d^py~e^{-i\vk\cdot\vy} e^{m{\g\over \g \Pi(\vy)}}\beta[\Pi]=\beta~(2\pi)^p\g(\vk)~\beta[\Pi]
	\\&\Rightarrow \mk \int d^py~e^{-i\vk\cdot\vy} \beta[\Pi-m\g_y]=\beta~(2\pi)^p\g(\vk)~\beta[\Pi]
	\\& \Rightarrow \beta[\Pi-m\g_y]={\beta\over \mk} \beta[\Pi]
\end{split}
\ee
Here we have used the notation $\beta[\Pi-m\g_y]=\beta[\Pi(\vy~')-m\g(\vy~'-\vy)]$. One can now go to the Fourier space to solve for $\beta[\Pi]$.
\be
\begin{split}
	&\beta[\Pi]=\int D\lm~e^{i\int d^py~\lm(\vy)\Pi(\vy)}\beta[\lm]
	\\&\beta[\Pi-m\g_y]={\beta\over \mk} \beta[\Pi]
	\Rightarrow e^{-im\lm(\vy)} \beta[\lm]={\beta\over \mk}~\beta[\lm]
	\\&\Rightarrow \beta[\lm]=\g\bigg[ \lm(\vy)-{i\over m}\ln\left( {\beta\over \mk} \right) \bigg]
\end{split}
\ee
So the solution of the functional difference equation is that $\beta[\lm]$ is a delta functional. One can put this back into the Fourier transform representation to get $\beta[\Pi]$. One gets,
\be
\beta[\Pi]=\exp{\ln{\left({\beta\over \mk}\right)}{1\over m}\int d^py~\Pi(\vy)}
\ee
As a consistency check, note that putting $\beta=\mk$, one gets $\mc[\Pi]=1$ which is the coherent state having eigenvalue $\mk$, as claimed. One can further see that the logarithm diverges when $\beta\rightarrow 0$, which is in accordance with our conclusion in the main text that the vacuum is not accessible in tachyon theory.

Let us now come to the question of equivalence between the states in wavefunctional representations corresponding to $J[\Pi]$ and $J^{\mk}[\Pi]$. Using the analysis given above we note that the state $|\beta\ra=e^{\beta a^{\da}_{0}}|0\ra$ is given the following two representations, which are related. (We put a subscript to distinguish the representations)
\be
\begin{split}
	\\& \beta_{\mk}[\Pi]=\exp{\ln{\left({\beta\over \mk}\right)}{1\over m}\int d^py~\Pi(\vy)};~~{\text{with Jacobian}}~J^{\mk}[\Pi]
	\\& \beta_{1}[\Pi]=\exp{\ln{\left({\beta\over 1}\right)}{1\over m}\int d^py~\Pi(\vy)};~~{\text{with Jacobian}}~J[\Pi]
	\\& \Rightarrow \beta_{\mk}[\Pi]=\exp{\ln\left({1\over\mk}\right){1\over m}\int d^py~\Pi(\vy)} \beta_{1}[\Pi]
\end{split}
\ee
Since both representations correspond to the same state $|\beta\ra=e^{\beta a^{\da}_{0}}|0\ra$, the inner product computation should match\footnote{Since we chose the constants $\beta$ and $\mk$ as real, we can avoid complex conjugation}. 
\be
\begin{split}
	&\la\beta|\beta\ra=\int D\Pi~J[\Pi]\beta_1[\Pi]\beta_1[\Pi]=\int D\Pi~J^{\mk}[\Pi]\beta_{\mk}[\Pi]\beta_{\mk}[\Pi]
	\\&\Rightarrow \int D\Pi~J[\Pi]\beta_1[\Pi]\beta_1[\Pi]=\int D\Pi~J^{\mk}[\Pi]\beta_{\mk}[\Pi]\beta_{\mk}[\Pi]
	\\&\Rightarrow \int D\Pi~J[\Pi]\beta_1[\Pi]\beta_1[\Pi]=\int D\Pi~J^{\mk}[\Pi] \exp{2\ln\left({1\over\mk}\right){1\over m}\int d^py~\Pi(\vy)} \beta_{1}[\Pi]\beta_{1}[\Pi]
\end{split}
\ee
This implies that the two Jacobians, $J[\Pi]$ and $J^{\mk}[\Pi]$, should be related by a multiplicative factor. One can check that this indeed is the case.
\be
J^{\mk}[\Pi] \exp{2\ln\left({1\over\mk}\right){1\over m}\int d^py~\Pi(\vy)}=J[\Pi]
\ee
\emph{This makes it clear that the theory remains the same even if we change the Jacobian to $J^{\mk}$, only the wavefunctional representation of the state changes.} This is presicely why the wavefunctional $\mc[\Pi]=1$ correponds to $e^{a^{\da}_{\vk}}|0\ra$ when we use the Jacobian $J[\Pi]$ whereas if we change the Jacobian to $J^{\mk}[\Pi]$, the wavefunctional $\mc[\Pi]=1$ corresponds to $e^{\mk a^{\da}_0}|0\ra$.

\section{All order matching of the Hamiltonian}
In this section, we derive the relativistic Hamiltonian \eqref{eq:free_H} using the collective field theory approach, expanding the square root to all orders, and show that it has the same form as Klein-Gordon Hamiltonian \eqref{eq:H_Klein_Gordon} in terms of creation and annihilation operators. To simplify the notation, we set $\hbar = 1$. This choice does not affect the validity of the proof, since $\hbar$ can be restored at any point if needed.
\subsection*{Collective field theory Hamiltonian}

In collective field theory, the Hamiltonian \eqref{eq:free_H} can be expanded to all orders in the square root as following
\begin{align}
	H = \sum_{i = 1}^N \sqrt{\vec{p}_i .\vec{p}_i+m^2} = Nm + \sum_{\ell= 1  }^{ \infty } {\footnotesize {\begin{pmatrix} {1\over2} \\ \ell \end{pmatrix}} } {1\over m ^{2 \ell -1}}\sum_{i = 1 }^N (\vec{p}_i .\vec{p}_i)^{\ell}~.
    \label{all-orders1}
\end{align}
Then, we change variables from the particle momenta to the collective field, as in \eqref{eq:sum_p_i_der}, to obtain
\begin{align}
    \vec{p}_i = (- i ) \int d^p y ~m\vec{\nabla}_{\vec{x}_i}  \sum_{i'} \delta(\vec{y} - \vec{x}_{i'}) {\delta \over \delta \phi(\vec{y})}~.
\end{align}
The expressions for $\ell = 1,2$ in \eqref{all-orders1} have already been given explicitly in \eqref{eq:H_first3_terms}. We need to understand the structure for generic $\ell$.
To simplify the expression, we first define  
\begin{align}
	\tr_{\text{pair}} \{	\otimes ^{2 \ell }\vec{p}_i  \}  \equiv \tr_{\text{pair}} \{ \underbrace{\vec{p}_i \otimes \vec{p}_i  \otimes \ldots \otimes \vec{p}_i }_{\text{ $2 \ell$ times }}	\} =(\vec{p}_i .\vec{p}_i)^{\ell} ~.
\end{align}
This allows us to treat the expression without keeping track of the specific inner product structure in the Hamiltonian.
To understand the full form of $\otimes^{2\ell}\vec{p}_i$, we need to consider the chain rule for higher-order derivatives. The resulting structure is determined by the integer partitions of the number $(2\ell)$. Suppose we express the number $(2\ell
)$ as a sum of natural numbers chosen from the set $\{1,2,\ldots,2\ell\}$ as
\begin{align}
	b_1 + 2b_2 + 3b_3 + \ldots + (2\ell)b_{2\ell} = 2\ell~.
    \label{partition}
%	\qquad \text{with} \qquad
%	b_1 + b_2 + \ldots + b_{2\ell} = q .
\end{align}
Here, $\{b_1, b_2, \ldots, b_{2\ell } \}$ is a set of natural numbers that specifies a partition of $(2\ell)$. If one computes the product $\otimes^{2\ell}\vec{p}_i$ by applying the chain rule recursively, one obtains the following expression, given by a sum over all such partitions $\{b_1, b_2, \ldots, b_{2\ell} \}$  (with no vector indices are contracted on either side)

\begin{align}
	&\otimes ^{2 \ell }\vec{p}_i=(- i  )^{2 \ell }  {(2\ell)! } \times\\& \sum_{
		\substack{
			\{ b_1 , b_2 ,\ldots ,b_{2 \ell }\} \\
			 1b_1+ 2 b_2 +\ldots + (2\ell) b_{2 \ell} = 2 \ell}}  \prod_{j= 1} ^{2 \ell  } \Bigg( {1 \over b_j!(j!) ^{b_j}}\prod_{n=1}^{b_j} \Big( \int d^p y_{n} \big(m (\otimes ^{j}\vec{\nabla}_{\vec{x}_i} )\sum_{i' }\delta(\vec{y}_n - \vec{x}_{i'}) \big)  {\delta \over \delta \phi (\vec{y}_n)} \Big) \Bigg) ~.
		 \label{faadi1}
\end{align}

In the above expression the derivatives are distributed according to partitions of the number $(2 \ell)$. To explain this, let us consider an integer $j$ from the set $\{1,2,\ldots,2\ell\}$. For a given $j$, we write the factor $\Big( \int d^p y \big(m  \vec{\nabla}^j_{\vec{x}_i}\sum_{i' }\delta(\vec{y} - \vec{x}_{i'}) \big)  {\delta \over \delta \phi (\vec{y})} \Big)$, which corresponds the $j^{\text{th}}$ derivative of the density. However, this number $j$ can appear $b_j$ times in the partition \eqref{partition}. Hence we need to put a product of $b_j$ copies of the same expression.
%that satisfy $(b_1 + b_2 + \cdots + b_{2\ell}) = q$. 
The numerical factor ${(2\ell)!}\prod_{j=1}^{2\ell}\frac{1}{b_j!(j!)^{b_j}}$ is uniquely determined and appears naturally in the derivation. We note that the above expression has a structure analogous to the Faà di Bruno formula, which extends the chain rule to higher-order derivatives.

We can further simplify the expression by interchanging $\vec{\nabla}_{\vec{x}_i}$ and $\vec{\nabla}_{\vec{y}_n}$, and then performing an integration by parts. The minus signs that arise in these two steps cancel each other.
\begin{align}
	\otimes ^{2 \ell }\vec{p}_i = (- i  )^{2 \ell }  {(2\ell)! }  \sum_{
		\substack{
			\{ b_1 , b_2 ,\ldots ,b_{2 \ell }\} \\
			\sum_{n=1}^{2 \ell }(n b_n) = 2 \ell}}     \prod_{j= 1} ^{2 \ell  } \Bigg( {m^{b_j} \over b_j!(j!) ^{b_j}} \prod_{n=1}^{b_j} \Big( \int d^p y_{n}  \delta(\vec{y}_n - \vec{x}_{i})  \big( (\otimes^{j}\vec{\nabla}_{\vec{y}_n})  {\delta \over \delta \phi (\vec{y}_n)} \big)\Big) \Bigg) ~.
\end{align}
We can perform most of the $y_n$ integrals to write everything in terms of a single $\int d^p{y}$ integral
\begin{align}
		\otimes ^{2 \ell }\vec{p}_i = (- i  )^{2 \ell } (2 \ell)!  \sum_{
			\substack{
				\{ b_1 , b_2 ,\ldots ,b_{2 \ell }\} \\
				\sum_{n=1}^{2 \ell }(n b_n) = 2 \ell }} 
                % {(2\ell)! \over b_1 ! b_2! \ldots b_{2 \ell}!}  
                \prod_{j= 1} ^{2 \ell } \Bigg( {1  \over b_j!} \int d^p y \delta( \vec{y} - \vec{x}_i)\Big({m \over j! } \big((\otimes ^j\vec{\nabla }_{\vec{y}} ) {\delta \over \delta \phi (\vec{y})}  \big) \Big)^{b_j} \Bigg)  ~.
\end{align}

Now we use the above results to compute the Hamiltonian \eqref{all-orders1}, which takes the form: 
\begin{align}
	H = 
	 & Nm \\
+&	 \sum_{\ell  = 1  }^{ \infty } {\footnotesize {\begin{pmatrix} {1\over2} \\ \ell \end{pmatrix}}}  { (- i  )^{2 \ell } (2\ell)! \over m ^{2 \ell -1}}\sum_{i = 1 }^N  \sum_{
	\substack{
		\{ b_1 , b_2 ,\ldots ,b_{2 \ell }\} \\
		\sum_{n=1}^{2 \ell }(n b_n) = 2 \ell } }   \tr_{\text{pair}} \Big\{ \prod_{j= 1} ^{2 \ell } {1 \over b_j!} \int d^p y \delta( \vec{y} - \vec{x}_i)\Big({m \over j! } \big((\otimes^j\vec{\nabla }_{\vec{y}} ) {\delta \over \delta \phi (\vec{y})}  \big) \Big)^{b_j} \Big\} ~.
\end{align}
We sum over the $N$ particles and write $\phi(\vec{y}) = m \sum_{i=1}^{N} \delta(\vec{y} - \vec{x}_i)$. 
Therefore, we can write the Hamiltonian as
\begin{equation}
	\begin{split}
		&H = N m+ \\
		& \sum_{\ell=1}^{\infty } {\footnotesize \begin{pmatrix} {1\over2} \\ \ell  \end{pmatrix} }  \big( -{ i \over m  } \big)^{2 \ell } (2\ell)! \int d^p y~ \phi(\vec {y}) \sum_{\substack{\{ b_1,b_2, \ldots, b_{2 \ell} \} \\ \sum_{n=1}^{2 \ell }(n b_n) = 2 \ell} }  \tr_{\text{pair}} \Big\{ \prod_{j= 1} ^{2 \ell } { 1\over b_j!}\Big ({m \over j! }{(\otimes^j \vec{\nabla }_{\vec{y}})~{\delta \over \delta \phi (\vec{y})}  } \Big)^{b_j} \Big\} ~.
        \label{col_gen}
	\end{split} 
\end{equation}
By setting $\ell = 1, 2$ in the above expression, one can verify that it exactly reproduces \eqref{eq:H_first3_terms}. This demonstrates that the numerical factors appearing in \eqref{eq:H_first3_terms} for $\ell = 1, 2$ are not accidental, but arise naturally from the partition structure of the integer $(2\ell)$.
\subsection*{Hamiltonian in terms of creation and annihilation operators.}
We claim that the collective field theory Hamiltonian \eqref{col_gen} is of the same form as the Klein–Gordon Hamiltonian. To show this, we start with
\begin{align}
	H = \int {d^p {k}\over (2 \pi)^p} \omega_{\vec{k}} a^{\dagger}_{\vec{k}} a_{\vec{k}}~.
\end{align}
Let us now express the Hamiltonian explicitly using the forms of the annihilation and creation operators given in \eqref{eq:a_and_a_da}.
\begin{equation}
	\begin{split}
		H =&\int {d ^p {k}\over (2 \pi)^p} \omega_{\vec{k}} a^{\dagger}_{\vec{k}} a_{\vec{k}}
		%line1
		\\
		=& \int {d ^p\vec{k}\over (2 \pi)^p} \sqrt{(  \vec{k} )^2 +m^2}\Big( {1 \over m} \int d^p y e^{i \vec{k}. \vec{y} } {\Pi (\vec{y}) }  e^{-{ m } {\delta \over \delta \Pi (\vec{y})}}\Big) \Big(  \int  d^{p}\vec{y'} e^{-i \vec{k}.\vec{y'} }   e^{+{m } {\delta \over \delta \Pi (\vec{y}')}}\Big) 
		%line2
		\\
		=&\int d^{p}y  \phi (\vec{y}) + \int {d ^p k \over (2 \pi)^p } \Big[ \Big( \sum_{\ell = 1  }^{ \infty } {\footnotesize \begin{pmatrix} {1\over2} \\ \ell  \end{pmatrix} }  { ( ^2\vec{k}^2)^{ \ell}\over m ^{2 \ell } }  \Big) \Big(  \int d^p y e^{i \vec{k}. \vec{y} } {\Pi (\vec{y}) }  e^{-{ m } {\delta \over \delta \Pi (\vec{y})}}\Big) %\\
		%& \qquad  \qquad\qquad \qquad\qquad\qquad\qquad \qquad\qquad
        \Big(  \int  d^{p}\vec{y'} e^{-i \vec{k}.\vec{y'} }   e^{+{m } {\delta \over \delta \Pi (\vec{y}')}}\Big) \Big] 
		%line3
		\\
		= & \int d^{p}y  \phi (\vec{y})+ \int {d ^pk \over (2 \pi)^p }\sum_{\ell = 1  }^{ \infty } \Big[  {\footnotesize \begin{pmatrix} {1\over2} \\ \ell  \end{pmatrix} }  {1\over m ^{2 \ell } } \Big( \int d^p y e^{i \vec{k}. \vec{y} } {\Pi (\vec{y}) }   e^{+{m } {\delta \over \delta \Pi (\vec{y})}} \Big)\\
		& \qquad \qquad  \qquad \qquad \qquad \qquad \qquad    \Big(  \int  d^{p}\vec{y'}  \big(  ((i  )^2  \vec{\nabla}_{\vec{y'}}. \vec{\nabla}_{\vec{y'}} )^\ell e^{-i \vec{k}.\vec{y'} }  \big) e^{+{m } {\delta \over \delta \Pi (\vec{y}')}}\Big) \Big] 
		%line4
		\\
		= & \int d^{p}y  \phi (\vec{y})+ \int {d ^pk \over (2 \pi)^p }\sum_{\ell = 1  }^{ \infty } \Big[  {\footnotesize \begin{pmatrix} {1\over2} \\ \ell  \end{pmatrix} }  {1\over m ^{2 \ell } } \Big( \int d^p y e^{i \vec{k}. \vec{y} } {\Pi (\vec{y}) }   e^{+{m } {\delta \over \delta \Pi (\vec{y})}} \Big)\\
		& \qquad \qquad  \qquad \qquad \qquad \qquad \qquad    \Big(  \int  d^{p}\vec{y'}    e^{-i \vec{k}.\vec{y'} }  \big(  ((-i  )^2  \vec{\nabla}_{\vec{y'}}. \vec{\nabla}_{\vec{y'}} )^\ell e^{+{m } {\delta \over \delta \Pi (\vec{y}')}} \big) \Big) \Big] 
	\end{split}
\end{equation}
In the above steps, we first expand the relativistic momenta, then substitute the momenta inside the integral of $a_{\vec{k}}$ as derivatives, and finally perform an integration by parts. To evaluate the expression, we need to use the Fa\`a di Bruno formula of differentiation
\begin{align}
	(\otimes^{2 \ell}\vec{\nabla}_{\vec{y}} ) e^{+{m } {\delta \over \delta \Pi (\vec{y})}}  =  e^{+{m } {\delta \over \delta \Pi (\vec{y})}} \sum_{
			\substack{
				\{ b_1 , b_2 ,\ldots ,b_{2 \ell }\} \\
				\sum_{n=1}^{2 \ell }(n b_n) = 2 \ell }} {(2\ell)! \over b_1 ! b_2! \ldots b_{2 \ell}!}  \prod_{j= 1} ^{2 \ell }\Big ({m \over j! }{(\otimes^j \vec{\nabla }_{\vec{y}}) ~{\delta \over \delta \Pi (\vec{y})}  } \Big)^{b_j}~.
    \label{faadi2}
\end{align} 
This expression has a structure similar to that of \eqref{faadi1}. To make this structure explicit, we consider writing the number $2\ell$ as a sum of natural numbers chosen from the set $\{ 1,2,\ldots,2\ell \}$ as following
\begin{align}
	2\ell = b_1 + 2b_2 + 3b_3 + \ldots + (2\ell) b_{2\ell}  ~.
\end{align}
Here, $\{b_1, b_2, \ldots, b_{2\ell}\}$ is a set of natural numbers describing a partition. When the operator is applied $2\ell$ times to a function, the resulting terms can be grouped according to the partitions of the integer $2\ell$. In the above notation, this corresponds to all possible choices of the set $\{b_1, b_2, \ldots, b_{2\ell} \}$. The operator we have can be expressed by taking a pairwise trace defined by
\begin{align}
	\tr_{\text{pair}} 	\Big\{ (\otimes ^{2 \ell } \vec{\nabla }_{\vec{y}})  e^{+{m } {\delta \over \delta \Pi (\vec{y})}} \Big\} =   (  \vec{\nabla}_{\vec{y}}.\vec{\nabla}_{\vec{y}}  )^\ell e^{+{m } {\delta \over \delta \Pi (\vec{y})}}~.
\end{align}
Hence we write
\begin{align}
	(\vec{\nabla}_{\vec{y}}.\vec{\nabla}_{\vec{y}}  )^\ell e^{+{m } {\delta \over \delta \Pi (\vec{y})}} =  e^{+{m } {\delta \over \delta \Pi (\vec{y})}}  (2\ell)!\sum_{
			\substack{
				\{ b_1 , b_2 ,\ldots ,b_{2 \ell }\} \\
				\sum_{n=1}^{2 \ell }(n b_n) = 2 \ell }}  \tr_{\text{pair}} \Big\{ \prod_{j= 1} ^{2 \ell } {1 \over b_j!}\Big ({m \over j! }{(\otimes^j\vec{\nabla }_{\vec{y}}) ~{\delta \over \delta \Pi (\vec{y})}  } \Big)^{b_j} \Big\} ~.
\end{align}
The Hamiltonian becomes
\begin{align}
		H =& \int d^p y \Pi(\vec{y}) \\
		+& \sum_{\ell=1}^{\infty } {\footnotesize \begin{pmatrix} {1\over2} \\ \ell  \end{pmatrix} }  \big( -{ i \over m  } \big)^{2 \ell } (2\ell)! \int d^p y \Pi(\vec {y}) \sum_{
			\substack{
				\{ b_1 , b_2 ,\ldots ,b_{2 \ell }\} \\
				\sum_{n=1}^{2 \ell }(n b_n) = 2 \ell }}   \tr_{\text{pair}} \Big\{ \prod_{j= 1} ^{2 \ell } {1 \over b_j!}\Big ({m \over j! }{(\otimes^j\vec{\nabla }_{\vec{y}}) ~{\delta \over \delta \Pi (\vec{y})}  } \Big)^{b_j} \Big\} ~.
	\end{align}
\noindent
We find that this expression is identical to the collective field theory Hamiltonian upon identifying the tachyon effective field $\Pi(\vec{y})$ with the collective field $\phi(\vec{y})$ and imposing the constraint $\int d^p y \Pi(\vec{y}) = N m$.

\bibliographystyle{JHEP}
\bibliography{biblio}
\end{document}